\documentclass{aa}

\usepackage{graphicx}
\usepackage[varg]{txfonts}
\usepackage{color}
\usepackage{xspace}

\newcommand{\rnoff}{\color{black}\xspace}

\usepackage{natbib}
\usepackage{longtable,lscape}
\usepackage{rotating}
\usepackage[normalem]{ulem}

\def\kms{km~s$^{-1}$}

\def\teff{$T_\mathrm{eff}$}

\newcommand{\BP}{$G_{\rm BP}$}
\newcommand{\RP}{$G_{\rm RP}$}
\newcommand{\RVS}{$G_{\rm RVS}$}
\newcommand{\Gaia}{\emph{Gaia}}
\newcommand{\uas}{$\mu$as}

\usepackage{verbatim}
\usepackage{color}

\bibpunct{(}{)}{;}{a}{}{,} 

\newcommand{\savefootnote}[2]{\footnote{\label{#1}#2}}
\newcommand{\repeatfootnote}[1]{\textsuperscript{\ref{#1}}}
\begin{document}

   \title{{\Gaia} photometry for white dwarfs\thanks{Tables~\ref{tab:photDA}~--~\ref{tab:photMix} are only available in electronic form at the CDS via anonymous ftp to cdsarc.u-strasbg.fr (130.79.128.5) or via http://cdsweb.u-strasbg.fr/cgi-bin/qcat?J/A+A/.}}


   \author{
          J.~M.~Carrasco\inst{1} \and
          S.~Catal\'an\inst{2,4} \and
          C.~Jordi\inst{1} \and
          P.-E.~Tremblay\inst{3} \and
          R.~Napiwotzki\inst{4} \and
          X.~Luri\inst{1} \and
          A.C.~Robin\inst{5} \and
              P.~M.~Kowalski\inst{6} 
          }

\offprints{J.M.~Carrasco (carrasco@am.ub.es)}

   \institute{
   Departament d'Astronomia i Meteorologia, Institut del Ci\`encies del Cosmos (ICC), Universitat de Barcelona (IEEC-UB), c/ Mart\'{\i} i Franqu\`es, 1, 08028 Barcelona, Spain, 
\email{carrasco@am.ub.es, carme@am.ub.es, xluri@am.ub.es}  
                \and
Department of Physics, University of Warwick, Gibbet Hill Road, Coventry, CV4 7AL, UK, \email{s.catalan-ruiz@warwick.ac.uk} 
          \and
Hubble Fellow, Space Telescope Science Institute, 700 San Martin
Drive, Baltimore, MD, 21218, USA, \email{tremblay@stsci.edu}
         \and
Centre for Astrophysics Research, University of Hertfordshire, Hatfield, AL10 9AB, UK, \email{r.napiwotzki@herts.ac.uk}         
                  \and
Universit\'e de Franche-Comt\'e, Institut Utinam, UMR CNRS 6213, OSU Theta,  BP1615, 25010 Besan\c{c}on Cedex, France, \email{annie@obs-besancon.fr}
\and
Institute of Energy and Climate Research (IEK-6), Forschungszentrum J\"{u}lich,
52425 J\"{u}lich, Germany, \email{p.kowalski@fz-juelich.de}
             }

   \date{\today, Received / Accepted}


  \abstract
    {
White dwarfs
can be used to study the structure and evolution of the Galaxy by analysing their luminosity function and initial mass function.
Among them, the very cool white dwarfs provide the information for the early ages of each population. Because white dwarfs are intrinsically faint 
only the nearby ($\sim 20$~pc) sample is reasonably complete. The {\Gaia} space mission will drastically increase the sample of known white dwarfs through its 5--6 years survey of the whole sky up to magnitude $V=20$--$25$.
    }
   {We provide a characterisation of {\Gaia} photometry for white dwarfs to better prepare for the analysis of the scientific output of the mission. Transformations between some of the most common photometric systems and {\Gaia} passbands are derived.
   We also give estimates of the number of white dwarfs of the different galactic populations that will be observed.
   }
   {Using synthetic spectral energy distributions and the most recent {\Gaia} transmission curves, we computed colours of three different types of white dwarfs (pure hydrogen, pure helium, and mixed composition with H/He$=0.1$). With these colours we derived transformations to other common photometric systems (Johnson-Cousins, Sloan Digital Sky Survey, and 2MASS). We also present numbers of white dwarfs predicted to be observed by {\Gaia}.
   }
   {We provide relationships and colour-colour diagrams among different photometric systems to allow the prediction and/or study of the {\Gaia} white dwarf colours. 
We also include estimates of the number of sources expected in every galactic population and with a maximum parallax error.
 {\Gaia} will increase the sample of known white dwarfs tenfold to about 200\,000. {\Gaia} will be able to observe thousands of very cool white dwarfs for the first time, which will greatly improve our understanding of these stars and early phases of star formation in our Galaxy.
}
   {}
   
  \maketitle
%

\rnoff
\section{Introduction}

White dwarfs (WDs) are the final remnants of low- and intermediate-mass stars. About 95\% of the main-sequence (MS) stars will end their evolutionary pathways as WDs and, hence, the study of the WD population provides details about the late stages of the life of the vast majority of stars. Their evolution can be described as a simple cooling process, which is reasonably well understood \citep{sal00,fon01}. WDs are very useful objects to understand the structure and evolution of the Galaxy because they have an
imprinted memory of its history \citep{ise01,lie05}. The WD luminosity function (LF) gives the number of WDs per unit volume and per bolometric magnitude \citep{winget87,ise98}. From a comparison of observational data with theoretical LFs important information on the Galaxy \citep{winget87} can be obtained (for instance, the age of the Galaxy, or the star formation rate). Moreover, the initial mass function (IMF) can be reconstructed from the LF of the relic WD population, that is, the halo/thick disc populations. The oldest members of these populations are cool high-mass WDs, which form from high-mass progenitors that evolved very quickly to the WD stage.

Because most WDs are intrinsically faint, it is difficult to detect them, and a complete sample is currently only available at very close distances. \cite{hol08} presented a (probably) complete sample of local WDs within 13~pc and demonstrated that the sample becomes incomplete beyond that distance. More recently, \citet{giammichele12} provided a nearly complete sample up to 20~pc. Completeness of WD samples beyond 20\,pc is still very unsatisfactory even though the number of known WDs has considerably increased thanks to several surveys. For instance, the Sloan Digital Sky Survey (SDSS), with a limiting magnitude of $g'=19.5$ \citep{fukugita96} and covering a quarter of the sky, has substantially increased the number of spectroscopically confirmed WDs\footnote{SDSS catalogue from \cite{eis06} added 9316 WDs to the 2249 WDs in \cite{mccook99}. 
A more recent publication \citep{kleinman13}, using data from DR7 release, almost doubles that amount, with of the order of $20\,000$ WDs.}.
This has allowed several statistical studies \citep{eis06}, and the consequent improvement of the WD LF and WD mass distribution \citep{kleinman13,tre11,kre09,deg08,hu07,har06}. However, the number of very cool WDs and known members of the halo population is still very low. A shortfall in the number of WDs below $\log (L/L_{\odot})=-4.5$ because of the finite age of the Galactic disc, called luminosity cut-off, was first observed in the eighties \citep[e.g.][]{lie80,winget87}. The {\Gaia} mission will be extremely helpful in detecting WDs close to the luminosity cut-off and even fainter, which is expected to improve the accuracy of the age determined from the WD LF.

{\Gaia} is the successor of the ESA Hipparcos astrometric mission \citep{hipparcos} and increases its capabilities drastically, both in precision and in number of observed sources, offering the opportunity to tackle many open questions about the Galaxy (its formation and evolution, as well as stellar physics). {\Gaia} will determine positions, parallaxes, and proper motions for a relevant fraction of stars ($10^9$ stars, $\sim1\%$ of the Galaxy). This census will be complete for the full sky up to $V=20-25$~mag (depending on the spectral type) with unprecedented accuracy (\citealt{perryman01}, \citealt{prusti11}). Photometry and spectrophotometry will be obtained for all the detected sources, while radial velocities will be obtained for the brightest ones (brighter than about 17$^{\mathrm{th}}$ magnitude). Each object in the sky will transit the focal plane about 70 times on average.

The {\Gaia} payload consists of three instruments mounted on a single optical bench: the astrometric instrument, the spectrophotometers, and one high-resolution spectrograph. The astrometric measurements
will be unfiltered to obtain the highest possible signal-to-noise ratio. The mirror coatings and CCD quantum efficiency define a broad (white-light) passband named $G$ \citep{jor10}. The basic shape of the spectral energy distribution (SED) of every source will be obtained by the spectrophotometric instrument, which will provide low-resolution spectra in the blue (330--680~nm) and red (640--1000~nm), BP and RP spectrophotometers, respectively \citep[see][for a detailed description]{jor10}. The BP and RP spectral resolutions are a function of wavelength. The dispersion is higher at short wavelengths. Radial velocities will also be obtained for more than 100 million stars 
through Doppler-shift measurements from high-resolution spectra ($R\sim 5000$--$11\,000$) obtained in the region of the IR Ca triplet around 860~nm by the Radial Velocity Spectrometer (RVS). Unfortunately, most WDs will show only featureless spectra in this region. The only exception are rare subtypes that display metal lines or molecular carbon bands (DZ, DQ, and similar).

The precision of the astrometric and photometric measurements will depend on the brightness and spectral type of the stars. At $G=15$~mag the end-of-mission precision in 
parallaxes will be $\sim25$~{\uas}. At $G=20$~mag the final precision will
drop to $\sim 300$~{\uas}, while for the brightest stars ($6<G<12$~mag) it
will be $\sim10$~{\uas}. The end-of-mission $G$-photometric performance will
be at the level of millimagnitudes. For radial velocities the
precisions will be in the range 1 to 15 {\kms} depending on the brightness and
spectral type of the stars \citep{katz11}. For a detailed description of
performances see the {\Gaia} website\savefootnote{foot:performances}{http://www.cosmos.esa.int/web/gaia/science-performance}.

An effective exploitation of this information requires a clear understanding of the potentials and limitations of {\Gaia} data. This paper aims to provide information to researchers on the WD field to 
obtain the maximum scientific gain from the {\Gaia} mission.

\cite{jor10} presented broad {\Gaia} passbands and colour-colour relationships for MS and giant stars, allowing the
prediction of {\Gaia} magnitudes and uncertainties from Johnson-Cousins \citep{bessell90} and/or SDSS \citep{fukugita96} colours. That article used the BaSeL-3.1 \citep{westera} stellar spectral library, which includes SEDs with $-1.0 < \log g < 5.5$, and thus excluded the WD regime ($7.0 < \log g < 9.0$). The aim of the present paper is to provide a similar tool for characterizing {\Gaia} observations of WDs. For that purpose, we used the most recent WD synthetic SEDs \citep[][see Sect.~\ref{sec:atmospheres}]{kil09,kil10,tre11,ber11} with different compositions to simulate {\Gaia} observations. In Sect.~\ref{sec:WDinGaia} we describe the conditions of WD observations by {\Gaia} (the obtained {\Gaia} spectrophotometry, the WD limiting distances, expected error in their parallaxes,
etc.). In Sect.~\ref{sec:transformations} we provide the colour-colour transformations between {\Gaia} passbands and other commonly used photometric systems like the Johnson-Cousins \citep{bessell90}, SDSS \citep{fukugita96}, and 2MASS \citep{cohen03}. In Sect.~\ref{sec:classification} relationships among {\Gaia} photometry and atmospheric parameters are provided.
Estimates of the number of WDs that {\Gaia} will potentially observe based on simulations by \cite{nap09} and {\Gaia} Universe Model Snapshot (GUMS, \citealt{rob12}) are provided in Sect.~\ref{sec:simulation}. 
Finally, in Sect.~\ref{sec:conclusions} we finish with a summary and conclusions.


\section{Model atmospheres}
\label{sec:atmospheres}

To represent the SED of WDs, we used grids of pure hydrogen (pure-H), pure helium (pure-He), and also mixed-composition models (H/He$=0.1$) with $7.0 < \log g < 9.0$ in steps of $0.5$~dex. These SEDs were computed from state-of-the-art model atmospheres and were verified in recent photometric and spectral analyses of WDs \citep{kil09,kil10,tre11,ber11}. 

\begin{figure}
\begin{center}
\includegraphics[angle=270,width=0.4\textwidth]{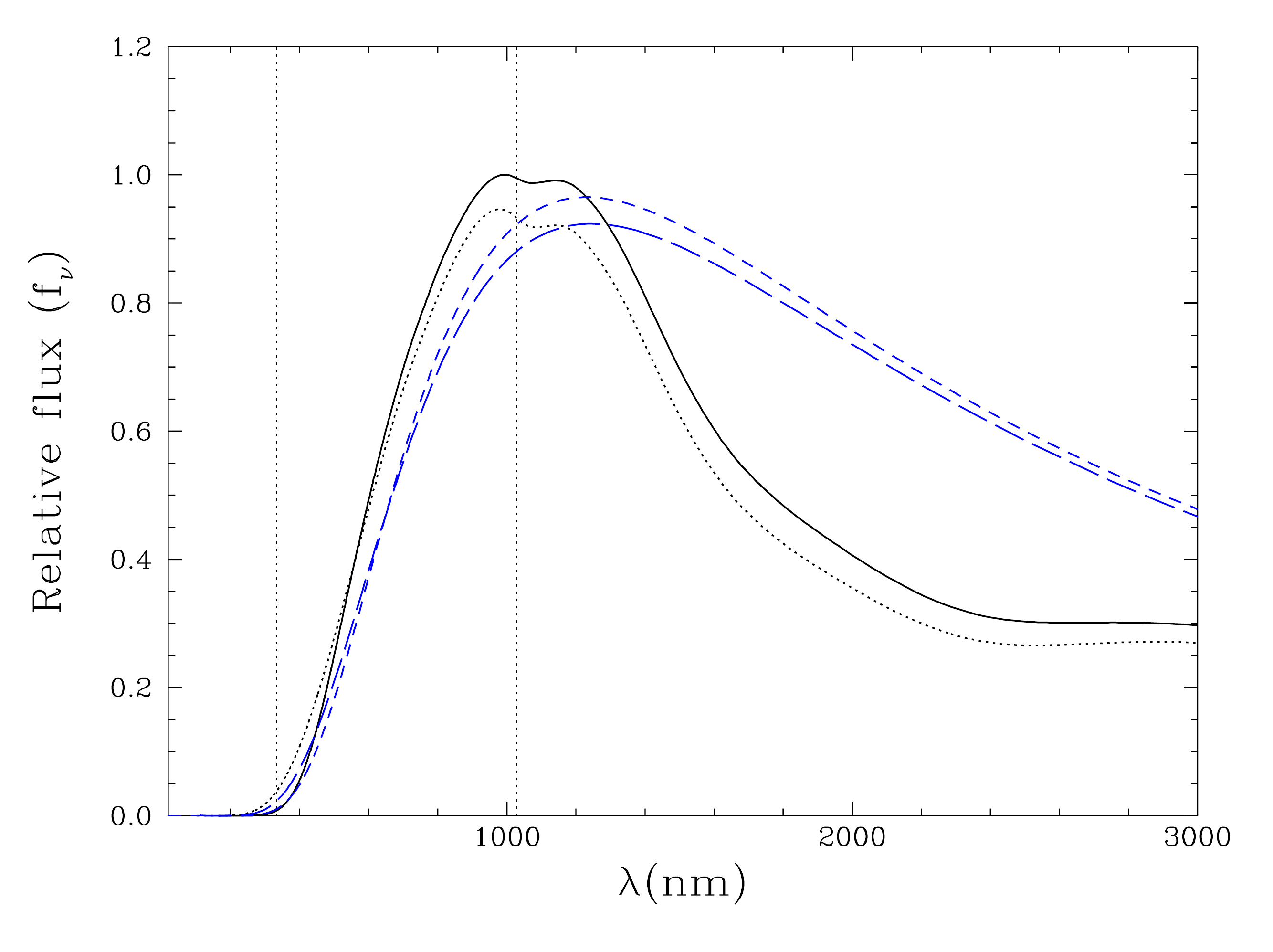}
\caption{Comparison of model SEDs with {\teff}~$=4000$~K and $\log g =
  8.0$. The pure-H models (in black) are computed with and without the
  Lyman-$\alpha$ opacity (solid and dotted lines).  The pure-He
  models (in blue) are drawn from the sequence using the equation of state of
  \citet{bergeron95} and using the improved high-density physics
  of \citealt{kowalski07} (medium and long dashed lines). The
  dotted vertical lines show the limits where the {\Gaia} transmission is
  higher than 5\%.
\label{fig:fXX}}
\end{center}
\end{figure}

The pure-H models are drawn from \citet{tre11} and cover the range\footnote{The temperature steps for the \citet{tre11} grid are $\Delta${\teff}~$= 250$~K for {\teff}~$< 5500$~K, $\Delta${\teff}~$= 500$~K for $6000$~K $<$~{\teff}~$< 17\,000$~K, $\Delta$~{\teff}~$= 5000$~K for $20\,000$~K $<$~{\teff}~$< 90\,000$~K, and $\Delta${\teff}~$= 10\,000$~K for {\teff}~$> 90\,000$~K.} of $1500$~K $<$~{\teff}~$< 140\,000$~K. These models were recently improved in the cool-temperature regime with updated collision-induced absorption (CIA) opacities \citep[see the discussion in][]{tremblay07}. In the present paper we also account for the opacity generated by the red wing of Lyman-$\alpha$ computed by \cite{kowalski06}. This opacity significantly changes the predicted flux in the $B$ passband at very cool temperatures ({\teff}~$\lesssim 6000$~K). Models with this opacity source have been successful in reproducing SEDs of many cool WDs \citep{kowalski06,kil09b,kil09,kil10b,durant2012}.
The colours for these improved models are accessible from Pierre Bergeron's webpage\footnote{http://www.astro.umontreal.ca/~bergeron/CoolingModels/}.
In Fig.~\ref{fig:fXX}, we compare the predicted spectra of cool ($4000$ K) pure-H atmosphere WDs using the former grid of \citet{tre11} with the present grid taking into account the Lyman-$\alpha$ opacity. 

We additionally used pure-He models drawn from \citet{ber11}, which cover a range\footnote{The temperature steps for the \citet{ber11} grid are $\Delta${\teff}~$= 250$~K for {\teff}~$< 5500$~K, $\Delta${\teff}~$=500$~K for $6000$~K $<$~{\teff}~$< 10\,000$~K and $\Delta${\teff}~$= 1000$~K for {\teff}~$> 10\,000$~K.} of $3500$~K $<$~{\teff}~$< 40\,000$~K.\rnoff The cooler pure-He DC\savefootnote{footnote:DC}{DC are WDs with featureless continuous spectra, which can have a pure-H, pure-He, or mixed atmosphere composition.} models
are described in more detail in \citet{kil10}, and their main feature is the non-ideal equation of state of \citet{bergeron95}. In recent years, new pure-He models of 
\cite{kowalski07}, which include improved description of non-ideal physics and chemistry of dense helium, have also been used in the analysis of the data \citep{kil09}. These models include a number of improvements in the description of pure-He atmospheres of very cool WDs. These include refraction \citep{kowalski04}, non-ideal chemical abundances of species, and improved models of Rayleigh scattering and He$^-$ free-free opacity \citep{iglesias02,kowalski07}. 
The SEDs of pure-He atmospheres are close to those of black bodies,
since the He$^-$ free-free opacity, which has a low dependence on wavelength,
becomes the dominant opacity source in these models. In Fig.~\ref{fig:fXX}, we
also present models at $4000$~K drawn from the two pure-He sequences. The blue flux in \cite{kowalski07} is slightly higher than in \cite{bergeron95} since the contribution of the Rayleigh scattering is diminished.
Note that the IR
wavelength domain, which is not covered by {\Gaia} detectors, is the range in which larger differences between the pure-H or pure-He composition are present. These differences result from strong CIA absorption by molecular hydrogen in hydrogen-dominated atmospheres.

\begin{figure}
   \centering
\includegraphics[width=0.4\textwidth]{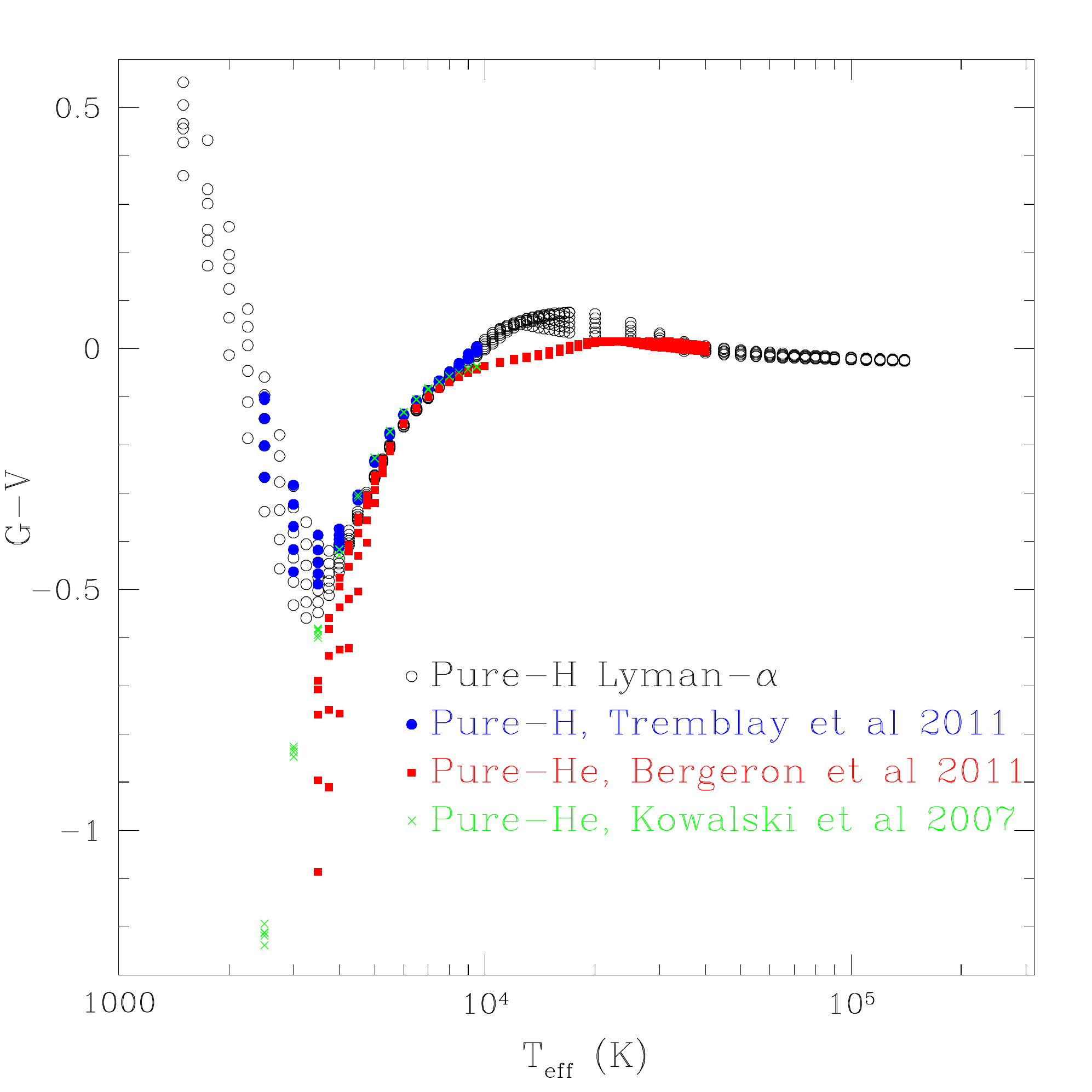}
\caption{($G-V$) vs. {\teff} diagram for WD SEDs  (red filled squares and black opened circles) compared with pure-H SEDs without the Lyman-$\alpha$ opacity
(blue filled circles) and pure-He SEDs from \citealt{kowalski07} (green star symbols).
  }
  \label{fig:TremblayvsKowalski}
\end{figure}

Figure~\ref{fig:TremblayvsKowalski} shows that in spite of the different
physics present in all these models, the colours (e.g. $G-V$) look quite
similar for the different SED libraries. For this reason, and because the purpose
of this paper is not to discuss the differences among
the WD SED libraries, but to provide a way to predict how WDs will be observed
by {\Gaia}, in Sect.~\ref{sec:transformations} we only included the
transformations derived using one sequence for each composition (\citealt{tre11} with Lyman-$\alpha$ for pure-H and \citealt{ber11} for pure-He).

The mixed model atmospheres used here cover a range of $2500$~K $<$~{\teff}~$< 6000$~K and are taken from \citet{kil10}. In the following, we use an abundance ratio of H/He $=0.1$ as a typical example for the composition of known mixed WDs \citep{kil09,kil10,leggett11,giammichele12}. It has to be kept in mind that because of the nature of the CIA opacities, which are dominant in the near-IR and IR in this mixed regime, the predicted spectra can vary considerably for different H/He ratios with the same {\teff} and $\log g$ \citep[see Fig.~10 of][]{kil10}. The colour space covered by our H/He $= 0.1$ sequence and the pure-H and He sequences illustrates the possible colour area where mixed-composition WDs can be found.

\section{WDs as seen by {\Gaia}}
\label{sec:WDinGaia}

\begin{figure}
   \centering
   \includegraphics[width=0.48\textwidth]{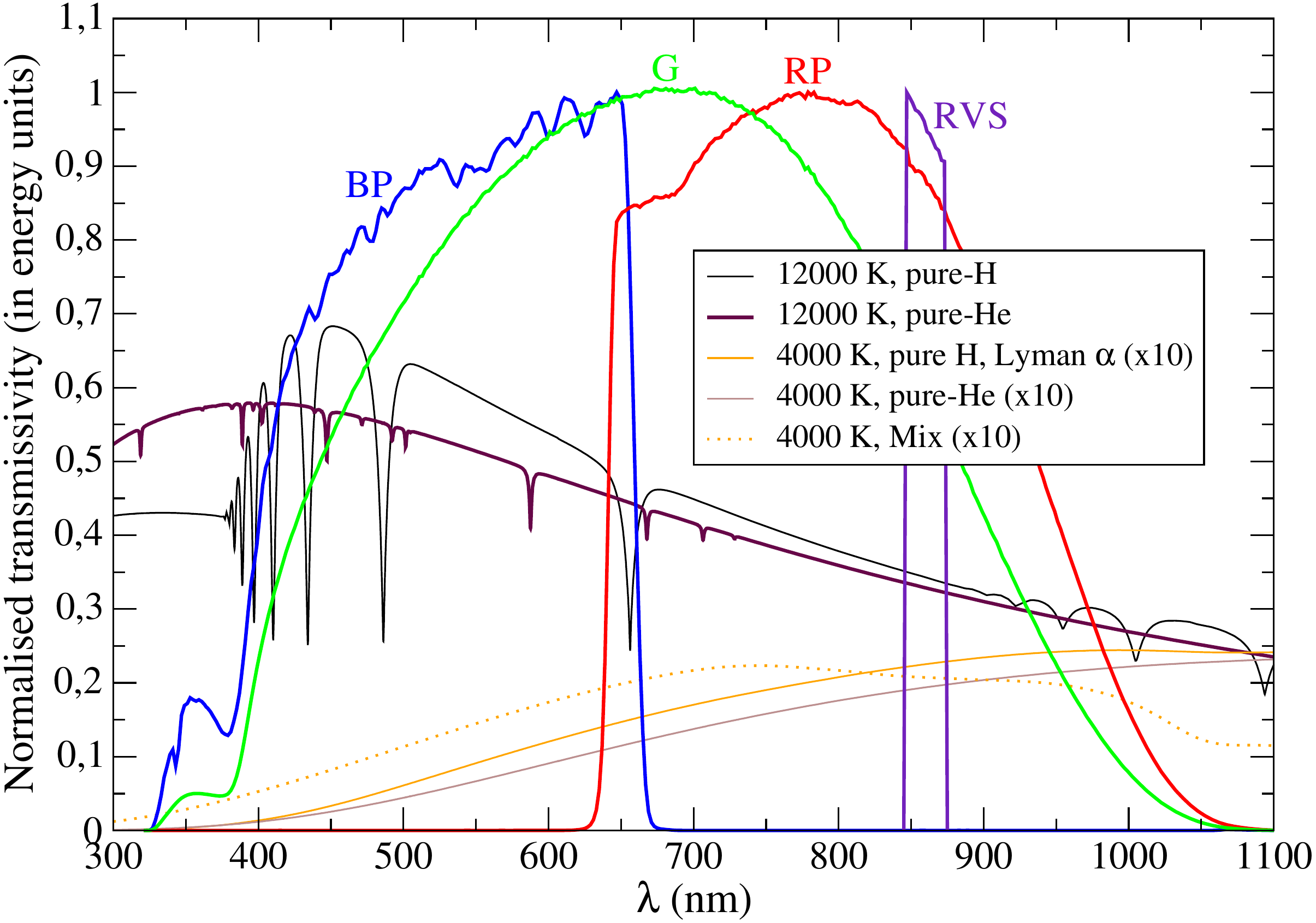}
     \caption{{\Gaia} passbands transmissivity. $G$ passband is depicted in green, {\BP} in blue, {\RP} in red, and {\RVS} in magenta lines. 
All WD SEDs plotted here (described in Sect.~\ref{sec:atmospheres})
correspond to $\log g=8.0$. 
     }
         \label{fig:gaiapassbands}
   \end{figure}

\begin{figure}
   \centering
  \includegraphics[width=.48\textwidth]{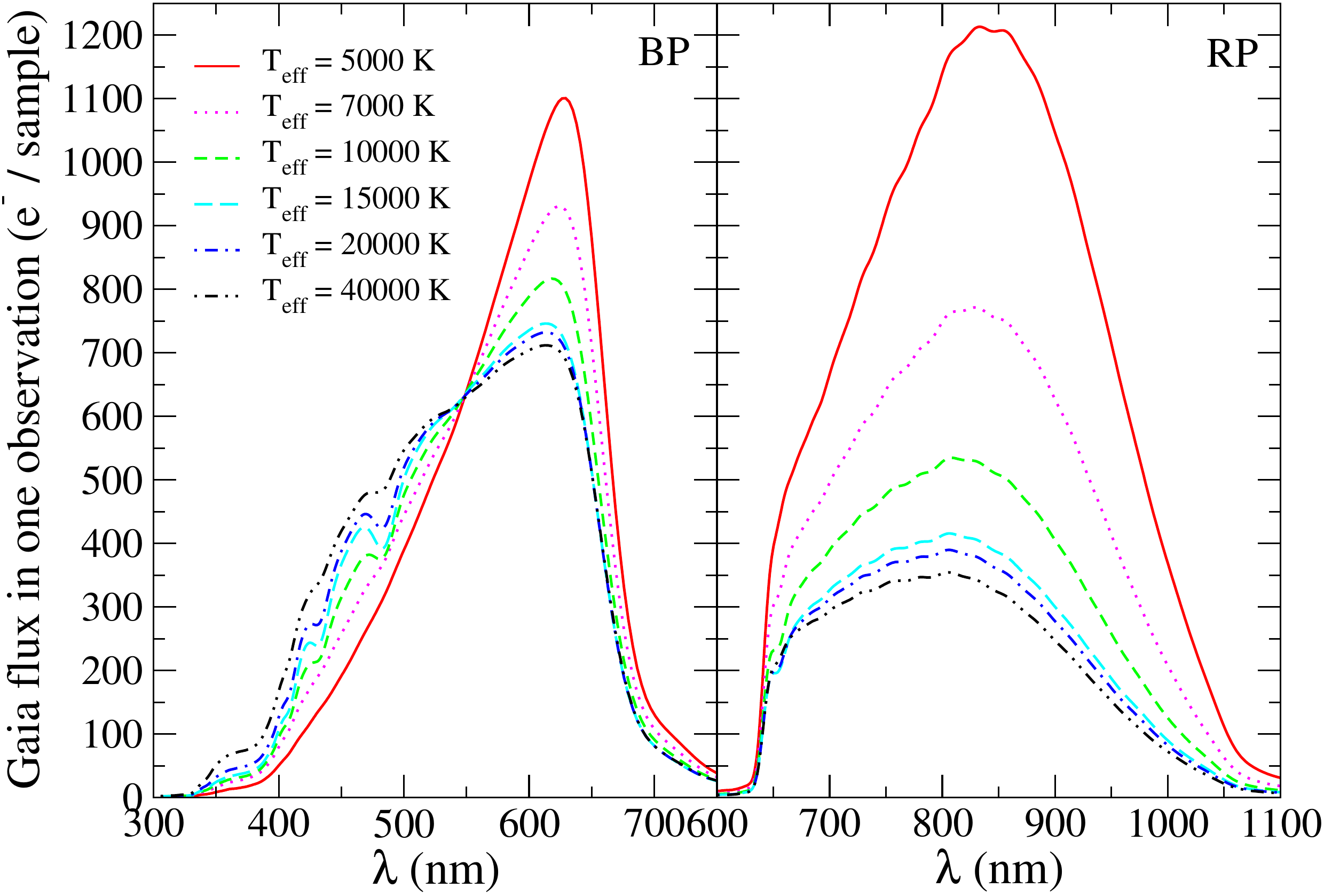}
  \caption{Examples of {\Gaia} BP/RP spectra for pure-H WDs \citep{tre11} with different temperatures
at $G=15$~mag. All SEDs plotted here have $\log g=8.0$.
}
  \label{fig:BPRPspectra}
\end{figure}

In Fig.~\ref{fig:gaiapassbands} the synthetic SEDs of selected WDs (see Sect.~\ref{sec:atmospheres}) are shown together with the transmission curves of the $G$ passband, BP/RP spectrophotometry, and
RVS spectroscopy. Low-resolution BP/RP spectra as will be obtained by {\Gaia} for pure-H WDs are shown in Fig.~\ref{fig:BPRPspectra}. If these spectra are re-binned, summing all their pixels together, we obtain their corresponding magnitudes, {\BP}, {\RP,} and {\RVS} (Fig.~\ref{fig:gaiapassbands}). In the same way, we can reproduce any other synthetic passband, if needed (e.g. Johnson-Cousins, SDSS, or 2MASS, etc.).

The faint {\Gaia} limiting magnitude 
will guarantee the detection of very cool WDs. 
In Table~\ref{tab:distancesWD} we compute the maximum distances at which WDs with
different temperatures and gravities will be detected with
{\Gaia}. Two limiting distances are provided, without considering
  interstellar absorption, $d$, and assuming an average absorption of 1 mag
  per kpc, $d_{\rm A_V}$, corresponding to an observation made in the
  direction of the Galactic disc \citep{odell00}. We also provide the absolute magnitudes ($M_G$) and consider two different compositions, pure-H and pure-He. 
All WDs with {\teff}~$>20\,000$~K will be detected within $270$~pc and all with {\teff}~$>10\,000$~K within $150$~pc, regardless of the atmospheric composition or interstellar absorption. The brightest unreddened WDs will even be observed farther away than 1.5 kpc. For the coolest regime the space volume of observation is smaller, especially at high $\log g$, with detections restricted to the nearest 50 pc (for {\teff}~$=5000$~K). 

\begin{table*}
\begin{center}
\caption{Maximum distances, $d$, at which unreddened WDs will be observed by {\Gaia} for pure-H \citep{tre11} and pure-He 
(\citealt{ber11}, which only covers {\teff}~$>3500$~K) models. $d_{\rm A_V}$ is the limiting distance when an average reddening corresponding to an observation in the disc direction has been applied (assuming an extinction of 1 mag$\cdot$kpc$^{-1}$). The ages of all these WDs can be consulted in the CDS online tables (Tables~\ref{tab:photDA}~--~\ref{tab:photMix}). $M_G$ is the absolute magnitude at $G$ passband.}
\tiny
\begin{tabular}{c|ccc|ccc|ccc}
\hline
& \multicolumn{3}{c|}{$\log g=7.0$} & \multicolumn{3}{c|}{$\log g=8.0$} & \multicolumn{3}{c}{$\log g=9.0$}\\ 
{\teff}  & $M_G$ & $d$ (pc) & $d_{\rm A_V}$ (pc)&  $M_G$ & $d$ (pc) & $d_{\rm A_V}$ (pc)& $M_G$ & $d$ (pc) & $d_{\rm A_V}$ (pc)\\
\hline
\multicolumn{10}{c}{Pure-H (Lyman-$\alpha$)}\\
\hline
$2000$    & $17.28$ &   $35$ &  $34$ & $18.22$ &  $23$ &  $22$ & $19.77$ &  $11$ &  $11$ \\
$3000$    & $15.80$ &   $69$ &  $67$ & $16.76$ &  $45$ &  $44$ & $18.38$ &  $21$ &  $21$ \\
$4000$    & $14.86$ &  $107$ & $102$ & $15.84$ &  $68$ &  $66$ & $17.44$ &  $33$ &  $32$ \\
$5000$    & $13.72$ &  $180$ & $167$ & $14.86$ & $107$ & $102$ & $16.55$ &  $49$ &  $48$\\
$10\,000$ & $10.74$ &  $711$ & $552$ & $12.18$ & $366$ & $317$ & $13.96$ & $161$ & $151$ \\
$20\,000$ &  $9.23$ & $1425$ & $929$ & $10.81$ & $689$ & $538$ & $12.59$ & $303$ & $268$ \\
\hline
\multicolumn{10}{c}{Pure-He}\\
\hline
 $4000$   & $15.16$ &   $93$ &  $89$ & $16.09$ &  $61$ &  $59$ & $17.77$ &  $28$ &  $27$ \\
 $5000$   & $13.76$ &  $177$ & $164$ & $14.86$ & $107$ & $102$ & $16.58$ &  $48$ &  $47$ \\
$10\,000$ & $10.92$ &  $655$ & $516$ & $12.28$ & $350$ & $304$ & $14.02$ & $157$ & $147$ \\
$20\,000$ &  $9.17$ & $1465$ & $947$ & $10.68$ & $731$ & $564$ & $12.44$ & $325$ & $285$ \\
\hline
\end{tabular}
\label{tab:distancesWD}
\end{center}
\end{table*}

\begin{figure}
  \begin{center}
  \includegraphics[width=.23\textwidth]{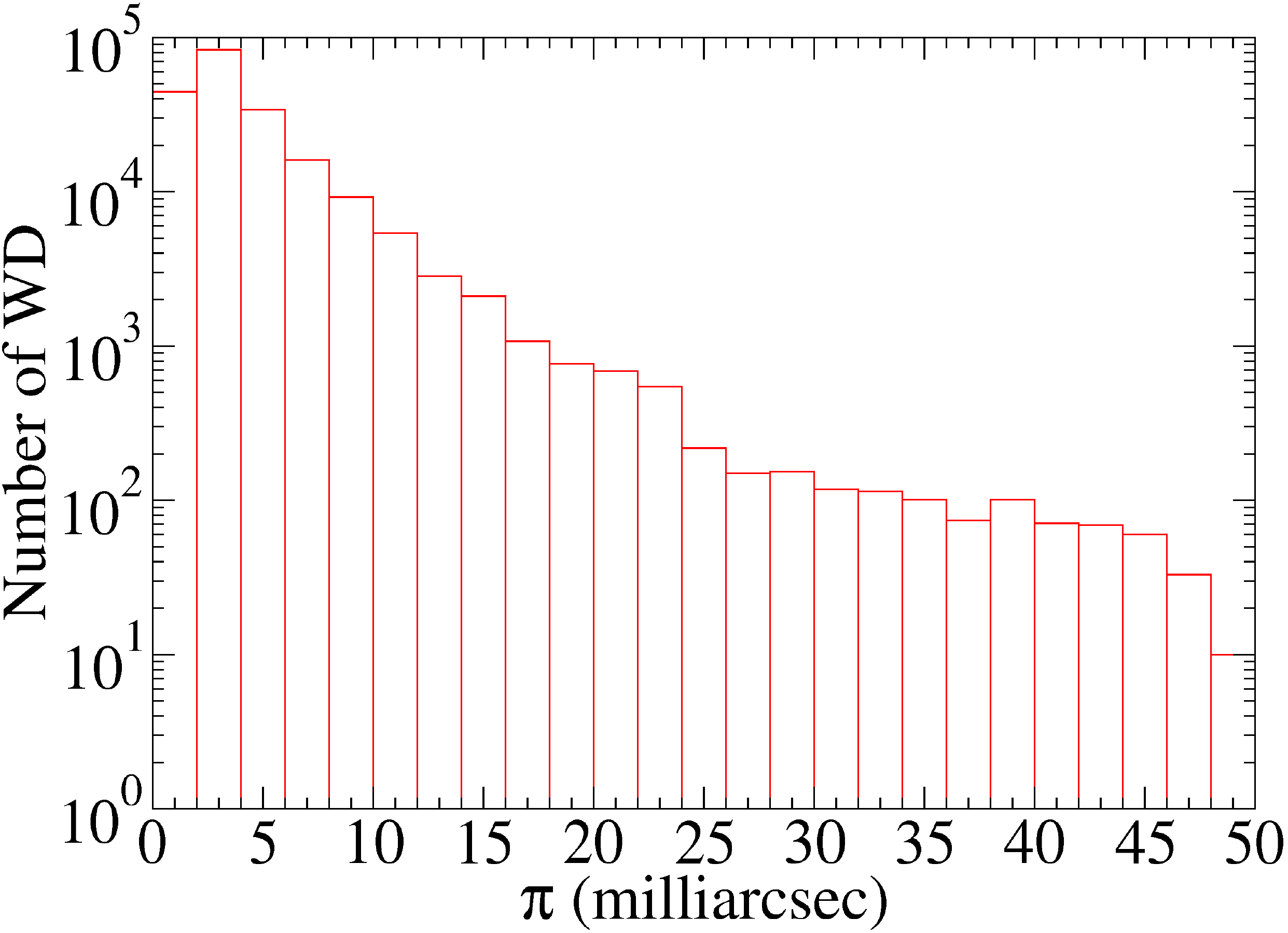}
  \includegraphics[width=.24\textwidth]{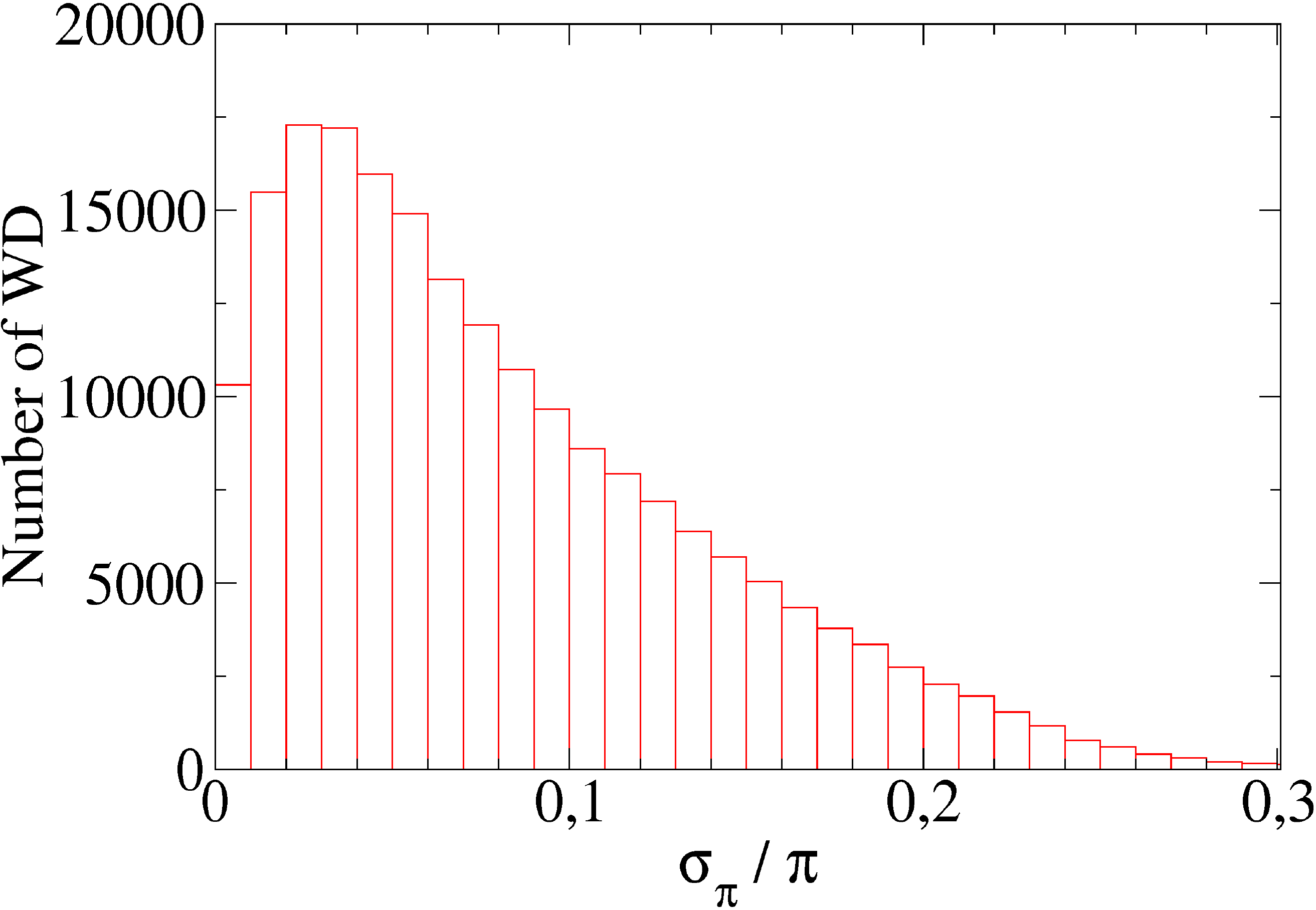}
   \end{center}   
        \caption{\small{Histogram of parallaxes (left) of single WDs with $G\leq20$ using data from GUMS. Histogram of parallax relative errors (right)
 computed using {\Gaia} performances\repeatfootnote{foot:performances}.
 }}
\label{fig:parallaxhistogrGUMS}    
\end{figure}

The {\Gaia} photometry combined with its extremely precise parallaxes will allow absolute magnitudes to be derived, which will provide precise locations in the Hertzsprung--Russell (HR) diagrams (see Sect.~\ref{sec:classification}). Estimates of the number of WDs observed by {\Gaia} with a given parallax and a given relative error in the parallax are provided in Fig.~\ref{fig:parallaxhistogrGUMS} (left and right, respectively), based on simulations performed with GUMS, see Sect.~\ref{sec:simulation}. The errors in parallaxes were computed using {\Gaia} performance prescriptions\repeatfootnote{foot:performances}. Estimates derived from Fig.~\ref{fig:parallaxhistogrGUMS} of the number of WDs with better parallax precision than a certain threshold are provided in the upper part of Table~\ref{tab:N_WDparallax}. About 95\% of the isolated WDs brighter than $G=20$ will have parallaxes more precise than 20\%.

\begin{table}
\begin{center}
\caption{Approximate number of WDs (all temperatures) with parallaxes better than a certain percentage, derived using the GUMS dataset and real WD SDSS datasets \citep[extracted from][]{kil10,tre11}.}
\begin{tabular}{ccccc}
\hline
&\multicolumn{4}{c}{GUMS}\\
\hline
\noalign{\smallskip}
&\multicolumn{2}{c}{All WDs}&\multicolumn{2}{c}{Single WDs}\\
$\sigma_{\pi}$ / $\pi$ & $N$ &  \% of observed & $N$ &  \% of observed\\   
\hline
$\leq$1\%  &  $20\,000$ &3.5\% & $10\,000$  & 5\% \\
$\leq$5\%  & $150\,000$ & 25\% & $75\,000$  & 40\% \\
$\leq$10\% & $300\,000$ & 50\% & $135\,000$ & 70\% \\
$\leq$20\% & $450\,000$ & 80\% & $190\,000$ & 95\% \\
\hline
&\multicolumn{4}{c}{WDs from SDSS samples}\\
\hline
\noalign{\smallskip}
&\multicolumn{2}{c}{\cite{kil10}}&\multicolumn{2}{c}{\cite{tre11}}\\
$\sigma_{\pi}$ / $\pi$ & $N$ &  \% of observed & $N$ &  \% of observed\\   
\hline
$\leq$1\%  & 76 & 61\% &  500  & 16\% \\
$\leq$5\%  & 125 & 100\% & $2275$  & 74\% \\
$\leq$10\% & 125 & 100\% & $2880$ & 94\% \\
$\leq$20\% & 125 & 100\% & $3048$ & 99\% \\
 \noalign{\smallskip}
\hline
\end{tabular}
\label{tab:N_WDparallax}
\end{center}
\end{table}

\begin{figure}
  \begin{center}
  \includegraphics[width=.45\textwidth]{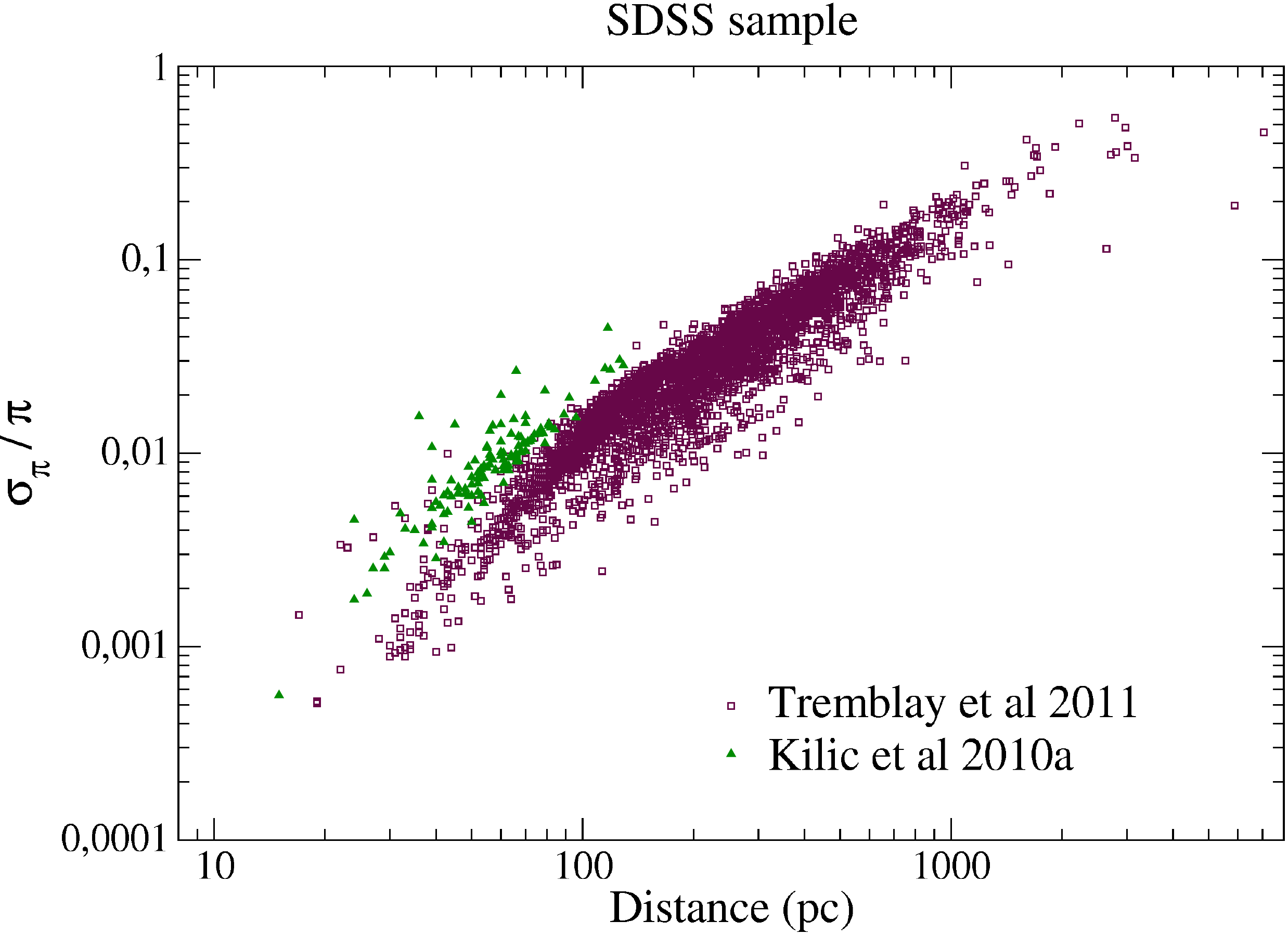}     
   \end{center}   
        \caption{\small{Relative error in {\Gaia} parallax as a function of the distance
for real WDs observed by SDSS, showing that {\Gaia} will be precise enough to easily cover the $100$~pc WD sample at a $\sim 1$\% level. 
        }}
\label{fig:sigmapiSDSS}    
\end{figure}

Expected end-of-mission parallax uncertainties were also computed for two
observational datasets extracted from SDSS data. The first sample includes 125
cool ({\teff}~$<7000$~K) WDs analysed by \cite{kil10}, the second data set
includes almost $3000$ hot ($6000$ K $<$~{\teff}~$< 140\,000$~K), non-magnetic and
single WDs from \cite{tre11}. The relative error of their parallaxes, derived
from predicted distances, are shown in Fig.~\ref{fig:sigmapiSDSS}. For these
samples, we computed the values quoted in the bottom part of
Table~\ref{tab:N_WDparallax}. As can be seen, 94\% of the
WDs will have parallax determinations better than 10\%, corresponding
to absolute magnitudes with uncertainties below $0.2$~mag, which allows a clear distinction between WDs and MS stars. While the masses for the sample of \cite{tre11} are known from spectroscopic fits, the masses of the cool WDs from \cite{kil10} are not constrained, and therefore {\Gaia} will be able to provide this information for the first time.

{\Gaia} will also observe (and discover new) binaries containing WDs (according to simulations by GUMS shown in Table~\ref{tab:numberWD}, only $36\%$ of WDs detected by {\Gaia} will be single) and orbital solutions will be achieved for a significant number of them. Therefore, {\Gaia} will provide independent mass determinations of WDs. These data are desirable to check and calibrate the currently available mass estimates of WDs based on the models and photometric/spectroscopic data. Among all these binaries, eclipsing binaries will be extremely useful to determine the radius of the WD. We currently know more
than $2000$ binary pairs composed of a WD and an MS star \citep{rebassa11}. Only 34 systems of this sample are eclipsing binaries. About a thousand more will be discovered by {\Gaia}.

\section{{\Gaia} photometric transformations}
\label{sec:transformations}

For each available WD SED in the libraries described in Sect.~\ref{sec:atmospheres}, we computed their {\Gaia} photometry as they would be observed with {\Gaia} and other commonly used photometric systems (Johnson-Cousins, SDSS and 2MASS) following the same strategy as \cite{jor10}. The results are listed in the CDS online Tables~\ref{tab:photDA}~--~\ref{tab:photMix}. The contents of these tables include the astrophysical parameters of the WDs (effective temperature, surface gravity, mass in solar masses, bolometric magnitude, bolometric correction in $V$, and age) as well as their simulated absolute magnitudes in Johnson-Cousins ($U$, $B$, $V$, $R_C$, $I_C$), 2MASS ($J$, $H$, $K_S$), SDSS ($u$, $g$, $r$ , $i$, $z$), and {\Gaia} passbands ($G$, $G_{\rm BP}$, $G_{\rm RP}$, and $G_{\rm RVS}$). The three different tables correspond to different compositions of the WDs (pure-H in Table~\ref{tab:photDA}, pure-He in Table~\ref{tab:photDB}, and mixed composition with H/He=0.1 in Table~\ref{tab:photMix}).

Because  WDs are intrinsically faint objects, they are observed close to us. Because of this, they are not much affected by extinction\footnote{The most reddened observable WDs in Table~\ref{tab:distancesWD} have $A_V<1$~mag, assuming an extinction of 1 mag$\cdot$kpc$^{-1}$.}. For this reason, the colour transformations presented in this section were obtained without reddening effects, which are considered negligible for most WDs observed by {\Gaia}. The hot WDs in the Galactic disc direction may, however, suffer from mild to considerable reddening, although this will have to be studied on a case-by-case basis, which is currently beyond the scope of this work.

\onltab{3}{
\begin{landscape}
\begin{table}
\begin{center}
\caption{{\Gaia}, Johnson, 2MASS, and SDSS absolute magnitudes derived for WD SEDs extracted from \cite{tre11} for pure-H composition WDs. Only the first records are shown. The full table is available at the CDS.}
\begin{scriptsize}
\begin{tabular}{ccccc|ccccc|ccc|cccccc|cccc|c}
{\teff} & $\log g$ & $M/M_{\odot}$ & $M_{\rm bol}$ & $BC_V$ & $M_U$ & $M_B$ & $M_V$ & $M_{R_C}$ & $M_{I_C}$ & $M_J$ & $M_H$ & $M_{K_S}$ & $M_u$ & $M_g$ & $M_r$ & $M_i$ & $M_z$ & $M_y$ & $M_G$ & $M_{\rm G_{\rm BP}}$ &  $M_{\rm G_{\rm RP}}$ &  $M_{\rm G_{\rm RVS}}$ & Age \\
\hline
  1500 &  7.0  &  0.150  & 19.064  &  1.043 &  22.926  & 20.527 &  18.021  & 17.746  & 19.438 &  19.578  & 19.584 &  21.336  & 23.940  & 19.241  & 17.619  & 19.668  & 19.352 & 17.980  & 18.380  & 18.414  & 18.660  & 19.037  & 1.623E+10\\
  1750 &  7.0 &   0.151 &  18.393 &   0.787 &  22.232 &  20.018 &  17.607 &  17.153 &  18.233 &  18.418 &  18.556 &  19.798 &  23.228 &  18.800 &  17.070 &  18.526 &  18.313 & 17.583 & 17.779 &  17.933 &  17.705 &  17.888 &  1.338E+10\\
  2000 &  7.0 &   0.151 &  17.812 &   0.524 &  21.626 &  19.588 &  17.288 &  16.687 &  17.240 &  17.456 &  17.705 &  18.621 &  22.599 &  18.437 &  16.673 &  17.580 &  17.441 & 17.273 & 17.275 &  17.575 &  16.924 &  16.946 &  1.078E+10\\
  2250 &  7.0 &   0.151 &  17.297 &   0.281 &  21.083 &  19.203 &  17.016 &  16.299 &  16.427 &  16.651 &  16.968 &  17.633 &  22.038 &  18.113 &  16.368 &  16.807 &  16.712 & 17.003 & 16.830 &  17.284 &  16.277 &  16.176 &  7.879E+09\\
  2500 &  7.0 &   0.152 &  16.837 &   0.059 &  20.590 &  18.855 &  16.778 &  15.974 &  15.757 &  15.963 &  16.330 &  16.817 &  21.529 &  17.822 &  16.128 &  16.191 &  16.094 & 16.764 & 16.440 &  17.039 &  15.740 &  15.538 &  5.479E+09\\
  2750 &  7.0 &   0.152 &  16.420 &  -0.136 &  20.123 &  18.526 &  16.556 &  15.696 &  15.220 &  15.353 &  15.742 &  16.103 &  21.046 &  17.546 &  15.922 &  15.710 &  15.578 & 16.541 & 16.099 &  16.816 &  15.297 &  15.019 &  3.943E+09\\
  3000 &  7.0 &   0.152 &  16.039 &  -0.295 &  19.654 &  18.195 &  16.334 &  15.451 &  14.807 &  14.794 &  15.173 &  15.439 &  20.563 &  17.269 &  15.727 &  15.342 &  15.158 & 16.318 & 15.802 &  16.596 &  14.939 &  14.613 &  3.472E+09\\
.\\
.\\
.\\
\end{tabular}
\end{scriptsize}
\label{tab:photDA}
\end{center}
\end{table}
\end{landscape}
}

\onltab{4}{
\begin{landscape}
\begin{table}
\begin{center}
\caption{{\Gaia}, Johnson-Cousins, 2MASS, and SDSS absolute magnitudes derived for WD SEDs extracted from \cite{ber11} for pure-He composition WDs. Only the first records are shown. The full table is available at the CDS.}
\begin{scriptsize}
\begin{tabular}{ccccc|ccccc|ccc|cccccc|cccc|c}
{\teff} & $\log g$ & $M/M_{\odot}$ & $M_{\rm bol}$ & $BC_V$ & $M_U$ & $M_B$ & $M_V$ & $M_{R_C}$ & $M_{I_C}$ & $M_J$ & $M_H$ & $M_{K_S}$ & $M_u$ & $M_g$ & $M_r$ & $M_i$ & $M_z$ & $M_y$ & $M_G$ & $M_{\rm G_{\rm BP}}$ &  $M_{\rm G_{\rm RP}}$ &  $M_{\rm G_{\rm RVS}}$ & Age \\
\hline
  3500 & 7.0 & 0.150 & 15.387 & -1.803 & 20.374 & 19.186 & 17.190 & 15.907 & 14.774 & 13.610 & 13.120 & 12.831 & 21.252 & 18.245 & 16.371 & 15.474 & 14.944 & 17.202 & 16.104 & 17.371 & 14.980 & 14.508 & 2.902E+09 \\
  3750 & 7.0 & 0.151 & 15.083 & -1.432 & 19.367 & 18.340 & 16.515 & 15.371 & 14.352 & 13.350 & 12.921 & 12.670 & 20.235 & 17.475 & 15.785 & 15.008 & 14.569 & 16.517 & 15.605 & 16.720 & 14.546 & 14.114 & 2.792E+09 \\
  4000 & 7.0 & 0.151 & 14.797 & -1.115 & 18.449 & 17.573 & 15.912 & 14.895 & 13.983 & 13.124 & 12.748 & 12.527 & 19.308 & 16.779 & 15.264 & 14.600 & 14.243 & 15.906 & 15.155 & 16.134 & 14.164 & 13.772 & 2.489E+09 \\
  4250 & 7.0 & 0.152 & 14.526 & -0.845 & 17.605 & 16.874 & 15.371 & 14.473 & 13.662 & 12.928 & 12.596 & 12.402 & 18.455 & 16.148 & 14.804 & 14.244 & 13.960 & 15.360 & 14.750 & 15.604 & 13.829 & 13.475 & 2.193E+09 \\
  4500 & 7.0 & 0.153 & 14.271 & -0.617 & 16.818 & 16.233 & 14.888 & 14.100 & 13.384 & 12.758 & 12.464 & 12.293 & 17.662 & 15.575 & 14.398 & 13.934 & 13.715 & 14.872 & 14.384 & 15.125 & 13.535 & 13.218 & 1.955E+09 \\
  4750 & 7.0 & 0.155 & 14.026 & -0.431 & 16.079 & 15.644 & 14.457 & 13.772 & 13.142 & 12.611 & 12.349 & 12.197 & 16.916 & 15.053 & 14.040 & 13.664 & 13.502 & 14.438 & 14.054 & 14.691 & 13.278 & 12.995 & 1.737E+09 \\
  5000 & 7.0 & 0.157 & 13.791 & -0.286 & 15.385 & 15.108 & 14.077 & 13.484 & 12.932 & 12.480 & 12.245 & 12.110 & 16.215 & 14.584 & 13.728 & 13.430 & 13.317 & 14.058 & 13.757 & 14.300 & 13.055 & 12.802 & 1.546E+09 \\
.\\
.\\
.\\
\end{tabular}
\end{scriptsize}
\label{tab:photDB}
\end{center}
\end{table}
\end{landscape}
}

\onltab{5}{
\begin{landscape}
\begin{table}
\begin{center}
\caption{{\Gaia}, Johnson-Cousins, 2MASS, and SDSS absolute magnitudes derived for WD SEDs extracted from \cite{kil10} for mixed (He/H$=0.1$) composition WDs. Only the first records are shown. The full table is available at the CDS.}
\begin{scriptsize}
\begin{tabular}{ccccc|ccccc|ccc|cccccc|cccc|c}
{\teff} & $\log g$ & $M/M_{\odot}$ & $M_{\rm bol}$ & $BC_V$ & $M_U$ & $M_B$ & $M_V$ & $M_{R_C}$ & $M_{I_C}$ & $M_J$ & $M_H$ & $M_{K_S}$ & $M_u$ & $M_g$ & $M_r$ & $M_i$ & $M_z$ &  $M_y$ & $M_G$ & $M_{\rm G_{\rm BP}}$ &  $M_{\rm G_{\rm RP}}$ &  $M_{\rm G_{\rm RVS}}$ & Age \\
\hline
  2000 & 7.0 & 0.147 & 17.840 & 1.023 & 18.395 & 17.917 & 16.817 & 17.047 & 18.728 & 19.978 & 21.054 & 23.900 & 19.233 & 17.308 & 16.982 & 18.455 & 19.574 & 16.741 & 17.307 & 17.207 & 18.037 & 19.059 & 1.017E+10 \\
  2250 & 7.0 & 0.147 & 17.331 & 0.894 & 17.960 & 17.561 & 16.437 & 16.392 & 17.764 & 18.949 & 20.028 & 22.339 & 18.794 & 16.974 & 16.351 & 17.550 & 18.563 & 16.387 & 16.771 & 16.762 & 17.113 & 18.016 & 9.629E+09 \\
  2500 & 7.0 & 0.147 & 16.874 & 0.749 & 17.566 & 17.239 & 16.125 & 15.870 & 16.862 & 18.002 & 19.086 & 20.968 & 18.398 & 16.674 & 15.886 & 16.675 & 17.649 & 16.091 & 16.299 & 16.400 & 16.323 & 17.096 & 8.280E+09 \\
  2750 & 7.0 & 0.147 & 16.454 & 0.598 & 17.206 & 16.944 & 15.856 & 15.449 & 16.008 & 17.115 & 18.183 & 19.570 & 18.036 & 16.399 & 15.535 & 15.927 & 16.830 & 15.832 & 15.880 & 16.100 & 15.664 & 16.262 & 4.873E+09 \\
  3000 & 7.0 & 0.148 & 16.068 & 0.441 & 16.884 & 16.681 & 15.627 & 15.126 & 15.300 & 16.357 & 17.403 & 18.354 & 17.712 & 16.156 & 15.270 & 15.379 & 16.057 & 15.608 & 15.530 & 15.852 & 15.140 & 15.484 & 3.673E+09 \\
  3250 & 7.0 & 0.149 & 15.714 & 0.285 & 16.597 & 16.447 & 15.429 & 14.872 & 14.738 & 15.670 & 16.713 & 17.425 & 17.424 & 15.939 & 15.062 & 14.974 & 15.389 & 15.411 & 15.234 & 15.643 & 14.715 & 14.816 & 3.173E+09 \\
  3500 & 7.0 & 0.150 & 15.387 & 0.148 & 16.321 & 16.221 & 15.239 & 14.655 & 14.312 & 15.009 & 16.031 & 16.564 & 17.146 & 15.730 & 14.876 & 14.667 & 14.861 & 15.223 & 14.977 & 15.446 & 14.371 & 14.299 & 2.902E+09 \\
.\\
.\\
.\\
\end{tabular}
\end{scriptsize}
\label{tab:photMix}
\end{center}
\end{table}
\end{landscape}
}

\subsection{Johnson-Cousins and SDSS colours}

\begin{figure}
  \includegraphics[width=.24\textwidth]{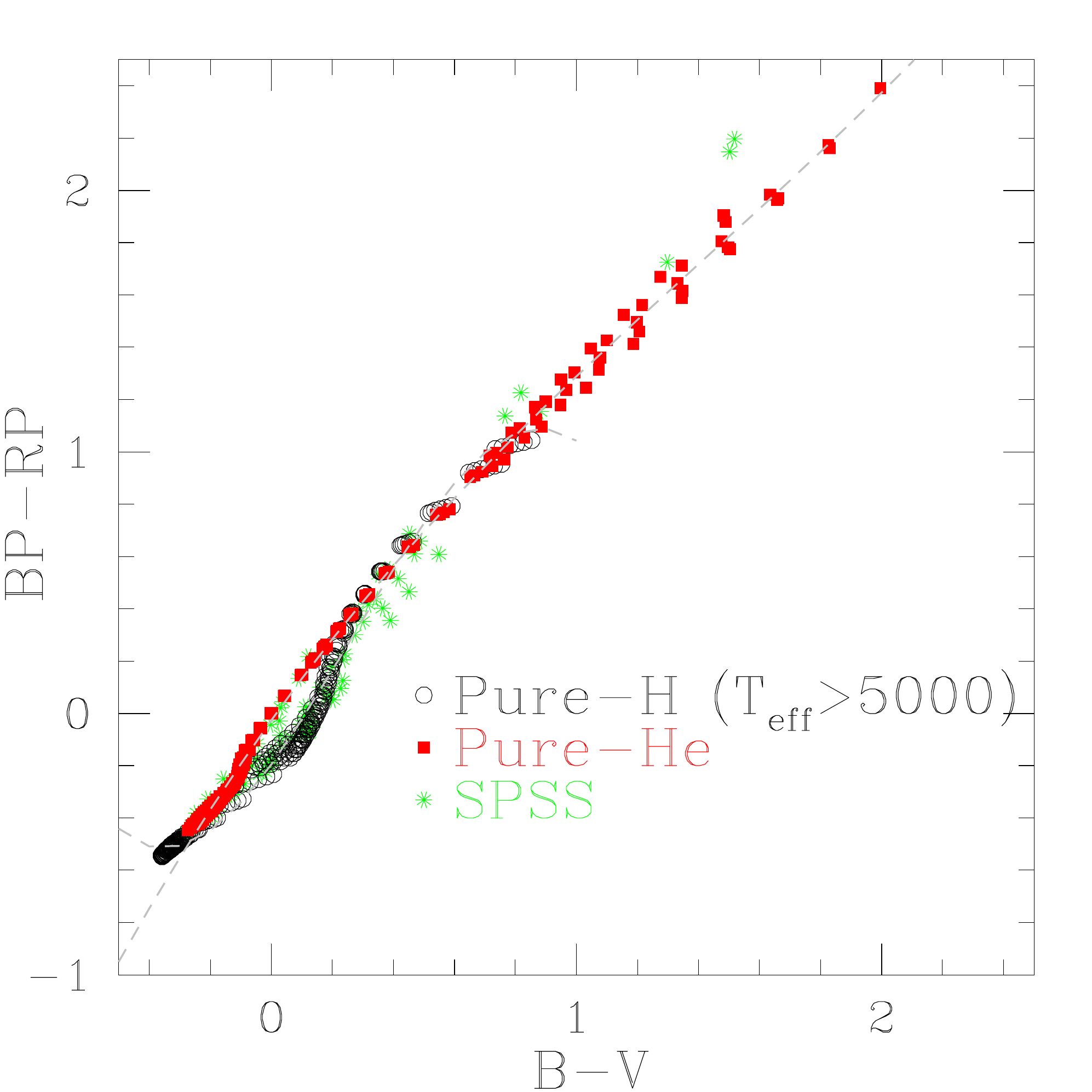}
  \includegraphics[width=.24\textwidth]{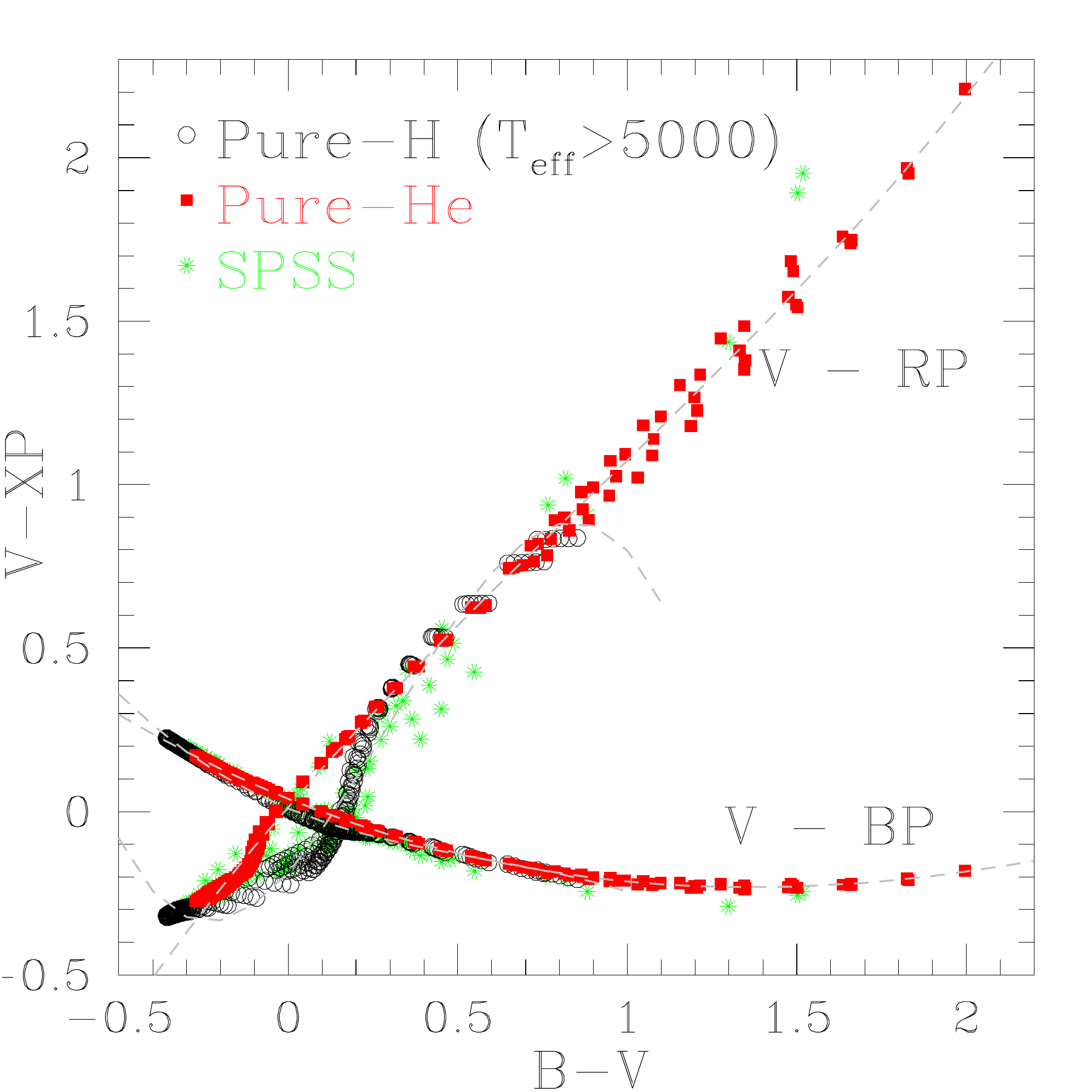}
  \includegraphics[width=.24\textwidth]{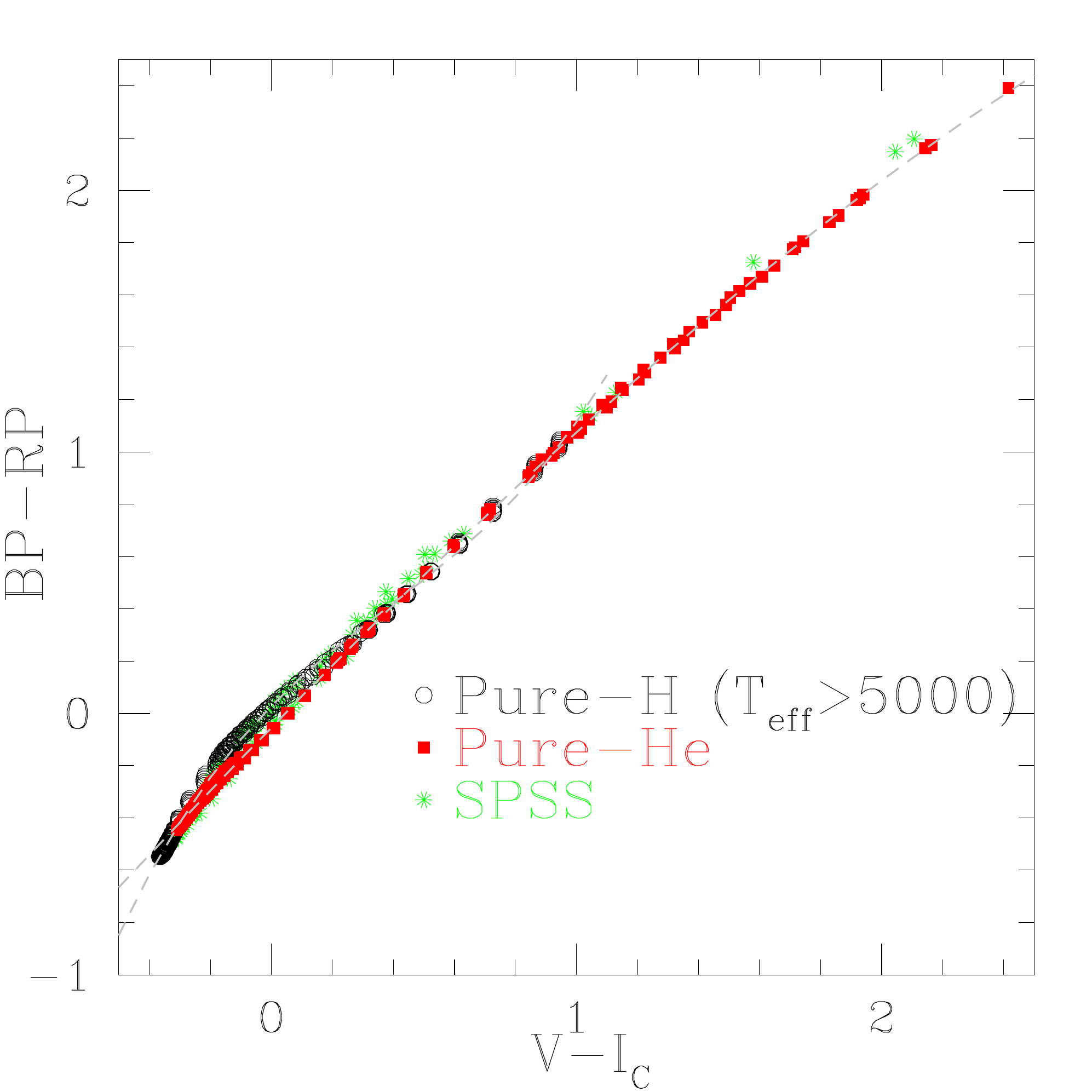}
  \includegraphics[width=.24\textwidth]{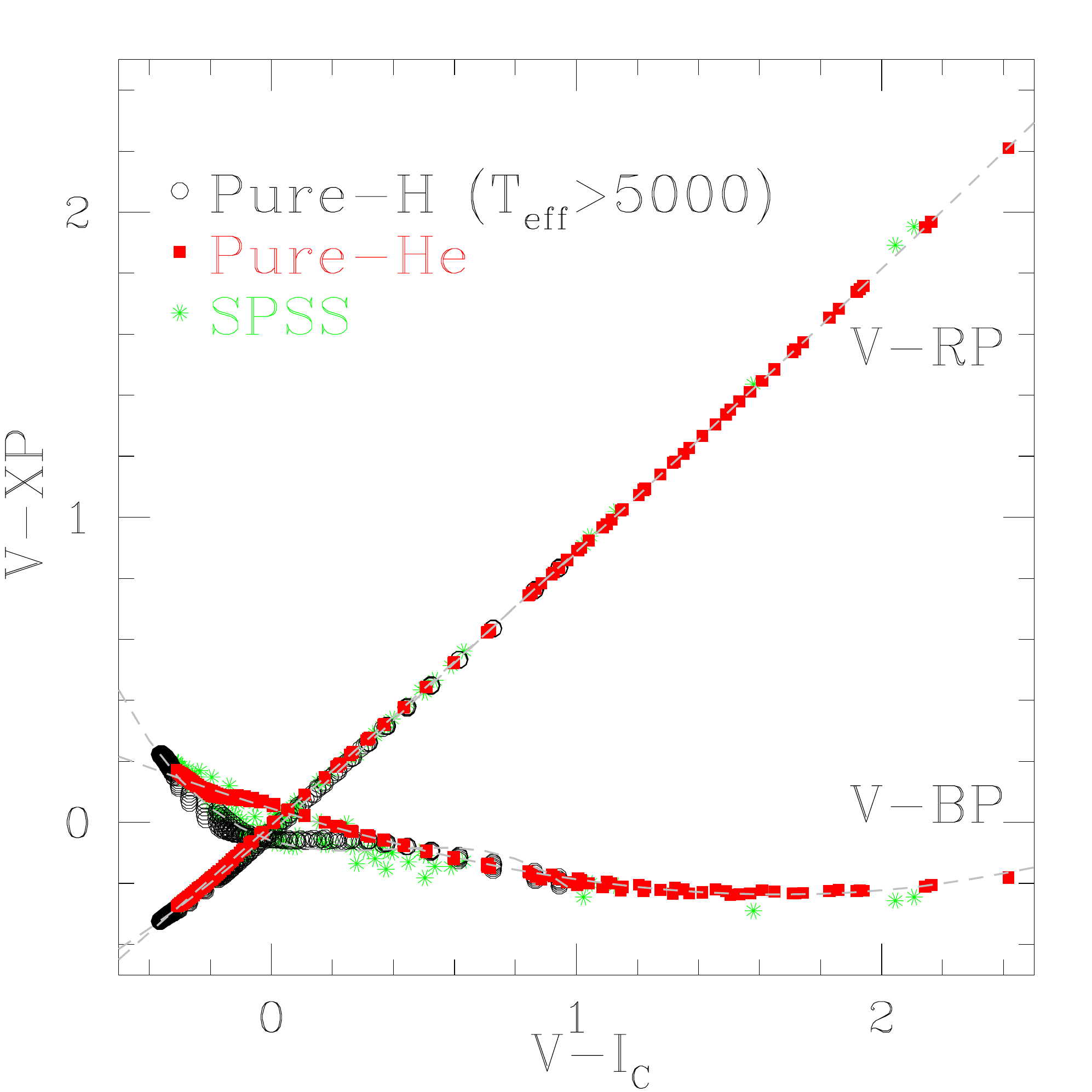}
  \includegraphics[width=.24\textwidth]{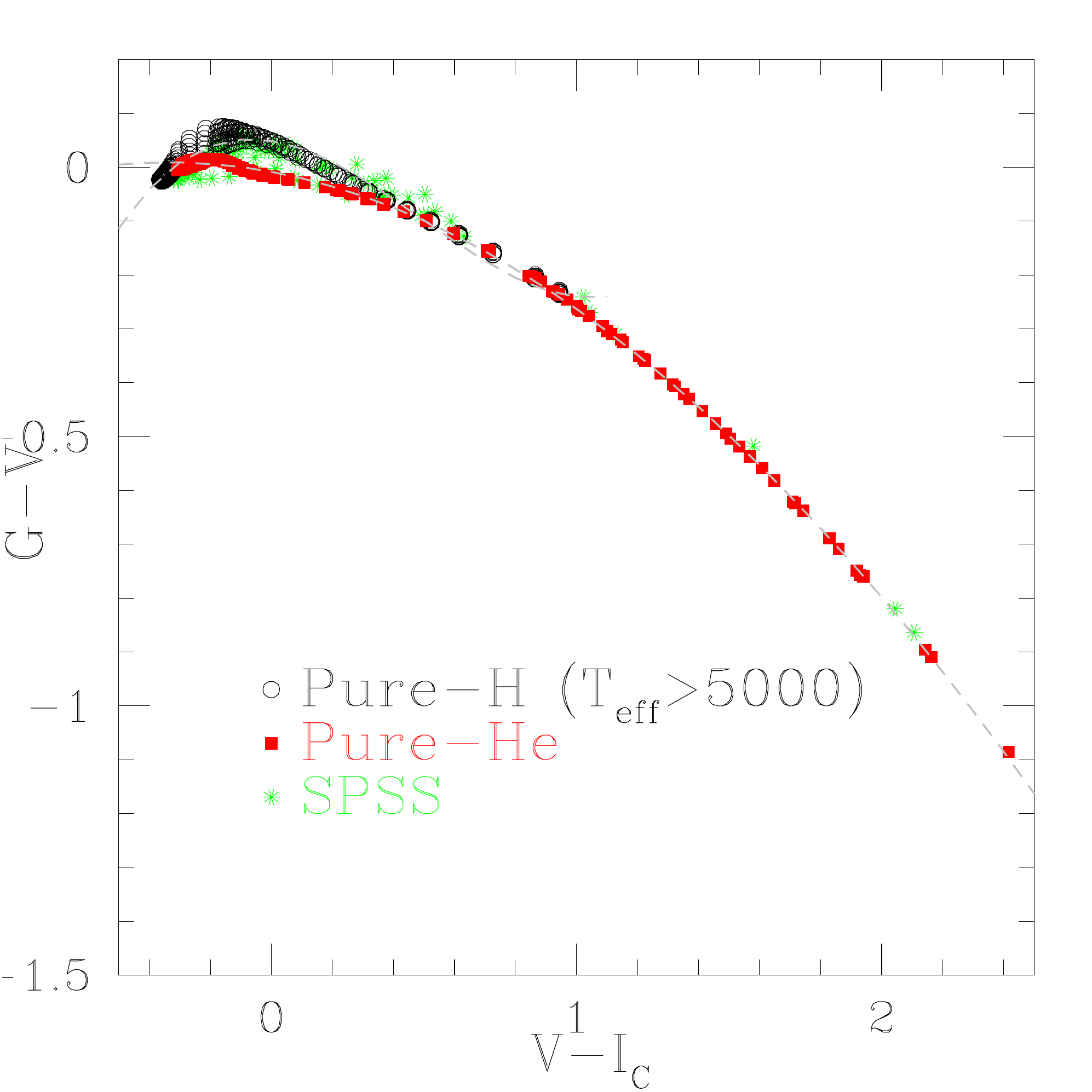}
  \includegraphics[width=.24\textwidth]{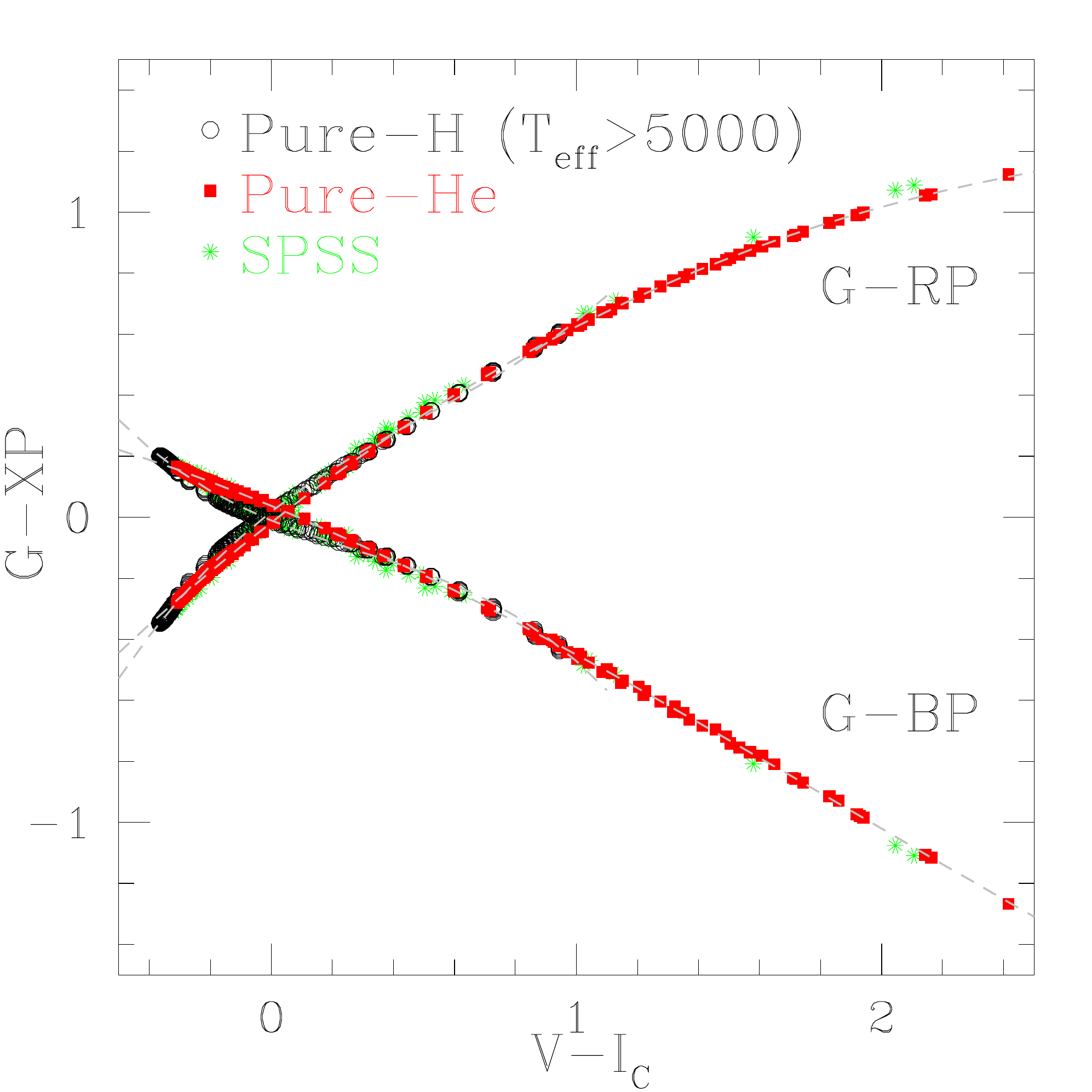}
  \caption{Several colour-colour diagrams obtained using {\Gaia} and Johnson-Cousins passbands 
for the 'normal' regime of pure-H ({\teff}~$> 5000$~K; black open circles) and for pure-He (red squares) WDs. Green star symbols correspond to real WDs selected from \cite{SPSSpaper}. Grey dashed curves are the fitted polynomials from Table~\ref{tab:coeficients}.
} 
  \label{fig:hotjohnsoncolours}
\end{figure}

\begin{figure}
  \includegraphics[width=.24\textwidth]{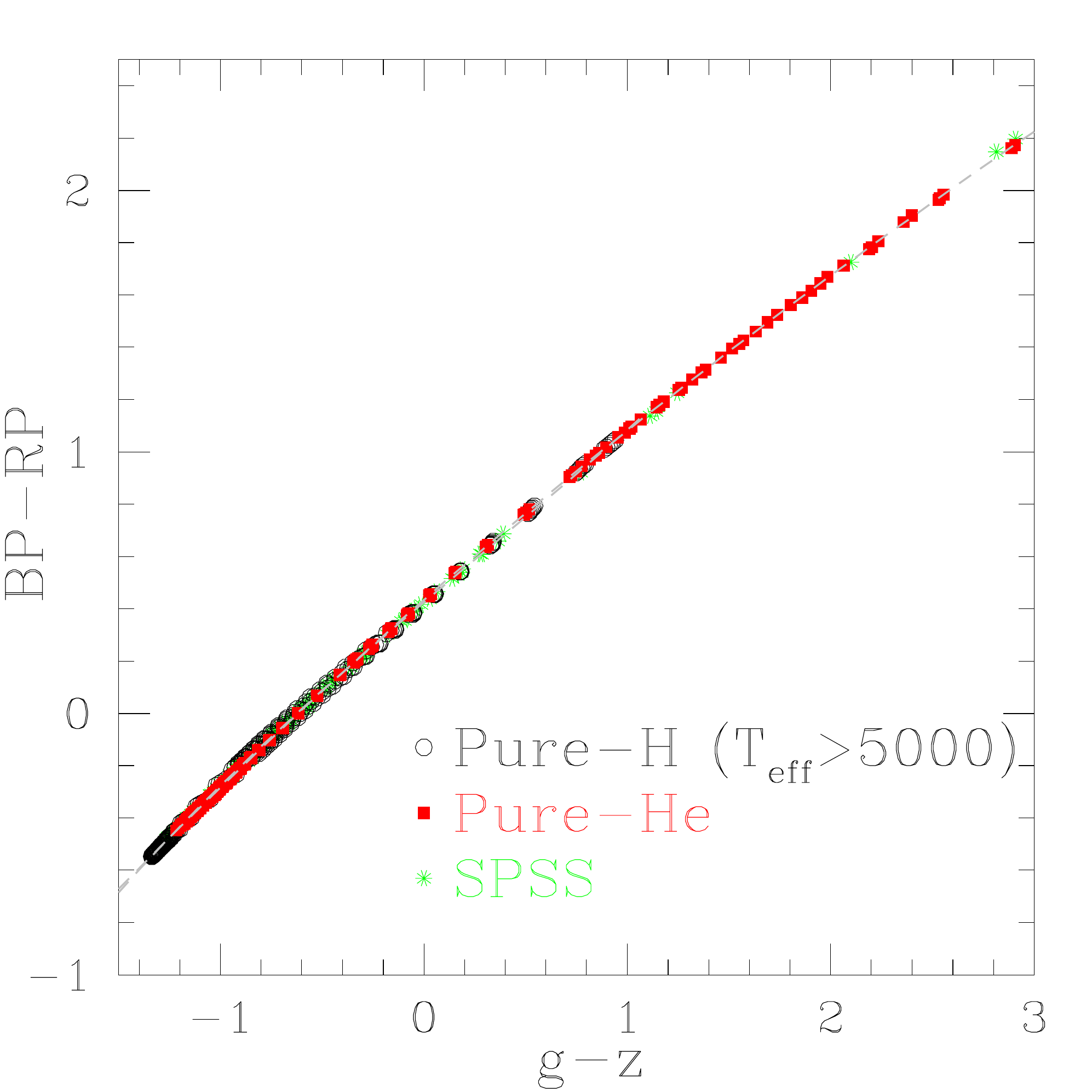}
  \includegraphics[width=.24\textwidth]{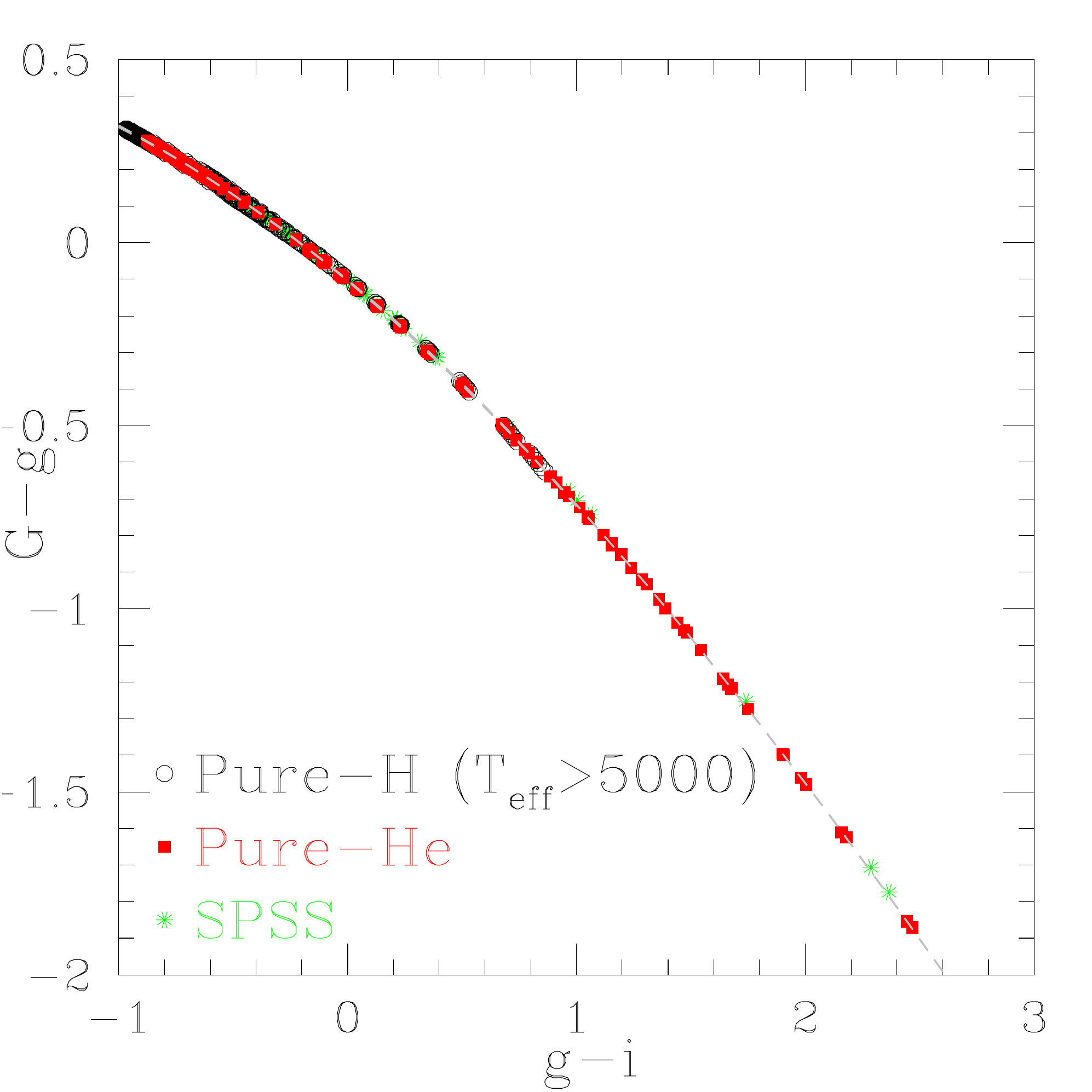}
  \includegraphics[width=.24\textwidth]{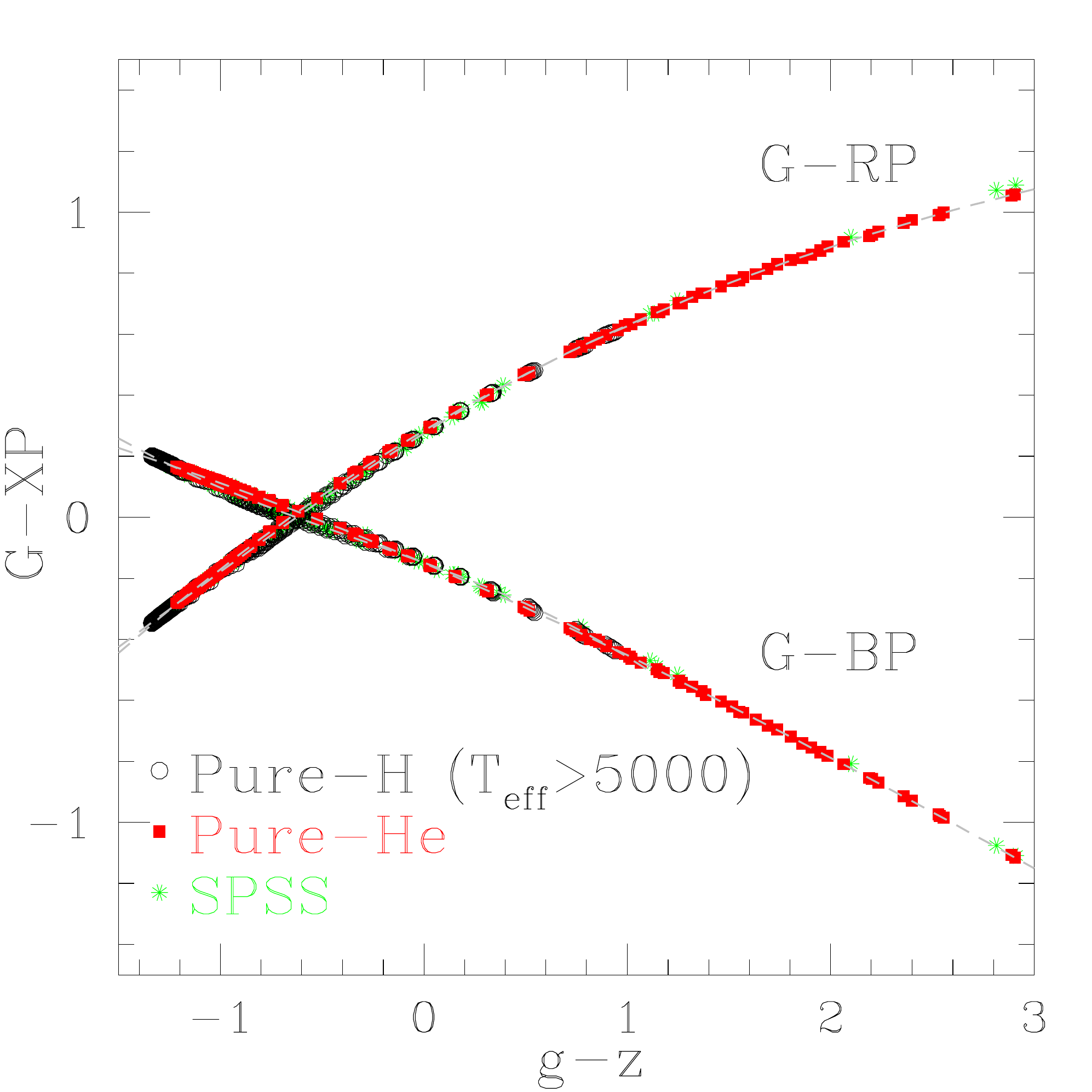}
  \includegraphics[width=.24\textwidth]{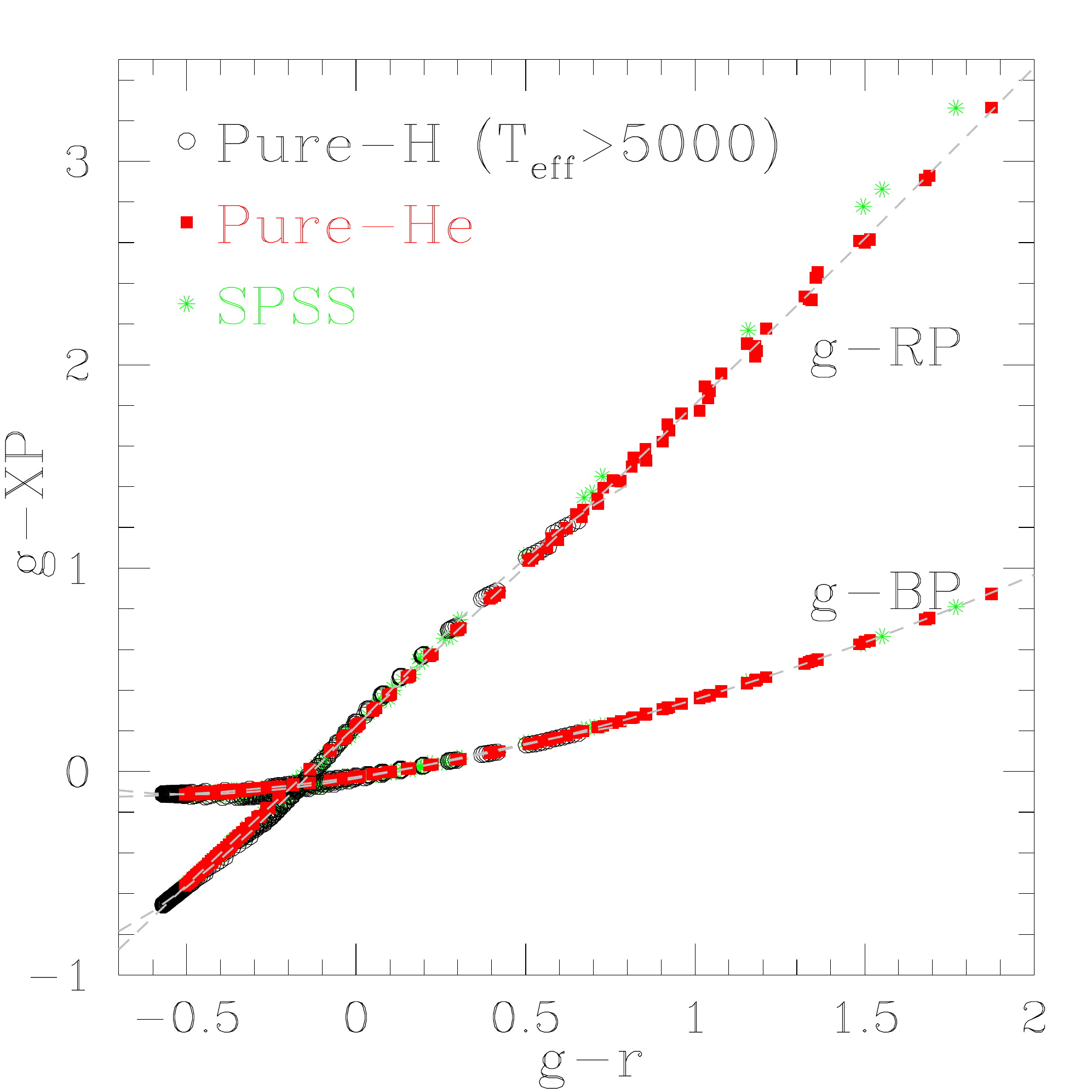}
  \caption{Same as Fig.~\ref{fig:hotjohnsoncolours}, but for SDSS colours.
}
  \label{fig:hotsloancolours}
\end{figure}

Figures~\ref{fig:hotjohnsoncolours}--\ref{fig:hotsloancolours} show several colour-colour
diagrams relating {\Gaia}, Johnson-Cousins \citep{bessell90}, and SDSS passbands \citep{fukugita96}. Only 'normal' pure-H ({\teff}~$> 5000$~K) and pure-He composition WDs are plotted. In this range of {\teff}, colours of mixed composition WDs coincide with those of pure-H WDs and are not overplotted for clarity. The relationship among colours is tight for each composition, that is, independent of the gravities. However, the $B$ and to a lesser degree the $V$ passband induce a distinction between the pure-H and the pure-He WDs at {\teff}~$\sim 13\,000$ K, where the Balmer lines and the Balmer jump are strong in pure-H WDs.

Synthetic photometry was also computed for 82 real WDs extracted from
the list of \cite{SPSSpaper}. They are SpectroPhotometric Standard Star (SPSS) candidates
for the absolute flux calibration of {\Gaia} photometric and spectrophotometric observations.
The whole list of SPSS is selected from calibration sources already used as flux standards 
for HST \citep{bohlin07}, some sources from CALSPEC standards 
(\citealt{oke90}, \citealt{hamuy92}, \citealt{hamuy94}, 
\citealt{stritzinger05}), and finally \cite{mccook99} but also SDSS, and other sources. The colours computed with the SEDs of the libraries used here (Sect.~\ref{sec:atmospheres}) agree very well with those of SPSS true WDs.

Because of the very tight relationship among colours, polynomial expressions were fitted.
240 synthetic pure-He and 276 synthetic pure-H WDs were used for the fitting.
We provide the coefficients for third-order polynomials and
the dispersion values in Table~\ref{tab:coeficients} (available online). Table~\ref{tab:coeficients} contains the following information. Column 1 lists the name of the source, Column 2 gives the bolometric luminosity, etc.

The dispersions are smaller than 0.02~mag for the SDSS passbands,
while for the Johnson-Cousins passbands they can reach 0.06~mag in some case, mainly for pure-H and when blue $B$ or {\BP} passbands are involved. 
The expressions presented here are useful to predict {\Gaia} magnitudes 
for WDs of different {\teff} and atmospheric
  compositions, for which colours in the
  Johnson-Cousins photometric system are known. These expressions should only be used in the {\teff} regimes indicated in Table~\ref{tab:coeficients}. In all other cases, individual values from the CDS online Tables~\ref{tab:photDA}--\ref{tab:photMix} can be used instead. 

\begin{figure}
  \includegraphics[width=.24\textwidth]{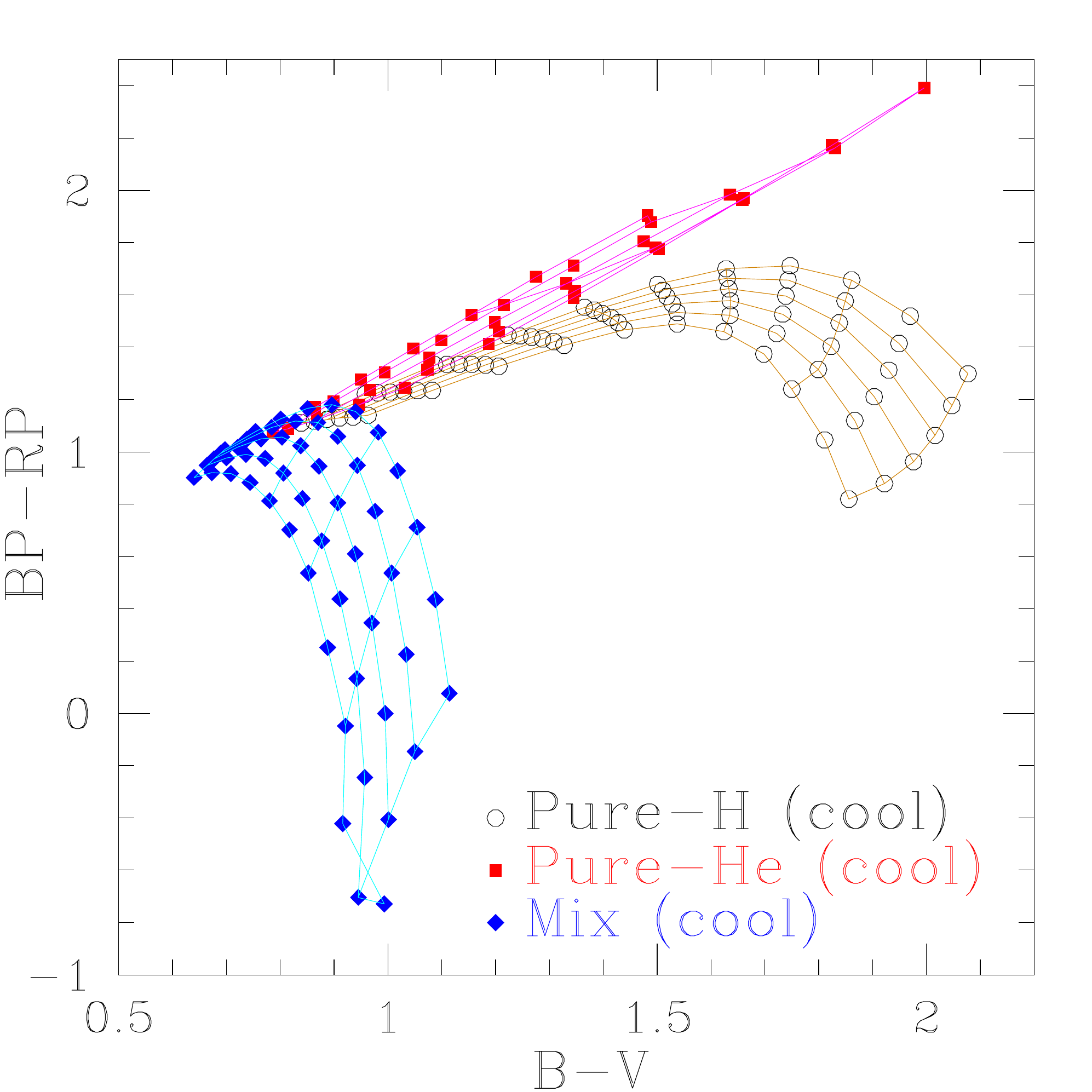}
  \includegraphics[width=.24\textwidth]{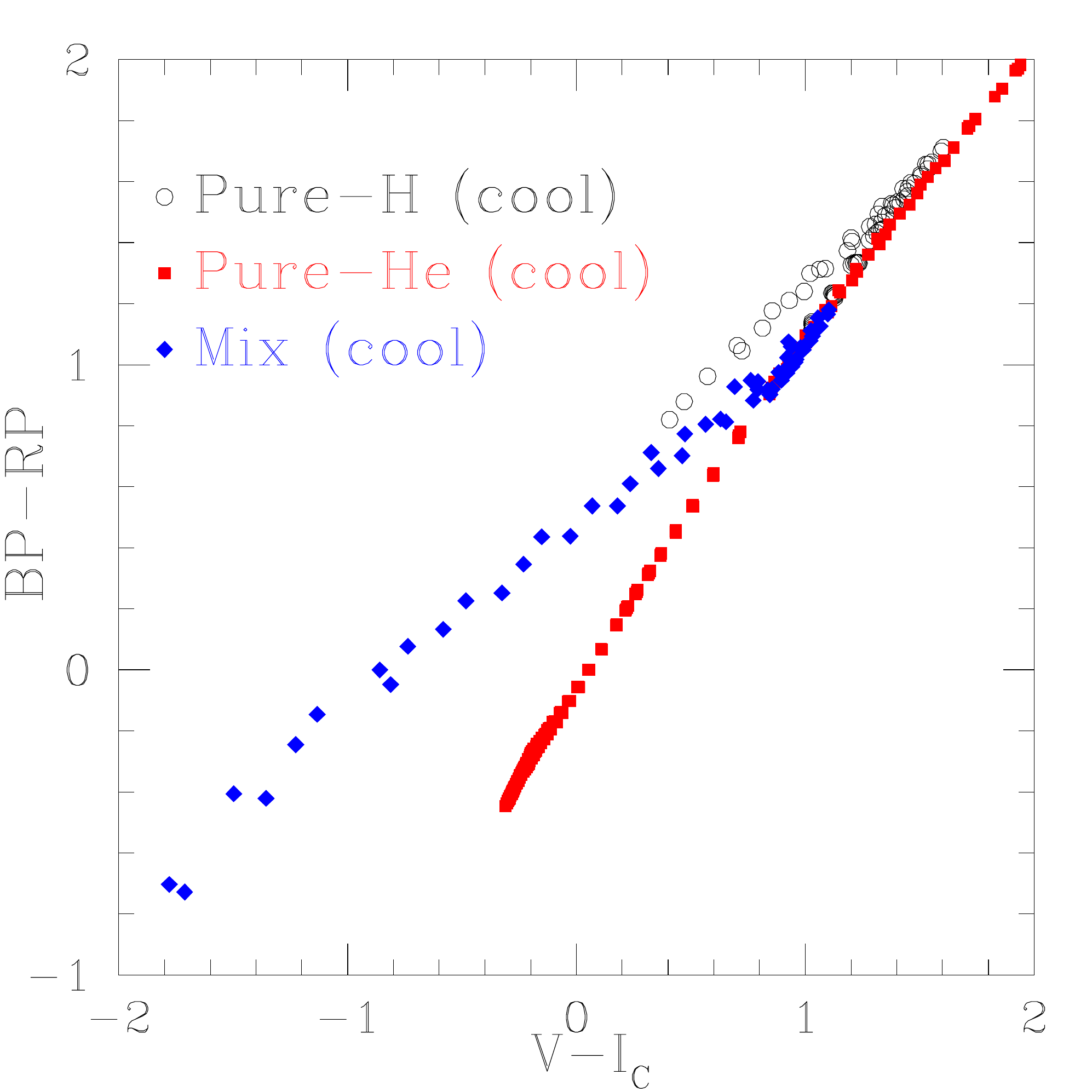}
  \includegraphics[width=.24\textwidth]{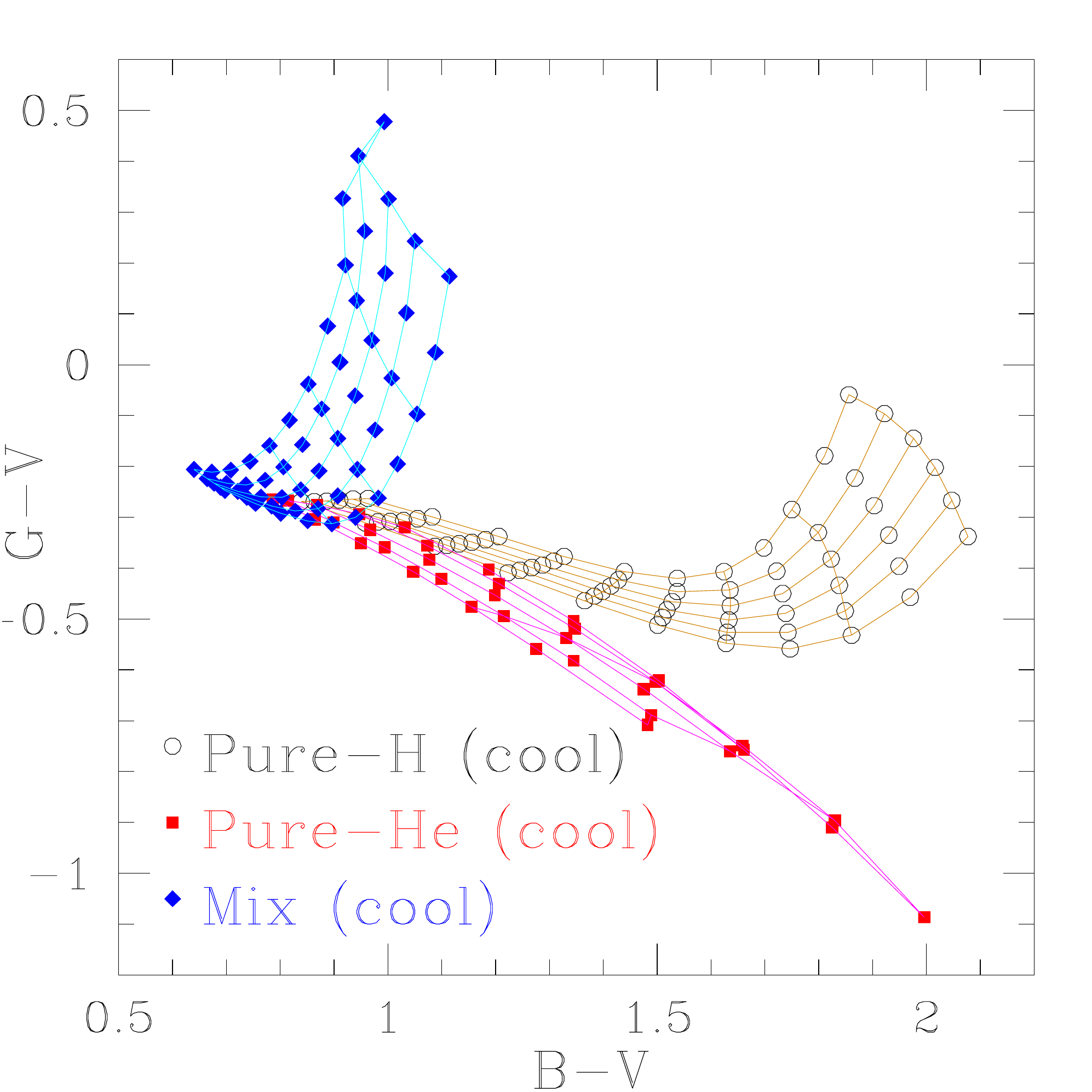}
  \includegraphics[width=.24\textwidth]{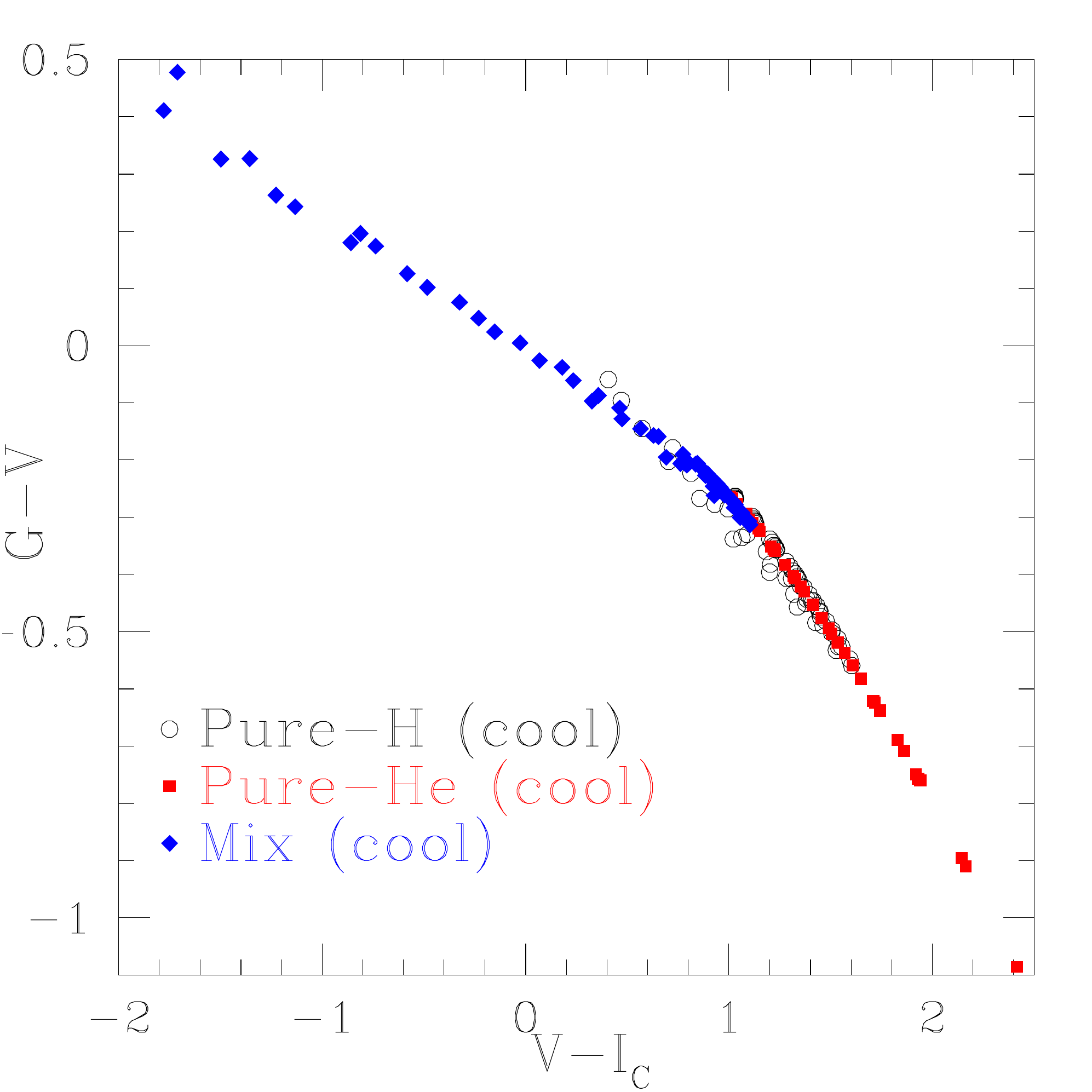}
  \caption{Several colour-colour diagrams obtained using {\Gaia} and Johnson-Cousins passbands in the cool regime ({\teff}~$< 5000$~K).
}
  \label{fig:cooljohnson}
\end{figure}

\begin{figure}
  \includegraphics[width=.24\textwidth]{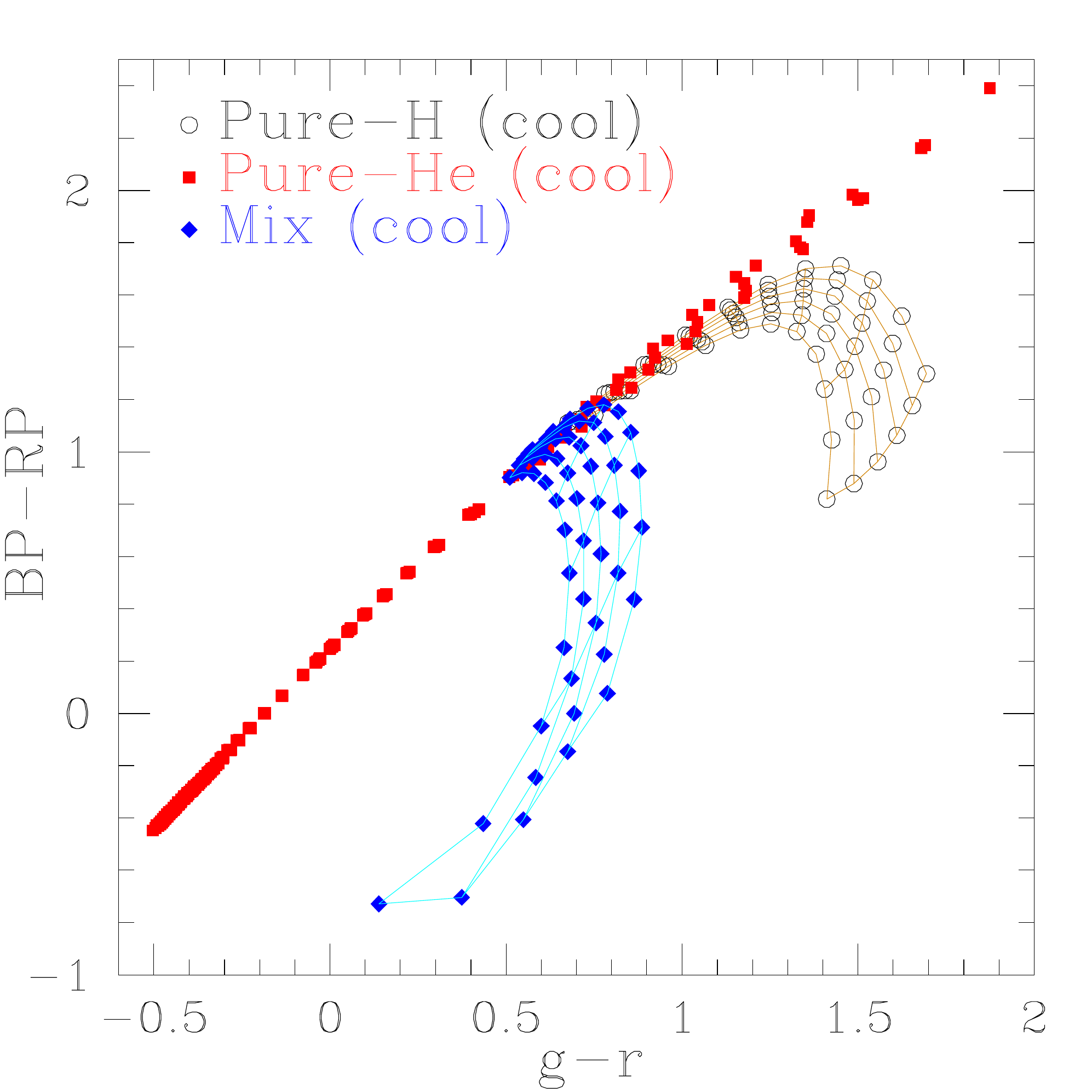}
  \includegraphics[width=.24\textwidth]{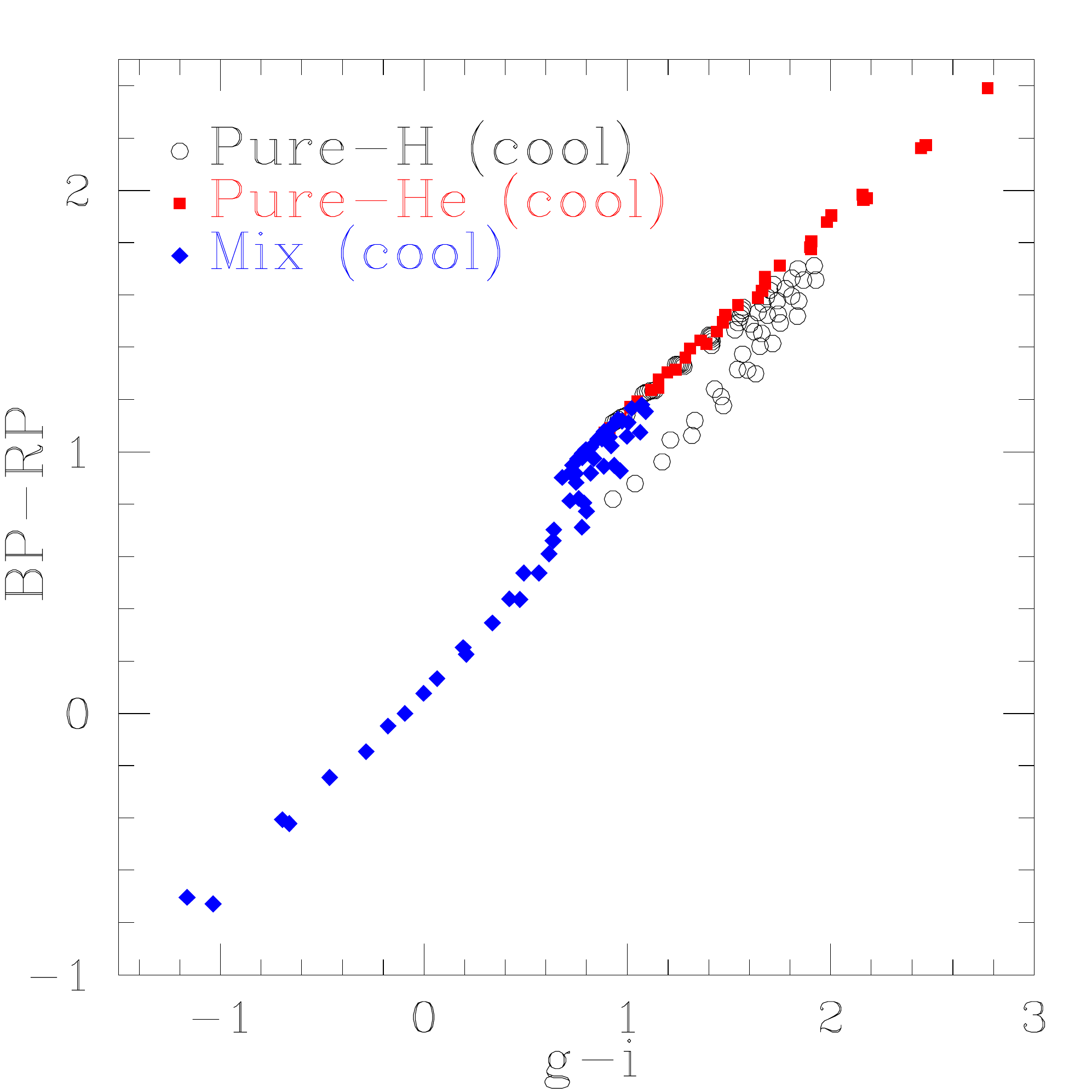}
  \includegraphics[width=.24\textwidth]{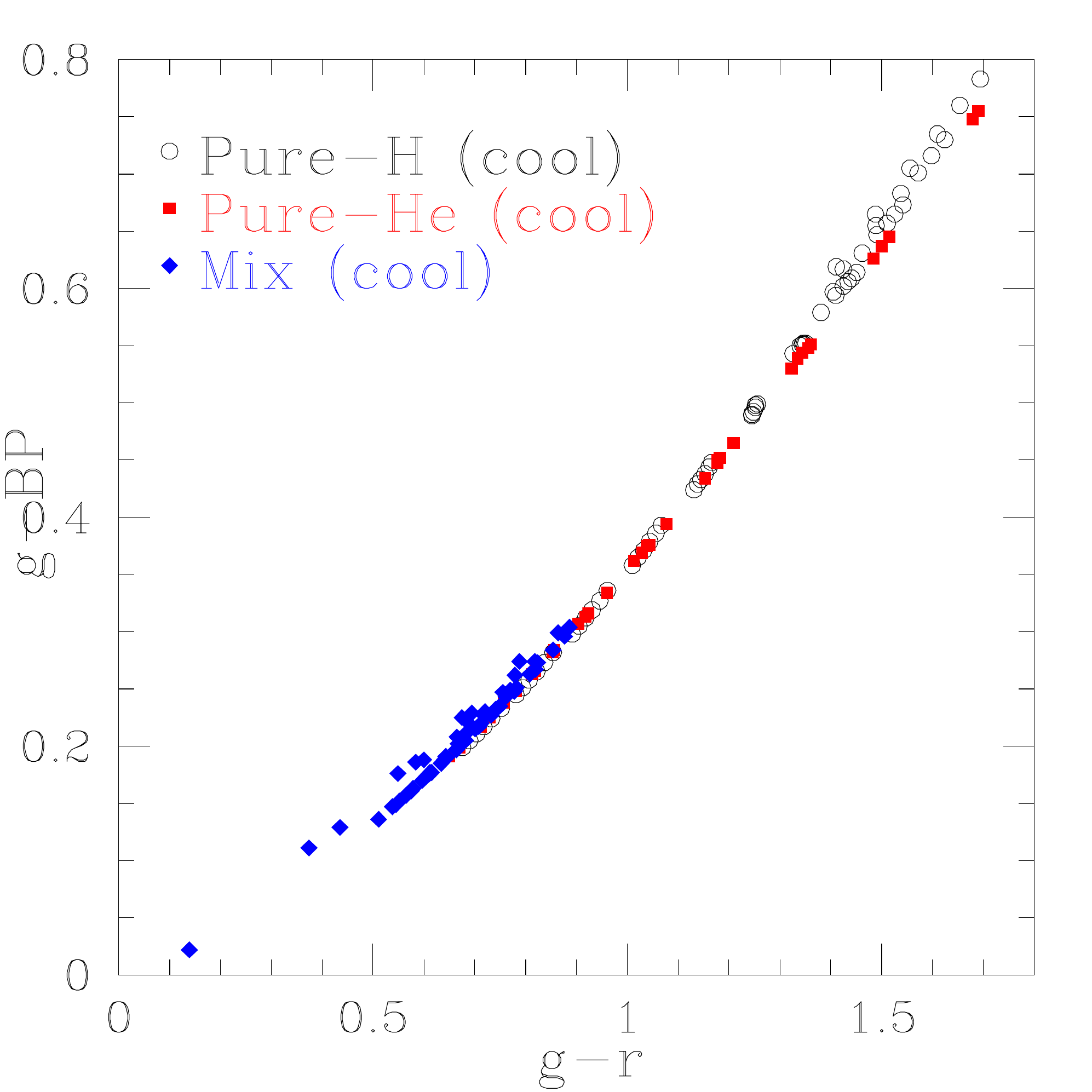}
  \includegraphics[width=.24\textwidth]{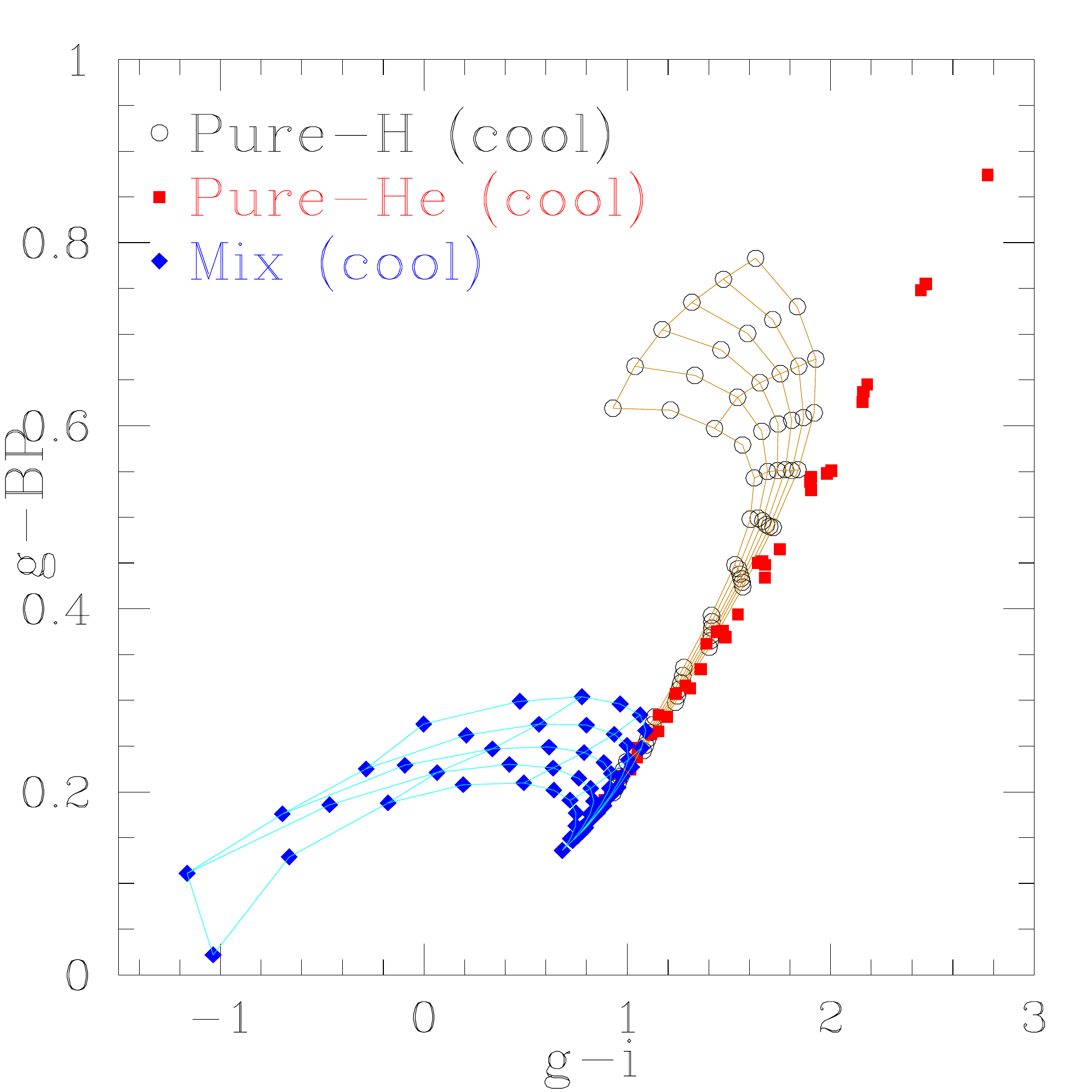}
  \caption{Same as Fig.~\ref{fig:cooljohnson}, but for SDSS passbands.}
  \label{fig:coolsloan}
\end{figure}

In the cool regime, {\teff}~$< 5000$~K, and for pure-H composition the colours depend considerably on the surface gravity, yielding a spread in the colour-colour diagrams 
(see Figs.~\ref{fig:cooljohnson} and \ref{fig:coolsloan}). 
Therefore, no attempt has been made to include
these cool pure-H WDs into the computation of the polynomial transformations. 
To derive the {\Gaia} magnitudes, we recommend the use of the individual values for the desired temperature and surface gravity listed in the CDS online Tables~\ref{tab:photDA}~--~\ref{tab:photMix}.

\onltab{6}{
\begin{table*}[p]
\begin{center}
\caption{Coefficients of the colour-colour polynomial fittings using the Johnson-Cousins and SDSS passbands.}
\label{tab:coeficients}
\tiny
\begin{tabular}{lccccc|lccccc}
\hline
\multicolumn{6}{c|}{\textbf{Johnson-Cousins}} & \multicolumn{6}{c}{\textbf{SDSS}}\\
\hline
\multicolumn{12}{c}{\textbf{Pure-H (\boldmath{{\teff}~$> 5000$}~K)}}\\
 \hline
Colour  &   Zero point  &   $V-I_C$  &$(V-I_C)^2$ &$(V-I_C)^3$ & $\sigma$ &   Colour  &   Zero point  &   $g-i$  &$(g-i)^2$ &$(g-i)^3$ & $\sigma$  \\
 \hline
$G-${\BP}   &   -0.0106 &  -0.4093 &   0.3189 &  -0.3699 &  0.009   &  $G-${\BP}   &   -0.1186 &  -0.3188 &  -0.0276 &  -0.0390 &  0.007   \\                                                    
$G-${\RP}   &    0.0187 &   0.7797 &  -0.4716 &   0.3166 &  0.009   &  $G-${\RP}   &    0.2214 &   0.5655 &  -0.0756 &  -0.0596 &  0.008   \\                                                    
$G-${\RVS}  &    0.0517 &   1.0443 &  -0.6142 &   0.4035 &  0.013   &  $G-${\RVS}  &    0.3246 &   0.7604 &  -0.1031 &  -0.0871 &  0.009   \\                                                    
{\BP}-{\RP} &    0.0292 &   1.1890 &  -0.7905 &   0.6865 &  0.019   &  {\BP}-{\RP} &    0.3400 &   0.8843 &  -0.0480 &  -0.0206 &  0.013   \\                                                    
$G-V$       &    0.0495 &  -0.0907 &  -0.6233 &   0.4240 &  0.013   &  $G-g$       &   -0.1020 &  -0.5132 &  -0.0980 &  -0.0077 &  0.003   \\                                                    
$V-${\RVS}  &    0.0022 &   1.1350 &   0.0091 &  -0.0205 &  0.001   &  $g-${\RVS}  &    0.4266 &   1.2736 &  -0.0051 &  -0.0794 &  0.007   \\                                                    
$V-${\BP}   &   -0.0601 &  -0.3186 &   0.9421 &  -0.7939 &  0.021   &  $g-${\BP}   &   -0.0166 &   0.1943 &   0.0704 &  -0.0313 &  0.009   \\                                                    
$V-${\RP}   &   -0.0308 &   0.8704 &   0.1516 &  -0.1073 &  0.004   &  $g-${\RP}   &    0.3234 &   1.0787 &   0.0224 &  -0.0519 &  0.005   \\                                                    
 \hline
 Color  &   Zero point &   $V-R$  &$(V-R)^2$ &$(V-R)^3$ & $\sigma$ &  Colour  &   Zero point  &   $g-r$  &$(g-r)^2$ &$(g-r)^3$ & $\sigma$ \\                                                             
 \hline
$G-${\BP}   &   -0.0232 &  -0.9322 &   1.9936 &  -3.8732 &  0.015   &  $G-${\BP}   &   -0.0891 &  -0.5172 &  -0.0306 &  -0.0206 &  0.007   \\                                                 
$G-${\RP}   &    0.0427 &   1.7388 &  -3.2195 &   4.3387 &  0.019   &  $G-${\RP}   &      0.1658 &   0.9376 &  -0.1314 &  -0.3685 &  0.011   \\                                                 
$G-${\RVS}  &    0.0838 &   2.3270 &  -4.2247 &   5.6185 &  0.026   &  $G-${\RVS}  &      0.2498 &   1.2602 &  -0.1749 &  -0.5183 &  0.014   \\                                                 
{\BP}-{\RP} &    0.0659 &   2.6710 &  -5.2132 &   8.2118 &  0.034   &  {\BP}-{\RP} &      0.2550 &   1.4548 &  -0.1008 &  -0.3480 &  0.016   \\                                                 
$G-V$       &    0.0473 &  -0.1712 &  -2.8314 &   4.0591 &  0.017   &  $G-g$         &   -0.0555 &  -0.8062 &  -0.2123 &   0.1398 &  0.003   \\                                                 
$V-${\RVS}  &    0.0366 &   2.4982 &  -1.3933 &   1.5594 &  0.010   &  $g-${\RVS}  &      0.3053 &   2.0664 &   0.0374 &  -0.6581 &  0.015   \\                                                 
$V-${\BP}   &   -0.0704 &  -0.7610 &   4.8250 &  -7.9323 &  0.032   &  $g-${\BP}   &   -0.0336 &   0.2889 &   0.1817 &  -0.1604 &  0.009   \\                                                 
$V-${\RP}   &   -0.0045 &   1.9100 &  -0.3881 &   0.2796 &  0.003   &  $g-${\RP}   &      0.2214 &   1.7438 &   0.0809 &  -0.5083 &  0.011   \\                                                 
 \hline                                                                   
Colour  &   Zero point &   $R-I$  &$(R-I)^2$ &$(R-I)^3$ & $\sigma$ &  Colour  &   Zero point &   $r-i$  &$(r-i)^2$ &$(r-i)^3$ & $\sigma$ \\                                                        
 \hline                                                                 
$G-${\BP}   &    0.0002 &  -0.7288 &   0.7736 &  -2.3657 &  0.007   &  $G-${\BP}   &   -0.1647 &  -0.8976 &  -0.9866 &  -2.5767 &  0.008   \\                                                 
$G-${\RP}   &   -0.0015 &   1.4227 &  -0.9942 &   1.4162 &  0.006   &  $G-${\RP}   &      0.3038 &   1.4093 &  -0.2031 &   0.8241 &  0.008   \\                                                 
$G-${\RVS}  &    0.0247 &   1.9071 &  -1.2742 &   1.7530 &  0.007   &  $G-${\RVS}  &      0.4355 &   1.8921 &  -0.3275 &   0.9117 &  0.008   \\                                                 
{\BP}-{\RP} &   -0.0017 &   2.1515 &  -1.7679 &   3.7820 &  0.013   &  {\BP}-{\RP} &      0.4685 &   2.3069 &   0.7835 &   3.4008 &  0.015   \\                                                 
$G-V$       &    0.0509 &  -0.1835 &  -2.1595 &   2.7057 &  0.010   &  $G-g$         &   -0.1796 &  -1.5031 &  -1.6721 &  -2.4821 &  0.009   \\                                                 
$V-${\RVS}  &   -0.0262 &   2.0906 &   0.8853 &  -0.9527 &  0.007   &  $g-${\RVS}  &      0.6150 &   3.3952 &   1.3446 &   3.3938 &  0.012   \\                                                 
$V-${\BP}   &   -0.0507 &  -0.5453 &   2.9331 &  -5.0714 &  0.015   &  $g-${\BP}   &      0.0149 &   0.6055 &   0.6855 &  -0.0945 &  0.010   \\                                                 
$V-${\RP}   &   -0.0524 &   1.6062 &   1.1653 &  -1.2894 &  0.009   &  $g-${\RP}   &      0.4834 &   2.9124 &   1.4690 &   3.3062 &  0.012   \\                                                 
 \hline                                                                 
 Colour  &   Zero point  &   $B-V$  &$(B-V)^2$ &$(B-V)^3$ & $\sigma$ &   Colour  &   Zero point  &   $g-z$  &$(g-z)^2$ &$(g-z)^3$ & $\sigma$ \\                                   
 \hline                                                                 
$G-${\BP}   &    0.0700 &  -0.5674 &  -0.4765 &   0.4891 &  0.015   &  $G-${\BP}   &   -0.1488 &  -0.2450 &  -0.0313 &  -0.0325 &  0.006   \\                                                 
$G-${\RP}   &   -0.1251 &   1.0288 &   0.8762 &  -1.2221 &  0.033   &  $G-${\RP}   &      0.2756 &   0.4128 &  -0.0513 &  -0.0099 &  0.004   \\                                                 
$G-${\RVS}  &   -0.1404 &   1.3779 &   1.1891 &  -1.6605 &  0.044   &  $G-${\RVS}  &      0.3973 &   0.5549 &  -0.0713 &  -0.0170 &  0.004   \\                                                 
{\BP}-{\RP} &   -0.1951 &   1.5962 &   1.3527 &  -1.7112 &  0.047   &  {\BP}-{\RP} &      0.4244 &   0.6579 &  -0.0201 &   0.0226 &  0.008   \\                                                 
$G-V$       &    0.0607 &  -0.1378 &  -0.8979 &   0.7349 &  0.020   &  $G-g$         &   -0.1508 &  -0.4044 &  -0.0731 &  -0.0219 &  0.006   \\                                                 
$V-${\RVS}  &   -0.2011 &   1.5157 &   2.0870 &  -2.3954 &  0.063   &  $g-${\RVS}  &      0.5481 &   0.9593 &   0.0018 &   0.0049 &  0.002   \\                                                 
$V-${\BP}   &    0.0093 &  -0.4297 &   0.4214 &  -0.2458 &  0.006   &  $g-${\BP}   &      0.0020 &   0.1594 &   0.0418 &  -0.0106 &  0.010   \\                                                 
$V-${\RP}   &   -0.1858 &   1.1666 &   1.7741 &  -1.9570 &  0.052   &  $g-${\RP}   &      0.4264 &   0.8173 &   0.0217 &   0.0120 &  0.003   \\                                                 
\hline
\multicolumn{12}{c}{\textbf{Pure-He (All \boldmath{{\teff}})}}\\
 \hline
  Colour  &   Zero point  &   $V-I_C$  &$(V-I_C)^2$ &$(V-I_C)^3$ & $\sigma$ & Colour  &   Zero point  &   $g-i$  &$(g-i)^2$ &$(g-i)^3$ & $\sigma$  \\
 \hline                                                             
$G-${\BP}   &    0.0372 &  -0.4155 &  -0.0864 &   0.0149 &  0.005   &  $G-${\BP}   &   -0.1127 &  -0.3463 &  -0.0320 &   0.0028 &  0.004  \\
$G-${\RP}   &   -0.0166 &   0.7803 &  -0.1451 &   0.0067 &  0.004   &  $G-${\RP}   &      0.2307 &   0.5106 &  -0.0860 &   0.0063 &  0.004  \\
$G-${\RVS}  &   -0.0076 &   1.0204 &  -0.1584 &   0.0035 &  0.006   &  $G-${\RVS}  &      0.3194 &   0.6817 &  -0.0984 &   0.0063 &  0.008  \\
{\BP}-{\RP} &   -0.0538 &   1.1958 &  -0.0587 &  -0.0082 &  0.009   &  {\BP}-{\RP} &      0.3434 &   0.8568 &  -0.0539 &   0.0034 &  0.007  \\
$G-V$       &   -0.0085 &  -0.1051 &  -0.1541 &   0.0046 &  0.006   &  $G-g$         &   -0.1051 &  -0.5219 &  -0.0949 &   0.0065 &  0.001  \\
$V-${\RVS}  &    0.0009 &   1.1255 &  -0.0043 &  -0.0011 &  0.002   &  $g-${\RVS}  &      0.4244 &   1.2036 &  -0.0035 &  -0.0002 &  0.007  \\
$V-${\BP}   &    0.0456 &  -0.3104 &   0.0676 &   0.0103 &  0.010   &  $g-${\BP}   &   -0.0076 &   0.1756 &   0.0629 &  -0.0036 &  0.004  \\
$V-${\RP}   &   -0.0082 &   0.8854 &   0.0089 &   0.0021 &  0.003   &  $g-${\RP}   &      0.3358 &   1.0324 &   0.0090 &  -0.0002 &  0.003  \\
 \hline                                                             
  Colour  &   Zero point &   $V-R$  &$(V-R)^2$ &$(V-R)^3$ & $\sigma$ &       Colour  &   Zero point  &       $g-r$  &$(g-r)^2$ &$(g-r)^3$ & $\sigma$   \\                                            
 \hline
$G-${\BP}   &    0.0410 &  -0.8360 &  -0.3441 &   0.1642 &  0.006   &  $G-${\BP}   &   -0.0758 &  -0.5153 &  -0.0698 &   0.0054 &  0.007  \\
$G-${\RP}   &   -0.0238 &   1.5787 &  -0.6202 &   0.0600 &  0.010   &  $G-${\RP}   &      0.1743 &   0.8064 &  -0.2102 &   0.0258 &  0.007  \\
$G-${\RVS}  &   -0.0169 &   2.0629 &  -0.6808 &   0.0292 &  0.015   &  $G-${\RVS}  &      0.2443 &   1.0710 &  -0.2412 &   0.0271 &  0.012  \\
{\BP}-{\RP} &   -0.0647 &   2.4147 &  -0.2761 &  -0.1042 &  0.016   &  {\BP}-{\RP} &      0.2501 &   1.3217 &  -0.1403 &   0.0204 &  0.013  \\
$G-V$       &   -0.0073 &  -0.2094 &  -0.6447 &   0.1165 &  0.006   &  $G-g$         &   -0.0500 &  -0.7598 &  -0.2141 &   0.0145 &  0.006  \\
$V-${\RVS}  &   -0.0096 &   2.2723 &  -0.0361 &  -0.0873 &  0.013   &  $g-${\RVS}  &      0.2943 &   1.8309 &  -0.0271 &   0.0126 &  0.018  \\
$V-${\BP}   &    0.0483 &  -0.6266 &   0.3006 &   0.0477 &  0.012   &  $g-${\BP}   &   -0.0258 &   0.2446 &   0.1443 &  -0.0091 &  0.003  \\
$V-${\RP}   &   -0.0164 &   1.7881 &   0.0245 &  -0.0565 &  0.008   &  $g-${\RP}   &      0.2243 &   1.5662 &   0.0039 &   0.0113 &  0.013  \\
 \hline                                                             
 Colour  &   Zero point   &   $R-I$  &$(R-I)^2$ &$(R-I)^3$ & $\sigma$ &      Colour  &   Zero point  &       $r-i$  &$(r-i)^2$ &$(r-i)^3$ & $\sigma$   \\
 \hline                                                            
$G-${\BP}   &    0.0334 &  -0.8238 &  -0.3379 &   0.0480 &  0.008   &  $G-${\BP}   &   -0.1866 &  -1.0602 &  -0.2816 &   0.1400 &  0.006  \\
$G-${\RP}   &   -0.0104 &   1.5351 &  -0.5160 &   0.0353 &  0.007   &  $G-${\RP}   &      0.3332 &   1.3942 &  -0.6660 &   0.0952 &  0.006  \\
$G-${\RVS}  &    0.0007 &   2.0093 &  -0.5587 &   0.0195 &  0.007   &  $G-${\RVS}  &      0.4570 &   1.8812 &  -0.7615 &   0.0749 &  0.007  \\
{\BP}-{\RP} &   -0.0438 &   2.3589 &  -0.1781 &  -0.0127 &  0.015   &  {\BP}-{\RP} &      0.5198 &   2.4544 &  -0.3844 &  -0.0449 &  0.012  \\
$G-V$       &   -0.0101 &  -0.2112 &  -0.5643 &  -0.0837 &  0.008   &  $G-g$         &   -0.2192 &  -1.6588 &  -0.8081 &   0.2847 &  0.014  \\
$V-${\RVS}  &    0.0107 &   2.2205 &   0.0056 &   0.1033 &  0.012   &  $g-${\RVS}  &      0.6762 &   3.5400 &   0.0466 &  -0.2098 &  0.016  \\
$V-${\BP}   &    0.0434 &  -0.6126 &   0.2264 &   0.1317 &  0.009   &  $g-${\BP}   &      0.0326 &   0.5986 &   0.5265 &  -0.1447 &  0.009  \\
$V-${\RP}   &   -0.0004 &   1.7463 &   0.0483 &   0.1190 &  0.013   &  $g-${\RP}   &      0.5524 &   3.0530 &   0.1421 &  -0.1895 &  0.017  \\
 \hline                                                             
Colour  &   Zero point &   $B-V$  &$(B-V)^2$ &$(B-V)^3$ & $\sigma$ &       Colour  &   Zero point &        $g-z$  &$(g-z)^2$ &$(g-z)^3$ & $\sigma$   \\
 \hline
$G-${\BP}   &    0.0247 &  -0.5733 &   0.0044 &  -0.0178 &  0.012   &  $G-${\BP}   &   -0.1499 &  -0.2838 &  -0.0201 &   0.0012 &  0.002  \\
$G-${\RP}   &   -0.0008 &   1.0490 &  -0.4004 &   0.0790 &  0.017   &  $G-${\RP}   &      0.2843 &   0.3961 &  -0.0547 &   0.0035 &  0.002  \\
$G-${\RVS}  &    0.0140 &   1.3753 &  -0.4842 &   0.0958 &  0.024   &  $G-${\RVS}  &      0.3910 &   0.5316 &  -0.0628 &   0.0036 &  0.005  \\
{\BP}-{\RP} &   -0.0255 &   1.6223 &  -0.4048 &   0.0967 &  0.028   &  {\BP}-{\RP} &      0.4342 &   0.6800 &  -0.0346 &   0.0023 &  0.003  \\
$G-V$       &   -0.0145 &  -0.1528 &  -0.1763 &  -0.0046 &  0.014   &  $G-g$         &   -0.1616 &  -0.4358 &  -0.0603 &   0.0030 &  0.007  \\
$V-${\RVS}  &    0.0285 &   1.5281 &  -0.3080 &   0.1004 &  0.038   &  $g-${\RVS}  &      0.5526 &   0.9674 &  -0.0025 &   0.0006 &  0.003  \\
$V-${\BP}   &    0.0392 &  -0.4205 &   0.1807 &  -0.0131 &  0.003   &  $g-${\BP}   &      0.0117 &   0.1520 &   0.0402 &  -0.0018 &  0.007  \\
$V-${\RP}   &    0.0137 &   1.2018 &  -0.2241 &   0.0836 &  0.030   &  $g-${\RP}   &      0.4459 &   0.8320 &   0.0056 &   0.0005 &  0.005  \\
\hline
\end{tabular}
\end{center}
\end{table*}
}

\subsection{2MASS colours}

\begin{figure}
  \includegraphics[width=.24\textwidth]{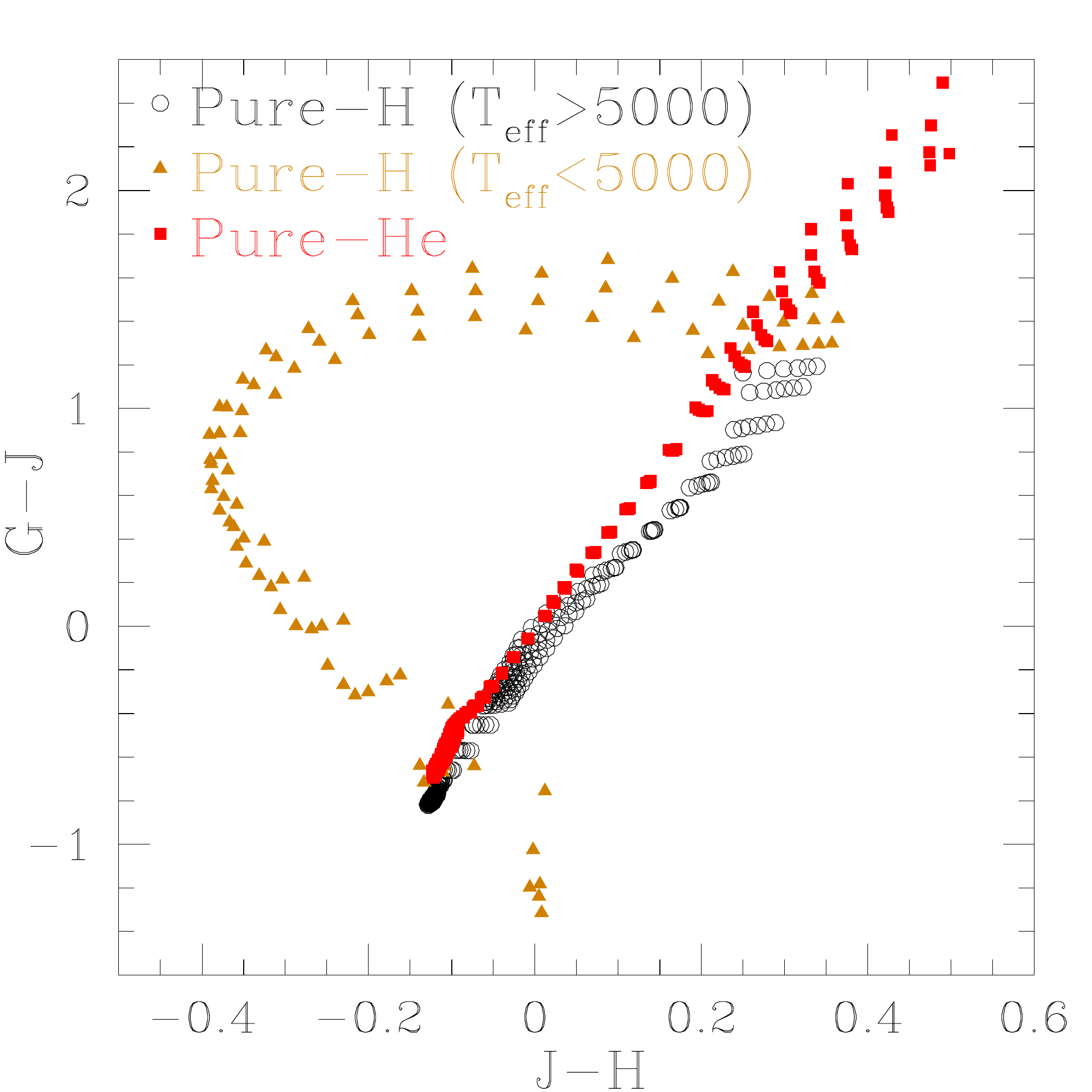}
  \includegraphics[width=.24\textwidth]{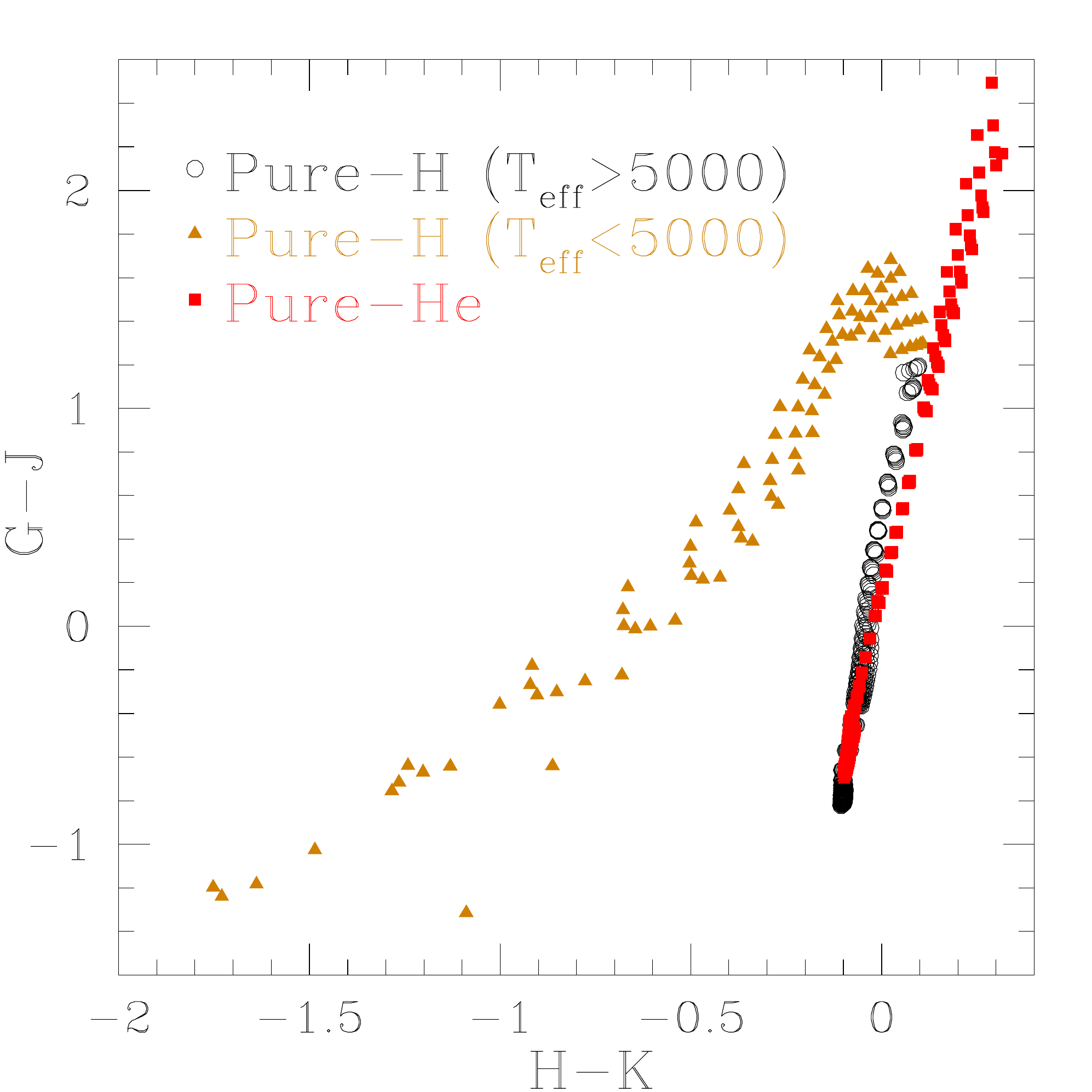}
  \includegraphics[width=.24\textwidth]{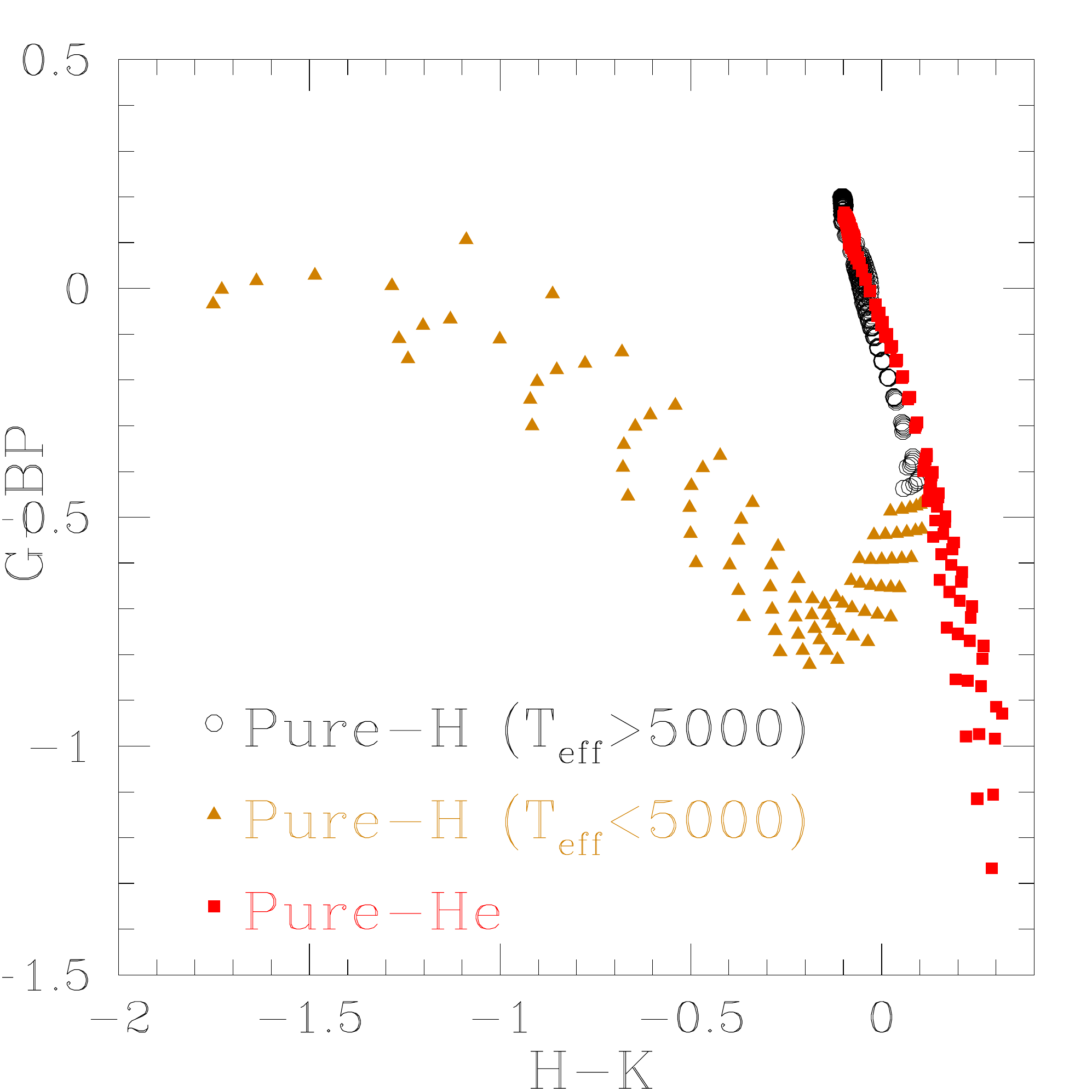}
  \includegraphics[width=.24\textwidth]{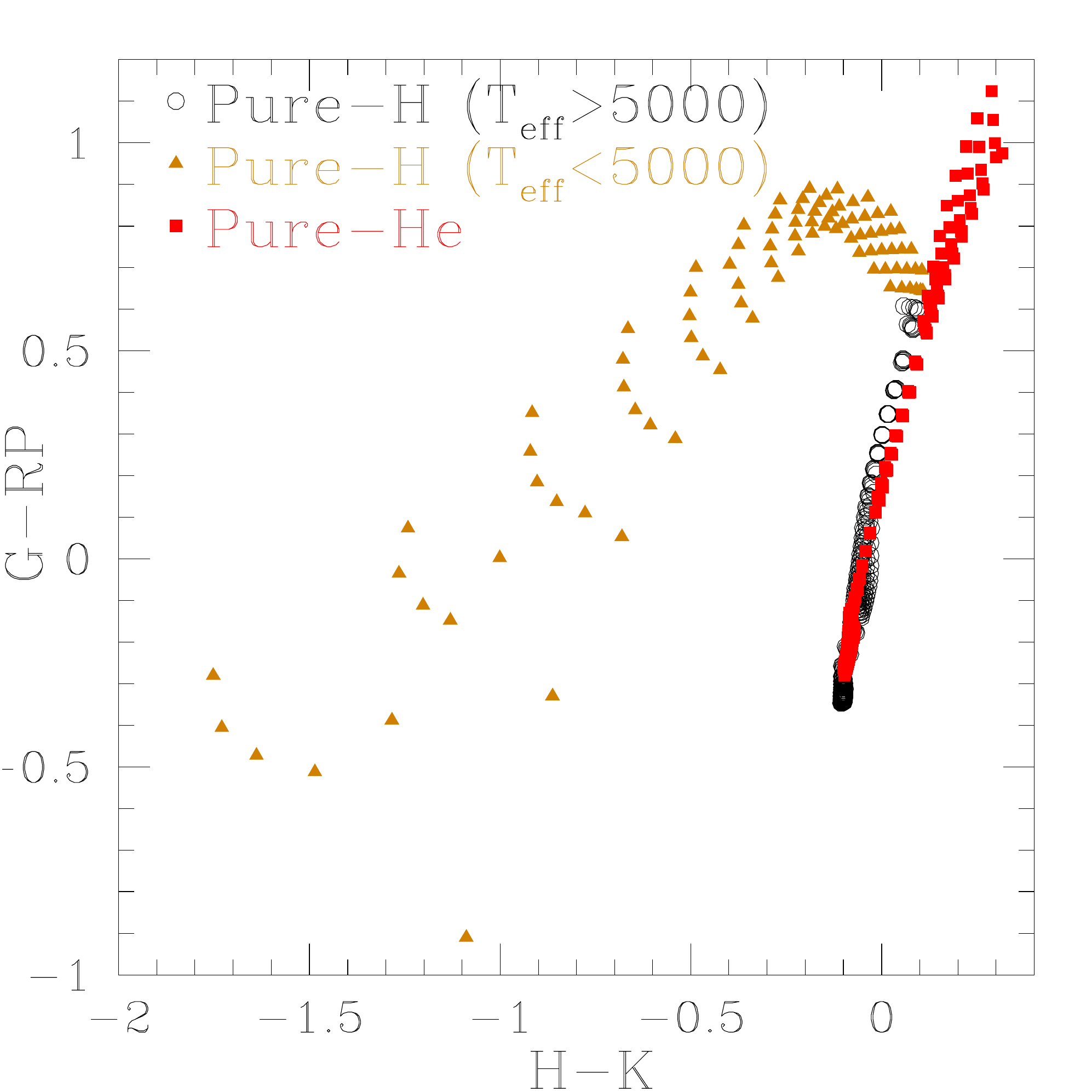}
  \caption{Several colour-colour diagrams obtained using {\Gaia} and 2MASS passbands for all {\teff}.}
  \label{fig:2MASScolours}
\end{figure}

Figure~\ref{fig:2MASScolours} shows some of the diagrams combining {\Gaia} and
2MASS \citep{cohen03} colours, in this case including both 'normal' and cool
WDs. The relationships are not as tight because $\Gaia$ and 2MASS passbands
are sampling different wavelength ranges of the SED and the
2MASS near-IR regime is more sensitive to the composition of WDs than
{\Gaia}'s optical range (see
Fig.~\ref{fig:fXX}). Table~\ref{tab:coeficients_2MASS} (available
  online) provides the coefficients of third-order fittings for 'normal'
pure-H with {\teff}~$>5000$~K and pure-He WDs. The dispersion values are
higher than for Johnson-Cousins or SDSS, as expected, and increase
up to $0.1$~mag in the worst cases. The user can employ the
individual values for the desired temperature and surface gravity listed in the CDS online
Tables~\ref{tab:photDA}~--~\ref{tab:photMix} if the dispersion is too large.

\onltab{7}{
\clearpage
\newpage
\begin{table}[p]
\begin{center}
\caption{Coefficients of the colour-colour polynomial fittings using 2MASS passbands. 
}
\label{tab:coeficients_2MASS}
\tiny
\begin{tabular}{lccccc}
\hline
\multicolumn{6}{c}{\textbf{Pure-H (with \boldmath{{\teff}~$>5000$}~K)}}\\
 \hline
  Colour  &   Zero point   &   $J-H$  &$(J-H)^2$ &$(J-H)^3$ & $\sigma$ \\
 \hline
$G-${\BP}   &   -0.0034 &  -1.1932 &   2.1269 &  -8.0313 &  0.028\\
$G-${\RP}   &    0.0065 &   2.2502 &  -3.2057 &   6.3488 &  0.029\\
$G-${\RVS}  &    0.0356 &   3.0105 &  -4.1872 &   8.1704 &  0.037\\
{\BP}-{\RP} &    0.0099 &   3.4434 &  -5.3326 &  14.3800 &  0.056\\
$G-J$       &   -0.1012 &   4.6781 &  -5.8656 &  12.5043 &  0.062\\
$G-H$       &   -0.1012 &   5.6781 &  -5.8656 &  12.5043 &  0.062\\
$G-K_S$       &   -0.1471 &   6.0017 &  -6.5462 &  15.6901 &  0.068\\
 \hline
 Colour  &   Zero point  &   $H-K_S$  &$(H-K_S)^2$ &$(H-K_S)^3$ & $\sigma$ \\
 \hline
$G-${\BP}   &   -0.1554 &  -3.3439 &   1.7800 &  17.6435 &  0.032\\
$G-${\RP}   &    0.2713 &   5.4984 & -12.2013 & -88.7226 &  0.050\\
$G-${\RVS}  &    0.3897 &   7.3579 & -16.1908 &-120.0339 &  0.069\\
{\BP}-{\RP} &    0.4267 &   8.8424 & -13.9813 &-106.3661 &  0.080\\
$G-J$       &    0.4739 &  11.9522 & -25.0546 &-216.7318 &  0.090\\
$G-H$       &    0.6107 &  14.8963 & -28.7366 &-293.7142 &  0.110\\
$G-K_S$       &    0.6107 &  15.8963 & -28.7366 &-293.7142 &  0.110\\
\hline
\multicolumn{6}{c}{\textbf{Pure-He (All \boldmath{{\teff}})}}\\
 \hline
 Colour  &   Zero point  &   $J-H$  &$(J-H)^2$ &$(J-H)^3$ & $\sigma$ \\
 \hline
$G-${\BP}   &   -0.0161 &  -1.5484 &  -1.7352 &   1.0170 &  0.039\\
$G-${\RP}   &    0.0854 &   2.6145 &  -1.1648 &  -0.3721 &  0.027\\
$G-${\RVS}  &    0.1256 &   3.4444 &  -1.1064 &  -0.6378 &  0.037\\
{\BP}-{\RP} &    0.1015 &   4.1628 &   0.5704 &  -1.3891 &  0.065\\
$G-J$       &   -0.0079 &   5.1327 &  -0.1668 &  -1.4542 &  0.049\\
$G-H$       &   -0.0079 &   6.1327 &  -0.1668 &  -1.4542 &  0.049\\
$G-K_S$       &   -0.0304 &   6.7716 &   0.1327 &  -1.9867 &  0.047\\
 \hline
 Colour  &   Zero point   &   $H-K_S$  &$(H-K_S)^2$ &$(H-K_S)^3$ & $\sigma$ \\
 \hline
$G-${\BP}   &   -0.0661 &  -2.5496 &  -3.4271 &   3.7898 &  0.046\\
$G-${\RP}   &    0.1692 &   3.9294 &  -4.1881 &   2.2276 &  0.036\\
$G-${\RVS}  &    0.2360 &   5.2135 &  -4.4738 &   2.0733 &  0.049\\
{\BP}-{\RP} &    0.2354 &   6.4790 &  -0.7610 &  -1.5622 &  0.080\\
$G-J$       &    0.1599 &   7.8890 &  -3.5933 &   2.1854 &  0.066\\
$G-H$       &    0.1933 &   9.4216 &  -4.3087 &   4.3196 &  0.070\\
$G-K_S$       &    0.1933 &  10.4216 &  -4.3087 &   4.3196 &  0.070\\
\hline
\end{tabular}
\end{center}
\end{table}
}

\section{Classification and parametrisation}
\label{sec:classification}

The spectrophotometric instrument onboard {\Gaia} has been designed to allow the classification of 
the observed objects and their posterior parametrisation. The classification and parametrisation will
be advantageous in front of other existing or planned photometric surveys because of the combination 
of spectrophotometric and astrometric {\Gaia} capabilities. The extremely precise parallaxes will permit one to 
decontaminate the WD population from cool MS stars or
subdwarfs. Parallaxes are especially important for very cool WDs ({\teff}~$< 5000$~K) since {\Gaia} will provide the necessary data to derive the masses of already known WDs (for the first time) and newly discovered cool WDs.

The classification and a basic parametrisation of the sources will be provided in the intermediate and final {\Gaia} data releases in addition to the spectrophotometry and integrated photometry. This classification and parametrisation process will be performed by the {\Gaia} Data Processing and Analysis Consortium, which
is in charge of the whole data processing, and will be based on all astrometric, spectrophotometric, and spectroscopic {\Gaia} data \citep{coryn13}. The purpose here is not to define or describe the methods to be used to determine the astrophysical parameters of WDs, but just to provide some clues on how to obtain from {\Gaia} the relevant information to derive the temperature, surface gravity, and composition of WDs.

\begin{description}
\item[\textbf{Effective temperature:}]
in Fig.~\ref{fig:bprpteff}, we can see the strong correlation between {\BP}~$-$~{\RP} colour and {\teff} valid for all compositions when {\teff}~$> 5000$~K, while for the cool regime the colour-temperature relationship
depends on the composition and the surface gravity.
The flux depression in the IR due to CIA opacities 
in pure-H and mixed compositions
causes that the relationship presents a turnaround at {\BP}~$-$~{\RP} $\sim 1.7$. WDs with {\BP}~$-$~{\RP} $>1.7$ 
most likely have pure-He composition and {\teff}~$\sim 3000$--$4000$~K. Therefore, {\Gaia} is expected to shed light on the atmospheric composition of the coolest WDs.

{\BP}~$-$~{\RP} colours for the WDs observed by SDSS and included in \cite{kil10} have been computed from
their 
$g-z$ colours using the polynomial expressions in Table~\ref{tab:coeficients} (available online) and are also included in Fig.~\ref{fig:bprpteff} for comparison purposes. Assuming that a prior classification in the 'normal' or cool regime has been performed from parallax information, the {\BP}$-${\RP} vs {\teff} relationship can be used to derive temperatures. The slope of the relationship for pure-He WDs and the expected errors of {\BP} $-$ {\RP} (estimated from {\Gaia} expected performances webpage\repeatfootnote{foot:performances}) were used to compute the {\teff} errors for the \cite{kil10} WDs (Fig.~\ref{fig:sigmaTvsT_SDSS}). 
From these SDSS WDs we found a mean $\sigma_{T_{\rm eff}}/T_{\rm eff}$ of $\sim 1$\%.

\begin{figure}
\hspace{0.5cm}
  \includegraphics[width=.4\textwidth]{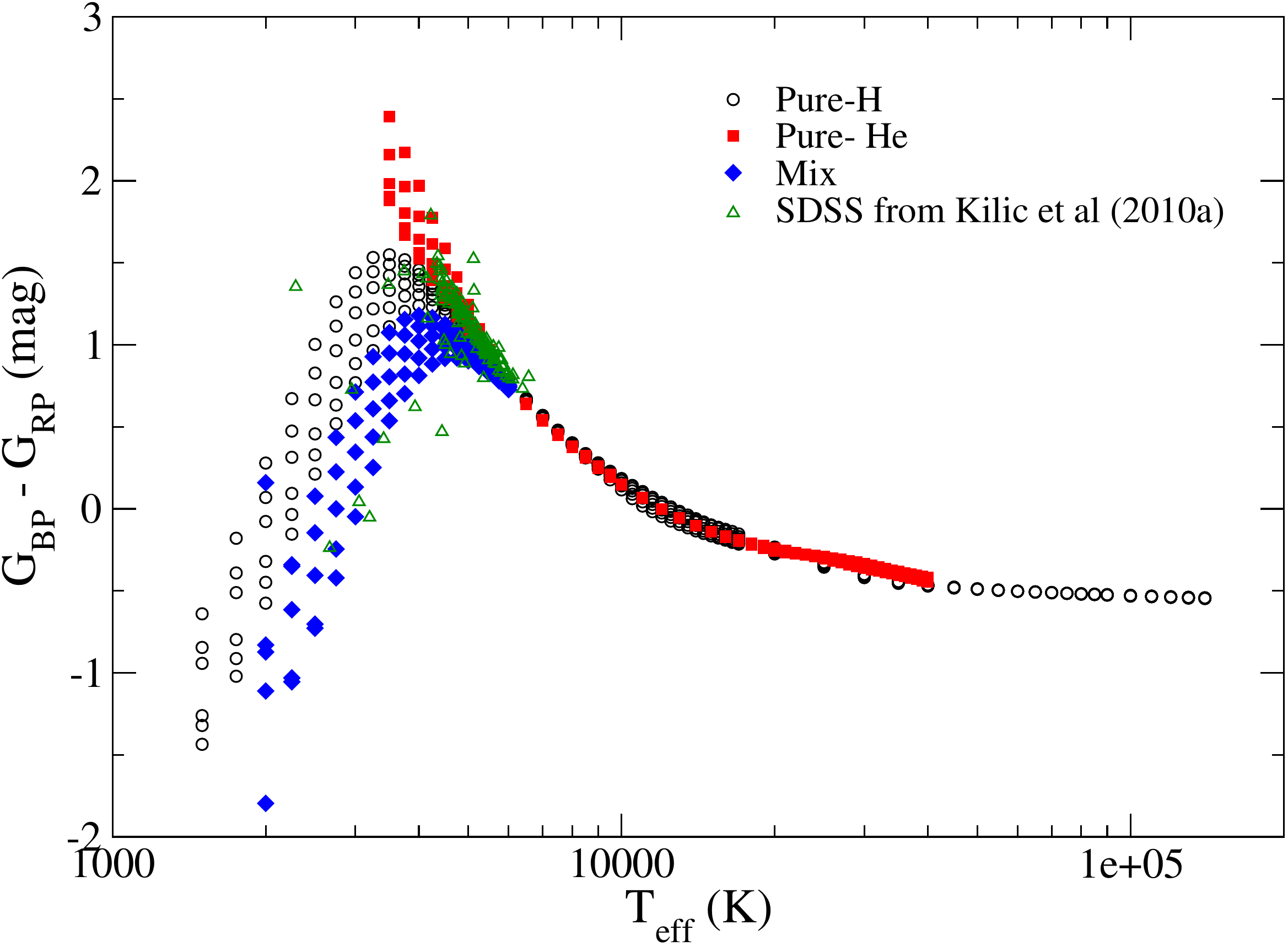}
  \caption{{\BP}-{\RP} colour dependency with  
{\teff} for pure-H (black), pure-He (red), or mixed composition (blue) WDs. WDs observed by SDSS and included in \cite{kil10} are also plotted (in green).
}
  \label{fig:bprpteff}
\end{figure}

 \begin{figure}
  \begin{center}
  \includegraphics[width=.45\textwidth]{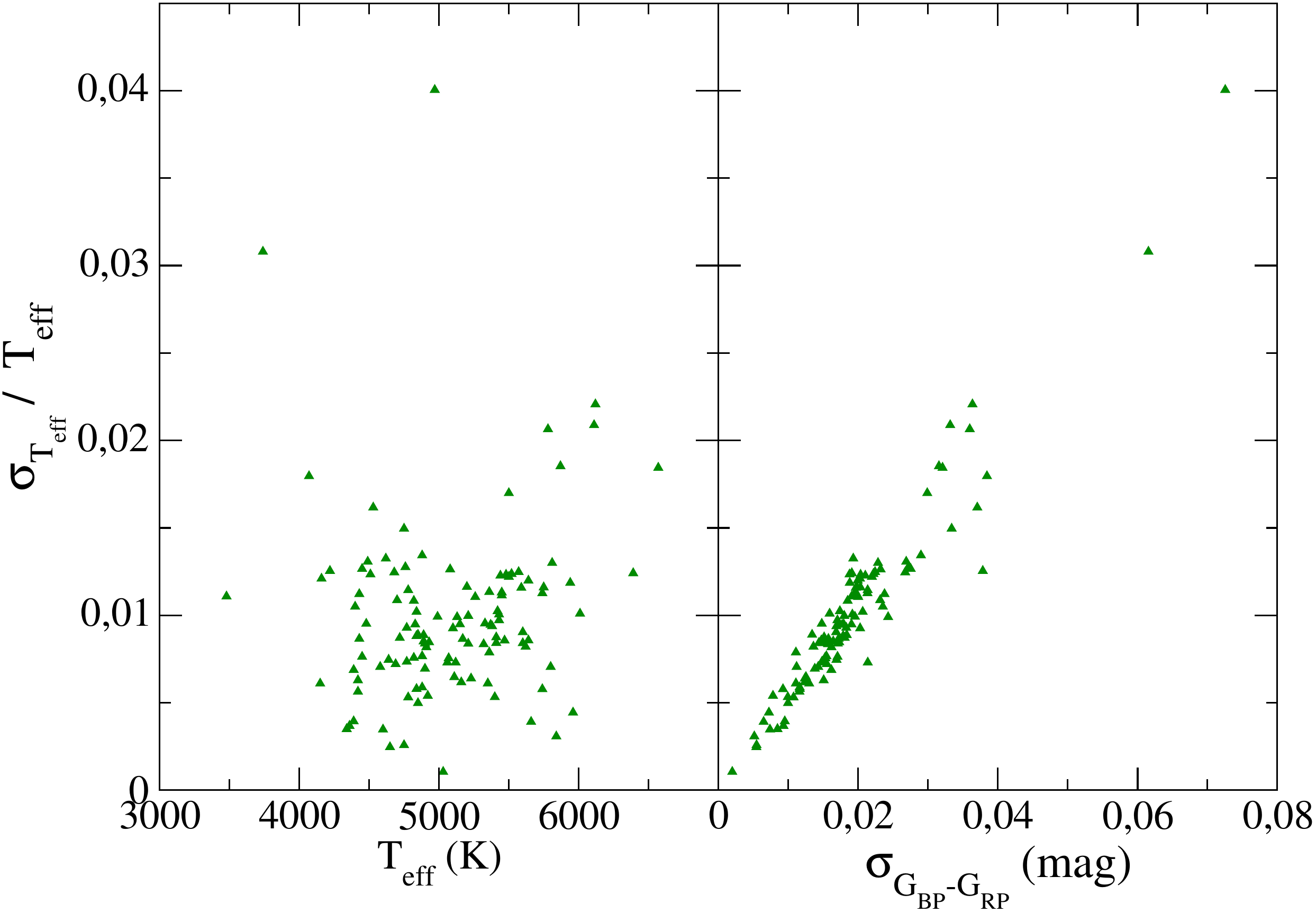}     
   \end{center}   
        \caption{\small{Relative error in {\teff} derived for real WDs observed by SDSS and extracted from \cite{kil10} as a function of {\teff} (left) and of the uncertainty in {\Gaia} {\BP}$-${\RP} colour (right).}
\label{fig:sigmaTvsT_SDSS}    
}   
\end{figure}

\item[\textbf{Surface gravity:}]

the spectral region around $\lambda=360$--$500$~nm with the hydrogen Balmer lines is particularly useful to derive the surface gravity. Although the transmission of {\BP} is low at this wavelength range, the end-of-mission signal-to-noise
ratio is better than 15 for $G\le16.5$ (if {\teff}~$= 5000$~K) or even $G\le17.7$ (if {\teff}~$=40\,000$~K). Unfortunately,
it is doubtful that precise atmospheric parameters
determination can be performed with {\Gaia} BP/RP spectra alone
because of the low resolution of the BP/RP instruments. We recall that by combining the {\Gaia} parallaxes and magnitudes with {\teff} estimates as described above and a theoretical mass-radius relationship for WDs, it is possible to derive fairly precise $\log g$ values.

\item[\textbf{Chemical composition:}]

the difference between pure-H and pure-He is visible in the Balmer jump
(with a maximum around {\teff}~$\sim 13\,000$~K) and the analysis of BP/RP of WD
spectra (not only their colours) will help to identify the differences in composition (see Fig.~\ref{fig:gaiapassbands}). However,
to determine atmospheric parameters, it will be better to use {\Gaia} photometry constrained by the parallax information. Additional observations may be necessary to achieve a better accuracy on the atmospheric parameters, such as additional photometry or higher resolution spectroscopic follow-up.
A comparison of the masses obtained from the {\Gaia} parallaxes with those determined from spectroscopic follow-up will allow one to test the mass-radius relations and the internal chemical composition for WDs. 

{The \Gaia} RVS range around 860~nm is not optimal to see features in WD SEDs (only DZ and similar rare WDs with metal lines will show some features in this region). Ground-based follow-up spectroscopic observations around the hydrogen Balmer lines need to be obtained to derive radial velocities and abundances.

The SED of the pure-H and pure-He WDs differ considerably in the IR wavelength range, particularly in the cool domain (see Fig.~\ref{fig:fXX}), and thus, the combination of {\Gaia} and IR photometry will allow one to disctinguish among compositions. Figure~\ref{fig:2MASScolours} shows {\Gaia}$-$2MASS colour-colour diagrams. The all-sky 2MASS catalogue can be used for WDs brighter than $J=16.5$, $H=16.5$ and $K=16$, which are the limiting magnitudes of the survey. {\Gaia} goes much fainter, and therefore near-IR surveys such as the UKIDSS Large Area Survey ($J_{\rm lim}=20$, $K_{\rm lim}=18.4$, \citealt{hew06}), VIKING (VISTA Kilo-Degree Infrared Galaxy Survey; $J_{\rm lim}=20.9$, \citealt{findlay2012}), and VHS (VISTA Hemisphere Survey, $J_{\rm lim}=21.2$, \citealt{arn10}) will be of great interest, although they only cover $4000$, $1500$, and $20\,000$ deg$^2$, respectively.

\end{description}

\section{White dwarfs in the Galaxy} 
\label{sec:simulation}

The currently known population of WDs amounts to $\sim$~$20\,000$ objects \citep{kleinman13}.
This census will be tremendously increased with the {\Gaia} all-sky deep
survey, and more significantly for the halo population and the cool regime
\citep{torres05}. We present estimates of the number of WDs
that will be observed (see Tables~\ref{tab:numberWD}~--~\ref{tab:numberWDvar}
and Figs.~\ref{fig:Mvhistogr_nap09}~--~\ref{fig:HRobs}) according to two
different simulations: one extracted from the {\Gaia} Universe Model Snapshot
\citep[GUMS,][]{rob12} and another one based on \cite{nap09}. These
simulations (which only include pure-H WDs) were
adapted to provide {\Gaia} -observed samples (limited to $G=20$). Both
simulations build their stellar content based on the galactic structure in the
Besan\c{c}on Galaxy Model, \citep[BGM, ][]{rob03}, but some differences were
considered in each simulation since then.

The BGM considers four different stellar populations: thin disc,
  thick disc, halo, and bulge, of which the latter has little relevance for our
  simulations and will thus not be discussed any further.  An age-dependent scale
  height is adopted for the thin disc and both WD simulations assume a
  constant formation rate of thin-disc stars.  The
  thick disc is modelled with a scale height of 800\,pc, and the stellar halo
  as a flattened spheroid.

Although the ingredients of the simulations have been detailed in
  the original papers, we briefly summarize in Sects.~\ref{sec:GUMS} and
\ref{sec:napiwotzki} some relevant information for each
  simulation to help understand their compared results in Sect.~\ref{sec:comparison}.

\subsection{Simulations using GUMS}
\label{sec:GUMS}

GUMS provides the theoretical Universe Model Snapshot seen by {\Gaia} and is being used for mission preparation and commissioning phases. It can be used to generate stellar catalogues for any given direction and returns information on each star, such as magnitude, colour, and distance, as well as kinematics and other stellar parameters. Although only WD simulations were extracted, GUMS also includes many Galactic and extra-galactic objects, and within the Galaxy it includes isolated, double, and multiple stars as well as variability and exoplanets. The ingredients of GUMS regarding the WD population are summarized here:

\begin{itemize}

\item Thin-disc WDs are modelled following the \cite{fon01} pure-H WD
  evolutionary tracks, assuming the \cite{wood92} LF for an age of 10 Gyr and an
  IMF from \cite{salpeter55}. It has been normalised to provide the same number
  of WDs of $M_V\le 13$ as derived by \cite{lie05} from the Palomar Green
  Survey. The photometry was calculated by \cite{ber11}.

\item For the thick disc, the WD models of \cite{chabrier99} were used assuming an age of 12 Gyr. For the normalisation, the ratio between the number of MS turn-off stars and the number of WDs depends on the IMF and on the initial-to-final mass relation, assuming that all stars with a mass greater than the mass at the turn-off ($M_{TO} = 0.83$~M$_{\odot}$) are now WDs. However, the predicted number of thick-disc WDs with this assumption is much higher than the number of observed WDs in the \cite{opp01} photographic survey of WDs. This led \cite{reyle01} to normalise the thick-disc WD LF to that of the \cite{opp01} sample, assuming it is complete: 
the ratio between the number of WDs (initial mass $> 0.83$~M$_{\odot}$) and the number of MS stars (initial mass $<0.83$~M$_{\odot}$) is taken to be 20\%.

Because the completeness of the \cite{opp01} sample is not certain, this value can be considered as a lower limit. This normalisation also agrees with the number of WDs in the expanded Luyten half-second proper motion catalogue (LHS) WD sample \citep{liebert99}.

\item The halo LF is derived from the truncated power-law initial mass function IMF2 of \cite{chabrier99} with an age of 14 Gyr. It was specifically defined to generate a significant number of MACHOS in the halo to explain the number of microlensing events towards the Magellanic Clouds. Hence the density of halo WDs is assumed to be 2\% of the density of the dark halo locally. Because of the assumed IMF, which is highest at M $= 2.4$~M$_{\odot}$, the generated WDs have cooled down with an LF peaking at $M_V\approx 16.5-17$. Hence the number of halo WDs observable at $G<20$ is rather small (see Table~\ref{tab:numberWD}), despite their large local density. This is one of the main differences (see Sect.~\ref{sec:comparison}) with respect to simulations obtained with \cite{nap09}. The total local density for each population is included in Table~\ref{tab:localdensities}.

\item GUMS adopts the 3D extinction model by \cite{drimmel03}.

\end{itemize}

\begin{table}
\begin{center}
\caption{Local densities used in both WD Galaxy simulations, expressed in M$_{\odot} \cdot$ pc$^{-3}$.}
\begin{tabular}{ccc}
\hline
\noalign{\smallskip}  
 & \cite{nap09} & GUMS \\
\hline
Thin  & $2.9\cdot 10^{-3}$ & $4.16 \cdot 10^{-3}$ \\ 
Thick & $1.7\cdot 10^{-3}$ & $5.06 \cdot 10^{-4}$ \\ 
Halo  & $2.7\cdot 10^{-4}$ & $2.80 \cdot 10^{-4}$ \\ 
 \noalign{\smallskip}
\hline
\end{tabular}
\label{tab:localdensities}
\end{center}
\end{table}

\begin{itemize}

\subsubsection{Variable and binary WDs}

\item Several types of intrinsic stellar variability were considered, and for WDs, they include pulsation of ZZ-Ceti type, cataclysmic dwarfs, and classical novae types. For WDs in close-binary systems (period shorter than 14 hours), half of them are simulated as dwarf novae and the other half as classical novae. WD$+$WD close-binary systems are also considered in these simulations.

\item Although the simulation of binaries is already quite realistic in GUMS, it still needs some further refinement in the specific case of WDs.  

  Work is on-going
  to improve the mass distribution in new versions of  the {\Gaia} simulator. For this reason, we are still unable to detail the number of WDs in binary systems in which orbital parameters can be obtained to provide independent mass estimates of the WDs. This study will be possible with future versions of the GUMS simulations.

\end{itemize}

\subsection{Simulations using \cite{nap09}}
\label{sec:napiwotzki}

  \cite{nap09} used the Galactic model structure of
  BGM to randomly assign positions of a large number of stars based
  on observed densities of the local WD population. Depending on population membership, each star is given a metallicity, an initial mass, and kinematical properties. The ingredients are summarized in the following items:

\begin{itemize}
\item The WD cooling sequences were created by combining the
  tracks of \cite{blo95} for hot WDs with the sequences of \cite{fon01} for
  cooler WDs.
\item The WD progenitor lifetime was calculated from the stellar tracks of \cite{gir00}.
\item The initial-final mass relationship by \cite{wei00} was considered.

\item The relative contributions of each Galactic population were calibrated using the kinematic study of \cite{pau06} of the brightness-limited SPY (ESO SN~Ia Progenitor Survey: \citealt{nap01}) sample. \cite{pau06} assigned population membership according to a set of criteria based on the measured 3D space velocity. An iterative correction for mis-assignment of membership of WDs with ambiguous kinematic properties was derived from the model \citep[see][]{nap09}.

\item Space densities were then calculated from the relative contributions and
the WD numbers in the 13\,pc sample of \cite{hol08}. 
The finally adopted local densities can be seen in Table~\ref{tab:localdensities}.

\item \citet{hol06} colour tables were used for WD simulations, and photometric transformation from Table~\ref{tab:coeficients} of the present paper were used to carry out the brightness selection for the {\Gaia} sample.

\item IMFs from \cite{salpeter55} were used to derive numbers in
  Table~\ref{tab:numberWD}, although other IMFs
  \citep[e.g.][]{chabrier99,bau05} were tested as well. These would give higher numbers of
  cool, high mass WDs.

\end{itemize}

\cite{nap09} compared the simulations with the proper-motion-selected sample of \cite{opp01}. Simulated numbers turned out to be slightly higher, which can easily be explained by a minor incompleteness of the observed sample.

\subsection{Results}
\label{sec:comparison}

Table~\ref{tab:numberWD} shows the number of WDs detectable for each population (about $200\,000$--$250\,000$ single WDs in total) according to the two simulations explained above. Table~\ref{tab:numberWDvar} shows the expected number and types of variable WDs observed by {\Gaia}, according to GUMS.

\begin{table}
\begin{center}
\caption{Total number of WDs with $G\leq 20$ expected in {\Gaia} for different {\teff} ranges.}
\begin{tabular}{lccc}
\hline
\noalign{\smallskip}  
{\teff}  &  $N_{\rm Thin}$ & $N_{\rm Thick}$ & $N_{\rm Halo}$ \\
\noalign{\smallskip}
\hline
\multicolumn{4}{c}{\cite{nap09}}\\
\hline
\noalign{\smallskip}      
All range, single & $196\,765$ & $48\,673$ & $9705$ \\
$< 5000$~K, single &  $925$ &  $1883$ & $347$ \\
\noalign{\smallskip}
\hline
\multicolumn{4}{c}{GUMS}\\
\hline
All range, single     & $198\,107$ & $3557$ & 63 \\
All range, Comp A     & $64\,905$  & $2340$ & 47 \\
All range, Comp B     & $296\,976$ & $1153$ &  4 \\
$< 5000$~K, single     & $8845$   & 
142  
& 63 \\
$< 5000$~K, Comp A     &  862   &  
95  & 47 \\
$< 5000$~K, Comp B     &  244   &  
12  &  4 \\
\hline
\end{tabular}
\label{tab:numberWD}
\end{center}
\end{table}

\begin{table}
\begin{center}
\caption{Number and types of variable WDs as predicted by GUMS.}
\begin{tabular}{lccc}
\hline
\noalign{\smallskip}  
Variables &  ZZCeti & Dwarf  & Classical  \\
          &         &  Novae &  Novae \\
\hline
\noalign{\smallskip} 
Isolated  & $18\,363$ & -- & -- \\
In binary & $6463$ & 340 & $10\,258$ \\
\hline
\end{tabular}
\label{tab:numberWDvar}
\end{center}
\end{table}

Most of the observed WDs belong to the thin disc because of the still on-going star formation. The number of thin-disc WDs in the two simulations agree remarkably well at the level of 1\% for single stars. We note that in GUMS binaries are added and produce about $3.6\cdot10^5$ additional thin-disc WDs (see Table~\ref{tab:numberWD}). 
However, this estimate includes WDs in binary systems that are
not resolvable by {\Gaia} alone. Just for comparison, previous estimates of the number of WDs in the disc reported by \cite{torres05} provided $3.9\cdot10^5$ WDs with $G<20$~mag.

The small number of WDs known in thick-disc and halo populations is the current limiting factor to constrain their properties. 
Due to these limitation, \cite{nap09} and GUMS arrive at different interpretations of the \cite{opp01} sample.
The local thick-disc density assumed by \cite{nap09} is three times higher than the one assumed by GUMS (see Table~\ref{tab:localdensities}). The local densities for the halo are more similar, but the more stringent difference comes from the shapes of the assumed LF and the IMFs. The IMF assumed by GUMS leads to a very low estimated number of thick-disc and halo WDs observable by {\Gaia}.

As expected, the number of observable cool WDs ({\teff}~$< 5000$~K) of all populations will be considerably smaller than the total sample. However, this will still represent a dramatic increase compared with any previous survey (e.g. only 35 cool WDs were found in the SDSS sample by \citealt{har06}).

In addition to this increase in the number of cool WDs, one of the main contributions of the {\Gaia} mission will also be the large portion of sources with good parallaxes (see Table~\ref{tab:N_WDparallax} for estimates of the number of WDs observed in all temperature regimes). For cool WDs ({\teff}~$< 5000$~K), GUMS estimates about $2000$ with $\frac{\sigma_{\pi}}{\pi}<1\%$ ($20\%$ of all predicted cool WDs). Furthermore, the complete sample of about 9000 observed cool WDs will have $\frac{\sigma_{\pi}}{\pi}<5.5\%$. Table~\ref{tab:N_WDparallax}
shows that all cool ({\teff}~$< 7000$~K) WDs in the \cite{kil10} sample extracted from SDSS will have $\frac{\sigma_{\pi}}{\pi}<5\%$ when observed with {\Gaia}.

\begin{figure}
\centering
  \includegraphics[width=.48\textwidth]{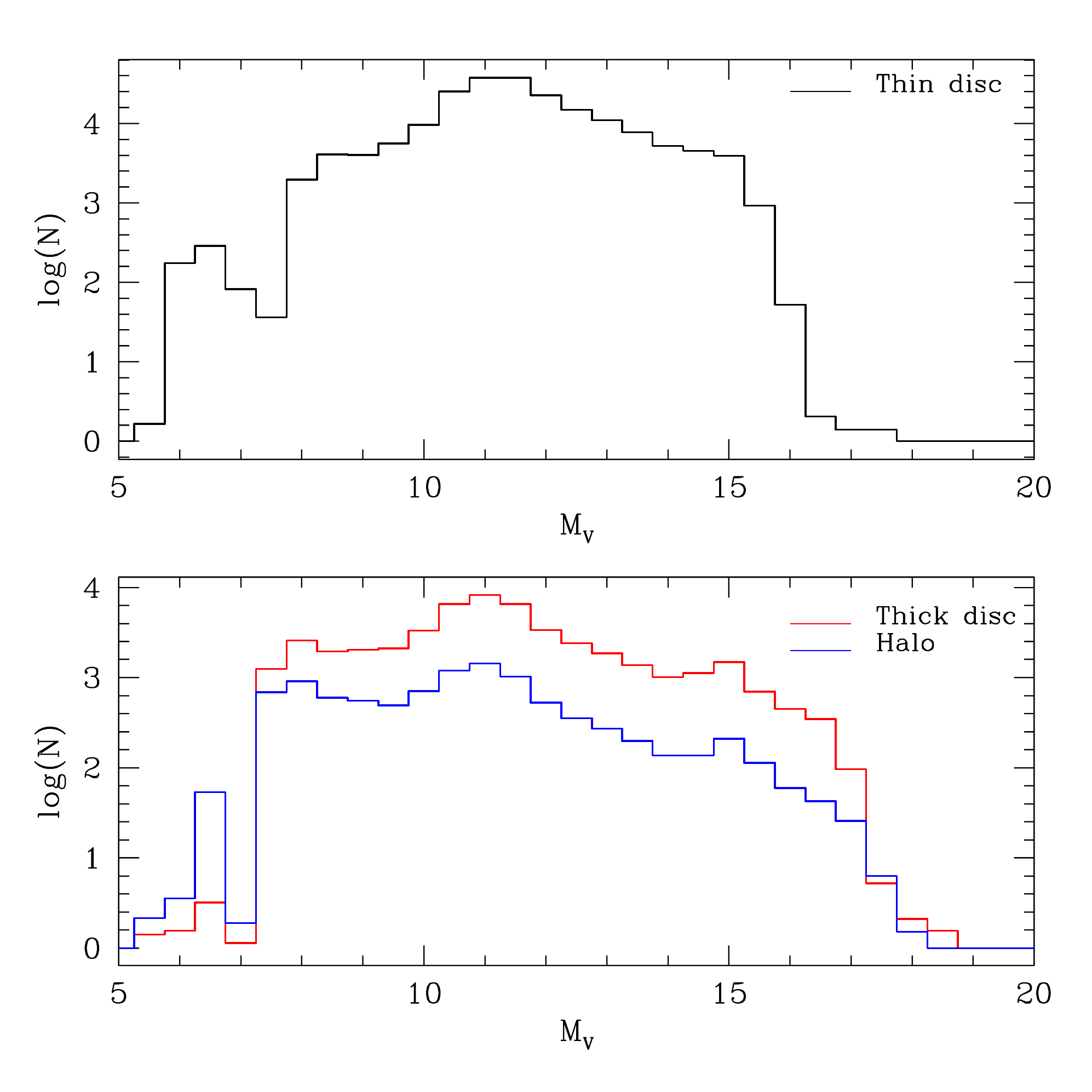}
  \caption{Simulated {\Gaia} $M_V$ distribution of a single WD population with $G<20$~mag from thin-disc (top panel), thick-disc and halo populations (bottom panel) based on \cite{nap09}.
  }
  \label{fig:Mvhistogr_nap09}
\end{figure}

\begin{figure}
\centering
  \includegraphics[width=.48\textwidth]{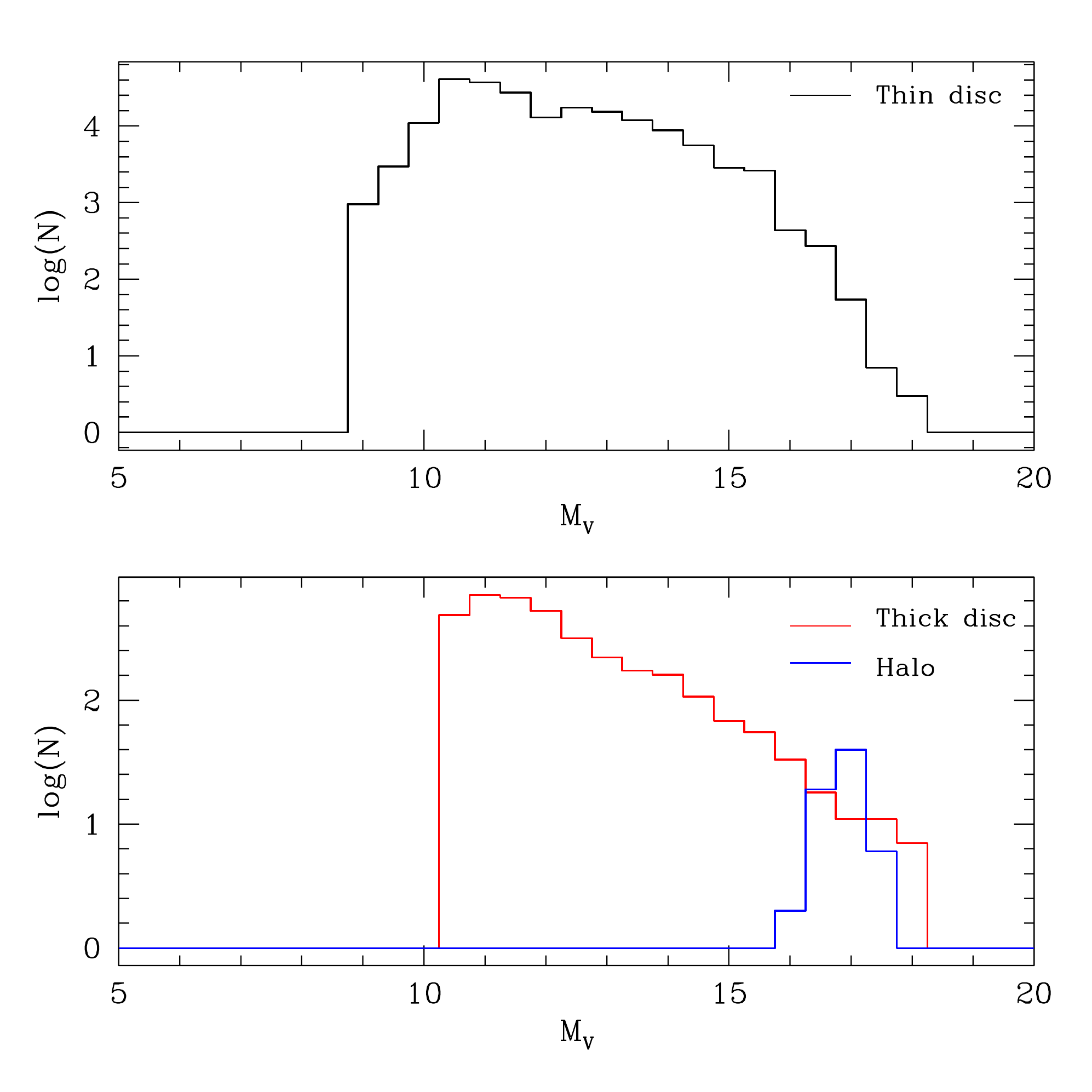}
\caption{Same as Fig.~\ref{fig:Mvhistogr_nap09}, now using GUMS simulations.
}
  \label{fig:Mvhistogr_GUMS}
\end{figure}

The resulting WDs $M_V$ distributions are plotted in Figs.~\ref{fig:Mvhistogr_nap09} and \ref{fig:Mvhistogr_GUMS} for \cite{nap09} and GUMS simulations, respectively. Values higher than average $M_V$ are more frequent for thick-disc and halo populations than for thin-disc populations, especially in the \cite{nap09} simulations. These represent the oldest WDs ($>10$~Gyr) produced from higher mass progenitors (with much higher masses than the $0.45$--$0.50$  M$_{\odot}$ WDs that dominate brightness-limited samples of halo WDs).
IMFs such as those from \cite{chabrier99,bau05} would enhance these tails resulting in much higher numbers of cool WDs.

\begin{figure}
\centering
  \includegraphics[width=.45\textwidth, height=170px]{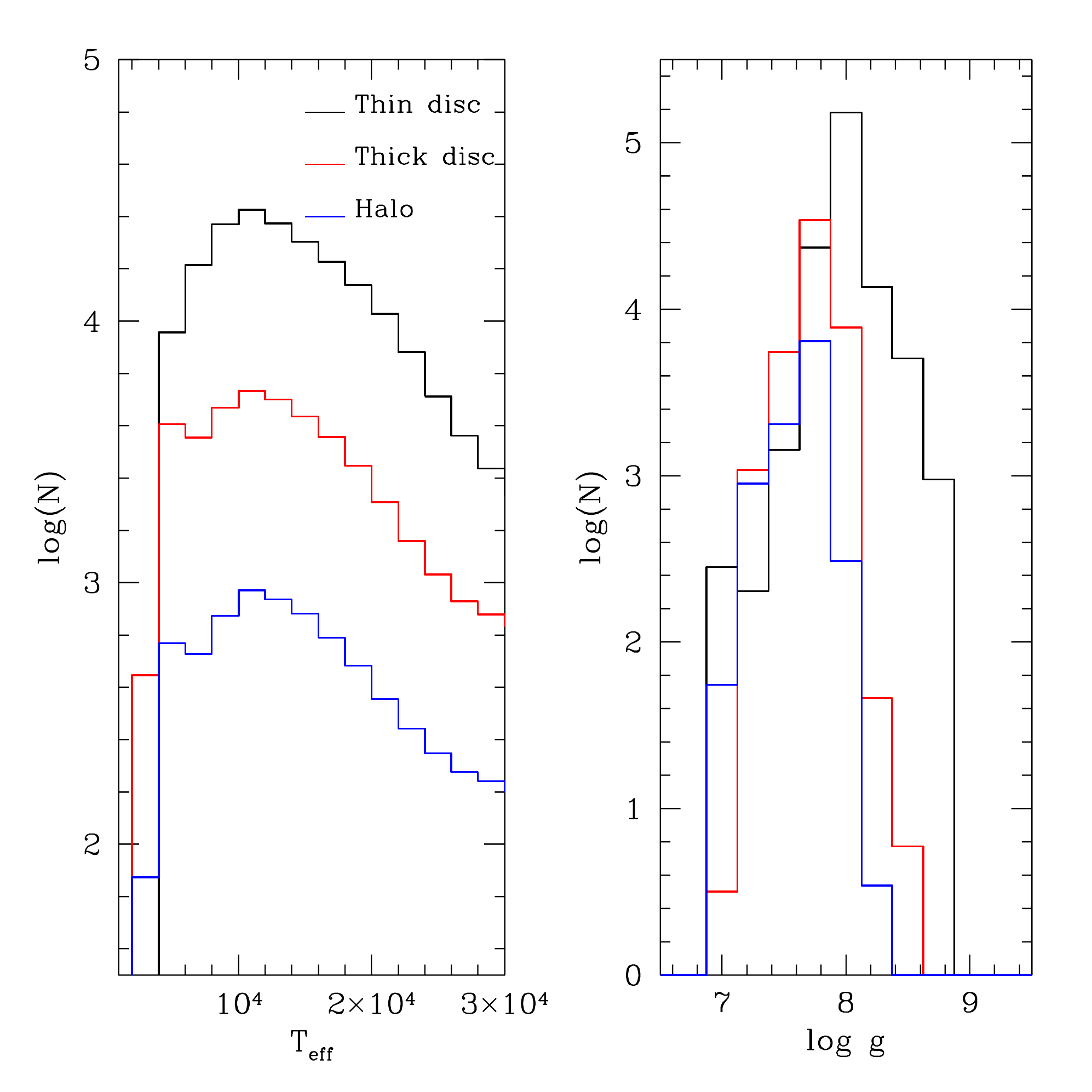}
  \includegraphics[width=.45\textwidth, height=170px]{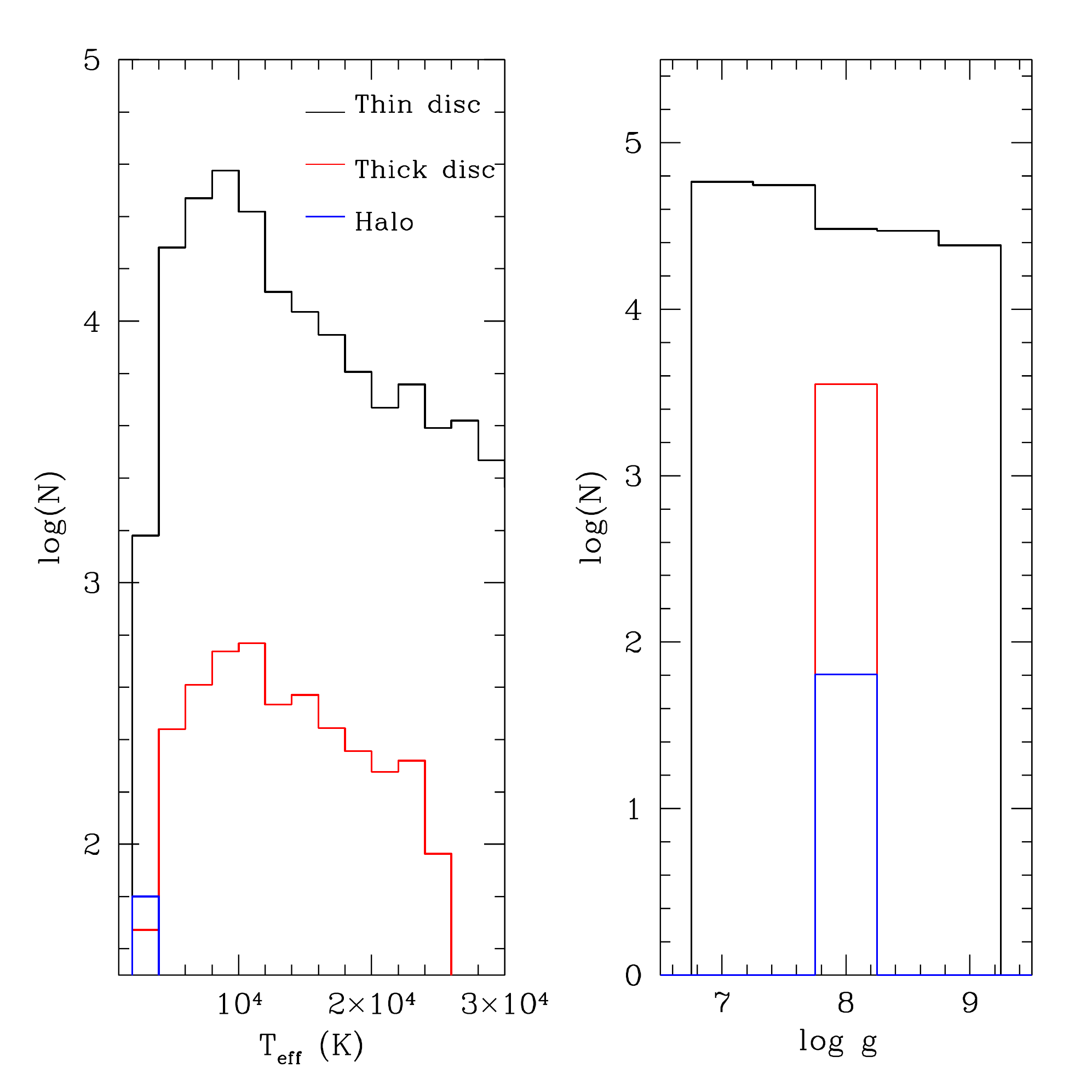}
  \caption{Comparison of {\teff} and $\log g$ histograms for single WDs in thin-disc, thick-disc, and halo populations obtained using \citealt{nap09} (upper panels) and GUMS simulations (bottom panels). GUMS halo and thick-disc WDs are assumed to have $\log g=8.0$.
}
\label{fig:TAndLogghistogrcompar}    
\end{figure}

Differences in the number of predicted WDs can also be seen in the {\teff} and $\log g$ histograms plotted in Fig.~\ref{fig:TAndLogghistogrcompar}.
In GUMS there are no halo WDs hotter than about 5000~K because of the type of IMF is assumed with an age of 14 Gyr. Thus, the detection and identification of these old massive WDs with {\Gaia} will be extremely helpful to obtain information about the IMF of the Galaxy. The peak of the {\teff} distribution for \cite{nap09} shown in Fig.~\ref{fig:TAndLogghistogrcompar} (top) is centred around $12\,000$~K, which is the result of the interplay between the WD cooling rate and the change of absolute brightness with temperature.

The present version of GUMS considers only discrete values for $\log g$ (from $7.0$ to $9.0$ in steps of $0.5$). In particular, for the thick-disc and halo WDs only $\log g=8.0$ was assumed (see Fig.~\ref{fig:TAndLogghistogrcompar}). In the simulations of \cite{nap09}, the peak tends to be lower for the thick-disc and halo populations than for the thin disc. This agrees with the results obtained by the SPY project \citep{pau06}, showing that hot WDs ({\teff}~$>10\,000$~K) that belongs to the halo population have masses in the range $0.45$--$0.50$ M$_{\odot}$.

\begin{figure}
  \begin{center}
\includegraphics[width=.39\textwidth, height=150px]{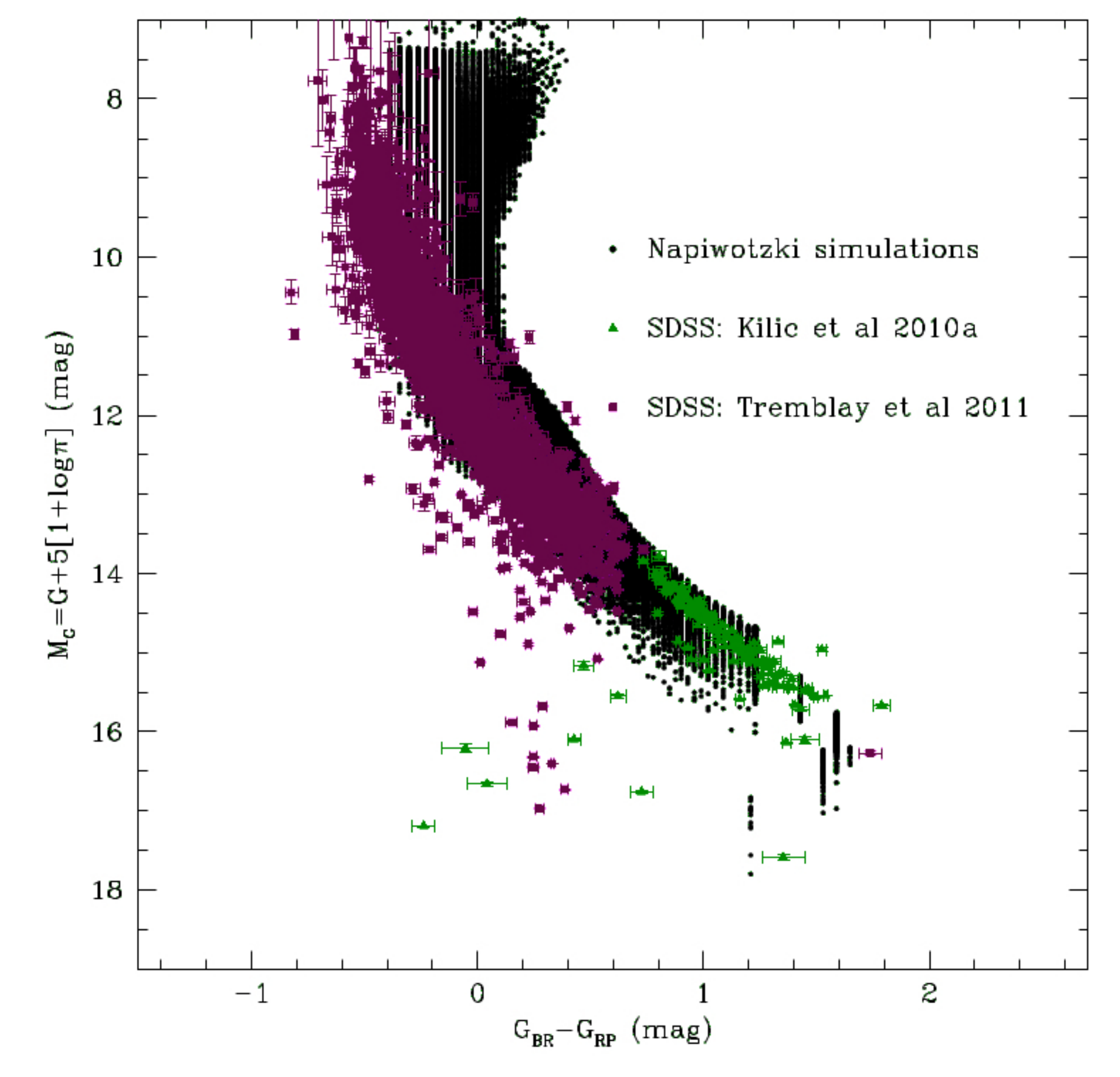}
  \includegraphics[width=.365\textwidth]{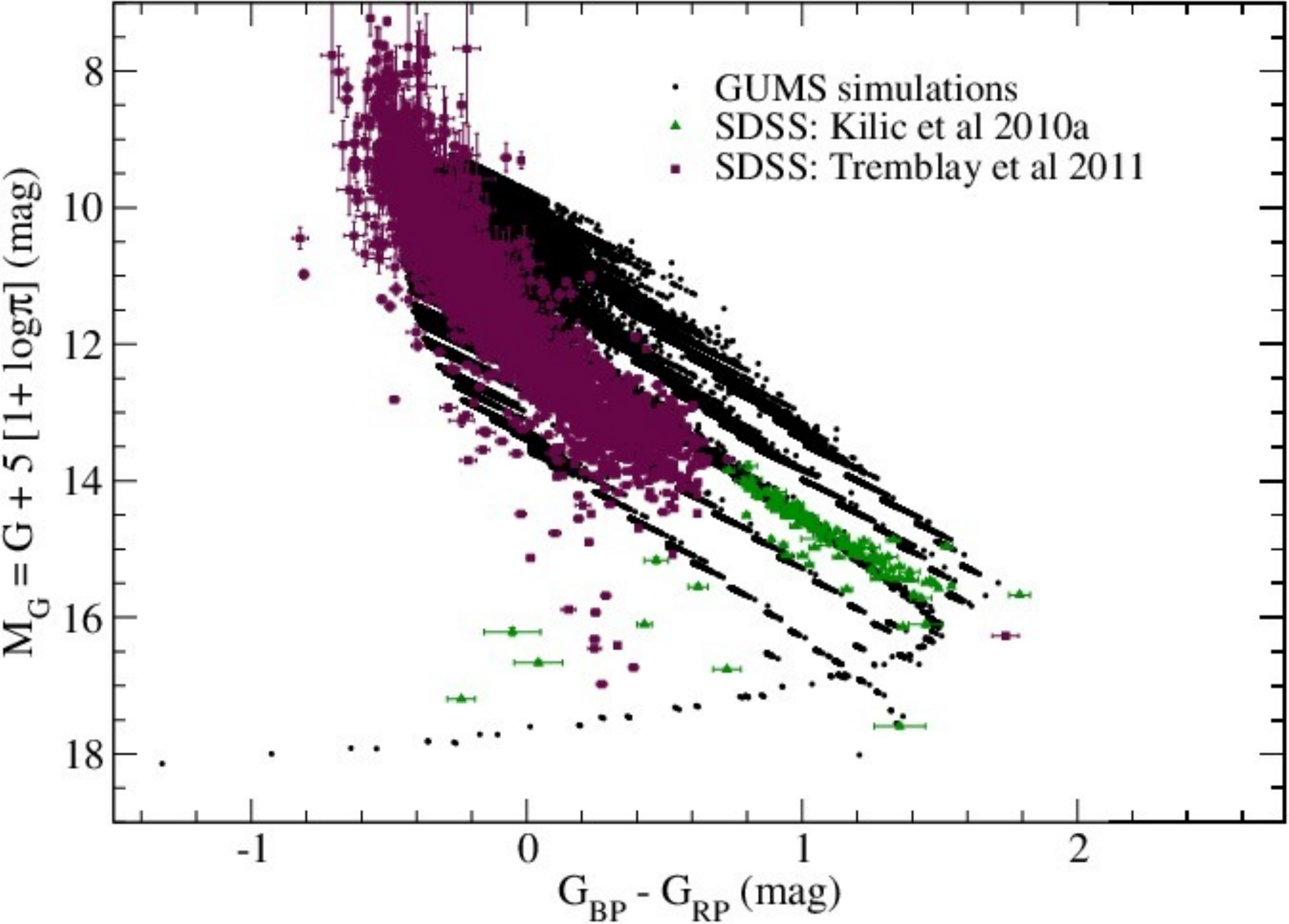}
   \end{center}   
        \caption{\small{
        Hertzsprung--Russell diagram built using only {\Gaia} observable quantities ($G$, $G_{BP}$, $G_{RP}$, and parallaxes) for single WDs with $G\leq20$ obtained from \cite{nap09}, top, and GUMS, bottom. The Hertzsprung--Russell GUMS diagram considers discrete values for $\log g$, which is reflected as quantised trends in this plot. SDSS WD samples extracted from \cite{tre11} and \cite{kil10} are overplotted with error bars as expected from {\Gaia} observations. Table~\ref{tab:N_WDparallax} shows that almost all SDSS WD plotted here have relative error of parallaxes better than 10\%.
}}
\label{fig:HRobs}    
\end{figure}

The predicted HR diagram in {\Gaia} observables is shown in Fig.~\ref{fig:HRobs}. According to Fig.~18 in \cite{rob12}, the low-mass MS can be perfectly separated from the WD branch (at least for {\teff}~$>3000$~K). Therefore, we will be able to separate cool WDs from cool MS stars in the {\Gaia} catalogues.

We did not aim to discuss which of the two simulations considered here better reproduces the reality, but only to provide an estimate of the number of WDs in different stellar populations predicted to be observed by {\Gaia}. On the other hand, a comparison of the true {\Gaia} data with our simulations will improve our knowledge about the Milky Way formation and validate the models and assumptions.

%

\section{Summary and conclusions}
\label{sec:conclusions}

We have presented colour-colour photometric transformations between {\Gaia} and other common optical and IR photometric systems (Johnson-Cousins, SDSS and 2MASS) for the case of WDs. To compute these transformations the most recent {\Gaia} passbands and WD SED synthetic libraries \citep{tre11,ber11} were used.

Two different behaviours were observed depending on the WD effective
temperature. In the 'normal' regime ({\teff}~$> 5000$~K) all WDs with the same
composition (pure-H or pure-He) could be fitted by a single law to transform
colours into the {\Gaia} photometric system. For the very cool regime, WDs
with different compositions, {\teff} and $\log g$, fall in different positions
in colour-colour diagram which produces a spread in
these diagrams. Colours with blue/UV information, like the $B$ Johnson passband,
seem to better disctinguish the different WD characteristics, but the
measurements in this regime will be rather noisy because of the low photon counts for
very cool sources and in practice might be hard to use. We therefore expect that observations in near-IR passbands, combined with {\Gaia} data, might be very helpful in characterising WDs, especially in the cool regime. 

Estimates of the number of WDs that {\Gaia} is expected to observe during its
five-year mission and the expected
precision in parallax were also provided. According to the number of sources predicted by
\cite{nap09} and by the {\Gaia} Universe Model Snapshot \citep{rob12}, we expect between $250\,000$ and $500\,000$ WDs detected by {\Gaia}. A few thousand of them will have {\teff}~$< 5000$~K, which will increase the statistics of these very cool WDs quite substantially, a regime in which only very few objects have been observed until now \citep{catalan12,har06}.

{\Gaia} parallaxes will be extremely important for the identification and characterisation of WDs. We provided estimates of the precision in WD parallaxes that {\Gaia} will derive, obtaining that about $95\%$ of WDs will have parallaxes better than $10\%$. For cool WDs ({\teff}$<5000$~K) they will have parallaxes better than $5\%$, and about $2000$ of them will have parallaxes better than $1\%$. 

Additional photometry or/and spectroscopic follow-up might be necessary to achieve a better accuracy on the atmospheric parameters. A comparison of the masses obtained from the {\Gaia} parallaxes with those determined from spectroscopic fits will allow testing the mass-radius relations for WDs. 
A better characterisation of the coolest WDs will also be possible since it will help to resolve the
discrepancy regarding the H/He atmospheric composition of these WDs that exist in the
literature \citep{kowalski06,kil09,kil10}. In addition,, the orbital
  solutions derived for the WDs detected in binary systems will provide
  independent mass determinations for them, and therefore will allow for
  stringent tests of the atmosphere models. This will improve the stellar population ages derived by means of the WD cosmochronology and our understanding of the stellar evolution.

\begin{acknowledgements}
      J.M.~Carrasco, C.~Jordi and X.~Luri were supported by the MINECO (Spanish Ministry of Economy) - FEDER through grant AYA2009-14648-C02-01, AYA2010-12176-E, AYA2012-39551-C02-01 and CONSOLIDER CSD2007-00050. GUMS simulations have been performed in the supercomputer MareNostrum at Barcelona Supercomputing
Center - Centro Nacional de Supercomputaci\'on (The Spanish National Supercomputing Center). 
          S.~Catal\'an acknowledges financial support from the European Commission in the form of a Marie Curie Intra 
European Fellowship (PIEF-GA-2009-237718).
      P.-E.~Tremblay was supported by the {\it Alexander von Humboldt Stiftung}.
        We would also like to thank F. Arenou and C. Reyl\'e for their comments on GUMS simulations that helped us to understand the results and the ingredients of the Galaxy model.
\end{acknowledgements}

\bibliographystyle{aa} 
\bibliography{bibliography} 

\end{document}